\documentclass{article} 
\usepackage{icais2025_conference,times}

\usepackage{amsmath,amsfonts,bm}

\def\eqref#1{equation~\ref{#1}}

\def\1{\bm{1}}

\DeclareMathAlphabet{\mathsfit}{\encodingdefault}{\sfdefault}{m}{sl}
\SetMathAlphabet{\mathsfit}{bold}{\encodingdefault}{\sfdefault}{bx}{n}

\usepackage{xurl}
\usepackage{hyperref}
\hypersetup{
    colorlinks=true,
    linkcolor=blue!70!black,
    citecolor=green!60!black,
    urlcolor=red!60!black
}
\usepackage{url}

\usepackage{xspace}
\usepackage{graphicx}
\usepackage{booktabs}
\usepackage{dirtytalk}
\usepackage{multirow}
\usepackage{longtable}
\usepackage{adjustbox}
\usepackage{tikz}
\usetikzlibrary{patterns}
\usepgflibrary{patterns}
\usepackage{pgfplots}
\usepackage{pgfplotstable}
\pgfplotsset{compat=1.18}
\usepackage{listings}
\usepackage{amssymb}
\usepackage[most]{tcolorbox}
\usepackage{fontawesome5}
\usepackage{minted}  
\definecolor{promptbg}{RGB}{248,249,250}
\definecolor{promptframe}{RGB}{33,82,126}
\definecolor{prompttitle}{RGB}{23,42,58}
\definecolor{sectionblue}{RGB}{30,58,95}

\newtcolorbox{llmprompt}[1]{
    breakable,
    enhanced,
    colback=promptbg,
    colframe=promptframe,
    coltitle=prompttitle,
    fonttitle=\bfseries\small,
    title={\faIcon{robot}\hspace{0.5em}#1},
    left=1.5em, right=1.5em, top=1em, bottom=1em,
    arc=4pt, 
    boxrule=0.8pt,
    titlerule=0pt,
    toptitle=0.8em, bottomtitle=0.6em,
    before upper={\setlength{\parindent}{0pt}},
    attach boxed title to top center={yshift=-3mm},
    boxed title style={
        colback=white,
        colframe=promptframe,
        boxrule=0.8pt,
        arc=4pt
    }
}

\newcommand{\promptsection}[1]{\vspace{0.8em}\noindent{\color{sectionblue}\faIcon{angle-right}\hspace{0.3em}\textbf{#1}}\vspace{0.4em}\par}

\definecolor{tcolorboxblue}{rgb}{0.2,0.3,0.6}
\definecolor{tcolorboxpink}{rgb}{1.0,0.75,0.8}
\definecolor{democraticblue}{RGB}{0,174,243}
\definecolor{republicanred}{RGB}{232,27,35}

\colorlet{jsonPunct}{red!60!black}
\definecolor{jsonBackground}{HTML}{EEEEEE}
\definecolor{jsonDelim}{RGB}{20,105,176}
\colorlet{jsonNumb}{magenta!60!black}

\lstdefinelanguage{json}{
    basicstyle=\normalfont\ttfamily,
    showstringspaces=false,
    breaklines=true,
    frame=lines,
    backgroundcolor=\color{jsonBackground!29},
    literate=
     *{0}{{{\color{jsonNumb}0}}}{1}
      {1}{{{\color{jsonNumb}1}}}{1}
      {2}{{{\color{jsonNumb}2}}}{1}
      {3}{{{\color{jsonNumb}3}}}{1}
      {4}{{{\color{jsonNumb}4}}}{1}
      {5}{{{\color{jsonNumb}5}}}{1}
      {6}{{{\color{jsonNumb}6}}}{1}
      {7}{{{\color{jsonNumb}7}}}{1}
      {8}{{{\color{jsonNumb}8}}}{1}
      {9}{{{\color{jsonNumb}9}}}{1}
      {`}{{{\color{jsonNumb}`}}}{1}
      {'}{{{\color{jsonNumb}'}}}{1}
      {"}{{{\color{jsonNumb}"}}}{1}
      {:}{{{\color{jsonPunct}{:}}}}{1}
      {,}{{{\color{jsonPunct}{,}}}}{1}
      {\{}{{{\color{jsonDelim}{\{}}}}{1}
      {\}}{{{\color{jsonDelim}{\}}}}}{1}
      {[}{{{\color{jsonDelim}{[}}}}{1}
      {]}{{{\color{jsonDelim}{]}}}}{1},
}

\newcommand{\ourApproach}{FlockVote\xspace}

\tikzset{USA map/.cd,
state/.style={fill, draw=white, ultra thick},
HI/.style={}, AK/.style={}, FL/.style={}, NH/.style={}, MI/.style={}, MI/.style={}, VT/.style={}, ME/.style={}, RI/.style={}, NY/.style={}, PA/.style={}, NJ/.style={}, DE/.style={}, MD/.style={}, VA/.style={}, WV/.style={}, OH/.style={}, IN/.style={}, IL/.style={}, CT/.style={}, WI/.style={}, NC/.style={}, DC/.style={}, MA/.style={}, TN/.style={}, AR/.style={}, MO/.style={}, GA/.style={}, SC/.style={}, KY/.style={}, AL/.style={}, LA/.style={}, MS/.style={}, IA/.style={}, MN/.style={}, OK/.style={}, TX/.style={}, NM/.style={}, KS/.style={}, NE/.style={}, SD/.style={}, ND/.style={}, WY/.style={}, MT/.style={}, CO/.style={}, ID/.style={}, UT/.style={}, AZ/.style={}, NV/.style={}, OR/.style={}, WA/.style={}, CA/.style={}}

\tikzset{
every state/.style={USA map/state/.style={#1}},
HI/.style={USA map/HI/.style={#1}}, AK/.style={USA map/AK/.style={#1}}, FL/.style={USA map/FL/.style={#1}}, NH/.style={USA map/NH/.style={#1}}, MI/.style={USA map/MI/.style={#1}}, VT/.style={USA map/VT/.style={#1}}, ME/.style={USA map/ME/.style={#1}}, RI/.style={USA map/RI/.style={#1}}, NY/.style={USA map/NY/.style={#1}}, PA/.style={USA map/PA/.style={#1}}, NJ/.style={USA map/NJ/.style={#1}}, DE/.style={USA map/DE/.style={#1}}, MD/.style={USA map/MD/.style={#1}}, VA/.style={USA map/VA/.style={#1}}, WV/.style={USA map/WV/.style={#1}}, OH/.style={USA map/OH/.style={#1}}, IN/.style={USA map/IN/.style={#1}}, IL/.style={USA map/IL/.style={#1}}, CT/.style={USA map/CT/.style={#1}}, WI/.style={USA map/WI/.style={#1}}, NC/.style={USA map/NC/.style={#1}}, DC/.style={USA map/DC/.style={#1}}, MA/.style={USA map/MA/.style={#1}}, TN/.style={USA map/TN/.style={#1}}, AR/.style={USA map/AR/.style={#1}}, MO/.style={USA map/MO/.style={#1}}, GA/.style={USA map/GA/.style={#1}}, SC/.style={USA map/SC/.style={#1}}, KY/.style={USA map/KY/.style={#1}}, AL/.style={USA map/AL/.style={#1}}, LA/.style={USA map/LA/.style={#1}}, MS/.style={USA map/MS/.style={#1}}, IA/.style={USA map/IA/.style={#1}}, MN/.style={USA map/MN/.style={#1}}, OK/.style={USA map/OK/.style={#1}}, TX/.style={USA map/TX/.style={#1}}, NM/.style={USA map/NM/.style={#1}}, KS/.style={USA map/KS/.style={#1}}, NE/.style={USA map/NE/.style={#1}}, SD/.style={USA map/SD/.style={#1}}, ND/.style={USA map/ND/.style={#1}}, WY/.style={USA map/WY/.style={#1}}, MT/.style={USA map/MT/.style={#1}}, CO/.style={USA map/CO/.style={#1}}, ID/.style={USA map/ID/.style={#1}}, UT/.style={USA map/UT/.style={#1}}, AZ/.style={USA map/AZ/.style={#1}}, NV/.style={USA map/NV/.style={#1}}, OR/.style={USA map/OR/.style={#1}}, WA/.style={USA map/WA/.style={#1}}, CA/.style={USA map/CA/.style={#1}}
}

\newcommand{\USA}[1][]{
    \begin{scope}[y=0.80pt,x=0.80pt,yscale=-1, inner sep=0pt, outer sep=0pt,
    #1
    ]

    \path[USA map/state, USA map/HI, local bounding box=HI] (233.0875,519.3095) -- (235.0274,515.7529) -- (237.2907,515.4296) --
      (237.6140,516.2379) -- (235.5124,519.3095) -- (233.0875,519.3095) --
      cycle(243.2722,515.5913) -- (249.4153,518.1778) -- (251.5169,517.8545) --
      (253.1335,513.9747) -- (252.4869,510.5798) -- (248.2837,510.0948) --
      (244.2421,511.8731) -- (243.2722,515.5913) -- cycle(273.9878,525.6143) --
      (277.7060,531.1107) -- (280.1309,530.7874) -- (281.2625,530.3024) --
      (282.7175,531.5957) -- (286.4357,531.4341) -- (287.4057,529.9791) --
      (284.4958,528.2009) -- (282.5558,524.4826) -- (280.4542,520.9261) --
      (274.6344,523.8360) -- (273.9878,525.6143) -- cycle(294.1954,534.5056) --
      (295.4887,532.5657) -- (300.1769,533.5357) -- (300.8236,533.0507) --
      (306.9667,533.6973) -- (306.6434,534.9906) -- (304.0568,536.4456) --
      (299.6919,536.1222) -- (294.1954,534.5056) -- cycle(299.5303,539.6788) --
      (301.4702,543.5587) -- (304.5418,542.4270) -- (304.8651,540.8104) --
      (303.2485,538.7088) -- (299.5303,538.3855) -- (299.5303,539.6788) --
      cycle(306.4817,538.5472) -- (308.7450,535.6373) -- (313.4331,538.0622) --
      (317.7980,539.1938) -- (322.1628,541.9421) -- (322.1628,543.8820) --
      (318.6063,545.6603) -- (313.7565,546.6302) -- (311.3315,545.1753) --
      (306.4817,538.5472) -- cycle(323.1328,554.0666) -- (324.7494,552.7734) --
      (328.1443,554.3900) -- (335.7424,557.9465) -- (339.1373,560.0481) --
      (340.7539,562.4730) -- (342.6938,566.8379) -- (346.7353,569.4245) --
      (346.4120,570.7178) -- (342.5321,573.9510) -- (338.3290,575.4059) --
      (336.8740,574.7593) -- (333.8024,576.5375) -- (331.3775,579.7708) --
      (329.1143,582.6807) -- (327.3360,582.5190) -- (323.7794,579.9324) --
      (323.4561,575.4059) -- (324.1028,572.9810) -- (322.4862,567.3229) --
      (320.3846,565.5446) -- (320.2229,562.9580) -- (322.4862,561.9880) --
      (324.5878,558.9165) -- (325.0727,557.9465) -- (323.4561,556.1682) --
      (323.1328,554.0666) -- cycle;

    \path[USA map/state, USA map/AK, local bounding box=AK] (158.0767,453.6750) -- (157.7534,539.0322) -- (159.3700,540.0021) --
      (162.4416,540.1638) -- (163.8965,539.0322) -- (166.4831,539.0322) --
      (166.6447,541.9420) -- (173.5962,548.7318) -- (174.0812,551.3184) --
      (177.4760,549.3785) -- (178.1227,549.2168) -- (178.4460,546.1452) --
      (179.9010,544.5286) -- (181.0326,544.3670) -- (182.9725,542.9120) --
      (186.0441,545.0136) -- (186.6907,547.9235) -- (188.6307,549.0551) --
      (189.7623,551.4801) -- (193.6422,553.2583) -- (197.0371,559.2398) --
      (199.7853,563.1197) -- (202.0486,565.8679) -- (203.5035,569.5861) --
      (208.5150,571.3644) -- (213.6882,573.4660) -- (214.6581,577.8308) --
      (215.1431,580.9024) -- (214.1732,584.2973) -- (212.3949,586.5605) --
      (210.7783,585.7522) -- (209.3233,582.6807) -- (206.5751,581.2257) --
      (204.7968,580.0941) -- (203.9885,580.9024) -- (205.4434,583.6507) --
      (205.6051,587.3689) -- (204.4735,587.8538) -- (202.5335,585.9139) --
      (200.4320,584.6206) -- (200.9169,586.2372) -- (202.2102,588.0155) --
      (201.4019,588.8238) .. controls (201.4019,588.8238) and (200.5936,588.5005) ..
      (200.1086,587.8538) .. controls (199.6236,587.2072) and (198.0070,584.4590) ..
      (198.0070,584.4590) -- (197.0371,582.1957) .. controls (197.0371,582.1957) and
      (196.7137,583.4890) .. (196.0671,583.1657) .. controls (195.4204,582.8423) and
      (194.7738,581.7107) .. (194.7738,581.7107) -- (196.5521,579.7708) --
      (195.0971,578.3158) -- (195.0971,573.3043) -- (194.2888,573.3043) --
      (193.4805,576.6992) -- (192.3489,577.1842) -- (191.3789,573.4660) --
      (190.7323,569.7478) -- (189.9240,569.2628) -- (190.2473,574.9209) --
      (190.2473,576.0526) -- (188.7923,574.7593) -- (185.2358,568.7778) --
      (183.1342,568.2928) -- (182.4876,564.5746) -- (180.8709,561.6647) --
      (179.2543,560.5331) -- (179.2543,558.2698) -- (181.3559,556.9765) --
      (180.8709,556.6532) -- (178.2844,557.2999) -- (174.8895,554.8750) --
      (172.3029,551.9650) -- (167.4531,549.3785) -- (163.4115,546.7919) --
      (164.7048,543.5587) -- (164.7048,541.9421) -- (162.9265,543.5587) --
      (160.0166,544.6903) -- (156.2984,543.5587) -- (150.6403,541.1338) --
      (145.1438,541.1338) -- (144.4972,541.6187) -- (138.0307,537.7389) --
      (135.9291,537.4155) -- (133.1809,531.5957) -- (129.6243,531.9191) --
      (126.0678,533.3740) -- (126.5528,537.9005) -- (127.6844,534.9906) --
      (128.6544,535.3139) -- (127.1994,539.6788) -- (130.4326,536.9306) --
      (131.0793,538.5472) -- (127.1994,542.9120) -- (125.9061,542.5887) --
      (125.4211,540.6488) -- (124.1279,539.8405) -- (122.8346,540.9721) --
      (120.0863,539.1938) -- (117.0148,541.2954) -- (115.2365,543.3970) --
      (111.8416,545.4986) -- (107.1534,545.3369) -- (106.6684,543.2353) --
      (110.3866,542.5887) -- (110.3866,541.2954) -- (108.1234,540.6488) --
      (109.0934,538.2238) -- (111.3566,534.3440) -- (111.3566,532.5657) --
      (111.5183,531.7574) -- (115.8831,529.4941) -- (116.8531,530.7874) --
      (119.6013,530.7874) -- (118.3081,528.2009) -- (114.5898,527.8775) --
      (109.5783,530.6258) -- (107.1534,534.0206) -- (105.3752,536.6072) --
      (104.2435,538.8705) -- (100.0403,540.3254) -- (96.9688,542.9120) --
      (96.6454,544.5286) -- (98.9087,545.4986) -- (99.7170,547.6002) --
      (96.9688,550.8334) -- (90.5023,555.0366) -- (82.7426,559.2398) --
      (80.6410,560.3714) -- (75.3062,561.5031) -- (69.9713,563.7663) --
      (71.7496,565.0596) -- (70.2947,566.5146) -- (69.8097,567.6462) --
      (67.0614,566.6762) -- (63.8282,566.8379) -- (63.0199,569.1011) --
      (62.0499,569.1011) -- (62.3733,566.6762) -- (58.8167,567.9695) --
      (55.9068,568.9395) -- (52.5119,567.6462) -- (49.6020,569.5861) --
      (46.3688,569.5861) -- (44.2672,570.8794) -- (42.6506,571.6877) --
      (40.5490,571.3644) -- (37.9624,570.2328) -- (35.6992,570.8794) --
      (34.7292,571.8494) -- (33.1126,570.7178) -- (33.1126,568.7778) --
      (36.1841,567.4845) -- (42.4889,568.1312) -- (46.8538,566.5146) --
      (48.9554,564.4130) -- (51.8653,563.7663) -- (53.6436,562.9580) --
      (56.3918,563.1197) -- (58.0084,564.4130) -- (58.9784,564.0896) --
      (61.2416,561.3414) -- (64.3132,560.3714) -- (67.7081,559.7248) --
      (69.0014,559.4015) -- (69.6480,559.8864) -- (70.4563,559.8864) --
      (71.7496,556.1682) -- (75.7911,554.7133) -- (77.7311,550.9951) --
      (79.9943,546.4686) -- (81.6110,545.0136) -- (81.9343,542.4270) --
      (80.3177,543.7203) -- (76.9228,544.3670) -- (76.2761,541.9421) --
      (74.9828,541.6187) -- (74.0129,542.5887) -- (73.8512,545.4986) --
      (72.3963,545.3369) -- (70.9413,539.5171) -- (69.6480,540.8104) --
      (68.5164,540.3254) -- (68.1931,538.3855) -- (64.1515,538.5472) --
      (62.0499,539.6788) -- (59.4634,539.3555) -- (60.9183,537.9005) --
      (61.4033,535.3139) -- (60.7566,533.3740) -- (62.2116,532.4040) --
      (63.5049,532.2424) -- (62.8582,530.4641) -- (62.8582,526.0993) --
      (61.8883,525.1293) -- (61.0800,526.5842) -- (54.9368,526.5842) --
      (53.4819,525.2909) -- (52.8352,521.4111) -- (50.7337,517.8545) --
      (50.7337,516.8846) -- (52.8352,516.0763) -- (52.9969,513.9747) --
      (54.1285,512.8430) -- (53.3202,512.3581) -- (52.0269,512.8430) --
      (50.8953,510.0948) -- (51.8653,505.0833) -- (56.3918,501.8501) --
      (58.9784,500.2335) -- (60.9183,496.5153) -- (63.6666,495.2220) --
      (66.2531,496.3536) -- (66.5765,498.7785) -- (69.0014,498.4552) --
      (72.2346,496.0303) -- (73.8512,496.6769) -- (74.8212,497.3236) --
      (76.4378,497.3236) -- (78.7010,496.0303) -- (79.5094,491.6654) .. controls
      (79.5094,491.6654) and (79.8327,488.7555) .. (80.4793,488.2705) .. controls
      (81.1260,487.7855) and (81.4493,487.3006) .. (81.4493,487.3006) --
      (80.3177,485.3606) -- (77.7311,486.1689) -- (74.4978,486.9772) --
      (72.5579,486.4923) -- (69.0014,484.7140) -- (63.9899,484.5523) --
      (60.4333,480.8341) -- (60.9183,476.9542) -- (61.5650,474.5293) --
      (59.4634,472.7511) -- (57.5234,469.0328) -- (58.0084,468.2245) --
      (64.7982,467.7396) -- (66.8998,467.7396) -- (67.8697,468.7095) --
      (68.5164,468.7095) -- (68.3547,467.0929) -- (72.2346,466.4463) --
      (74.8212,466.7696) -- (76.2761,467.9012) -- (74.8212,470.0028) --
      (74.3362,471.4578) -- (77.0844,473.0744) -- (82.0959,474.8526) --
      (83.8742,473.8827) -- (81.6110,469.5178) -- (80.6410,466.2846) --
      (81.6110,465.4763) -- (78.2161,463.5364) -- (77.7311,462.4047) --
      (78.2161,460.7881) -- (77.4078,456.9083) -- (74.4978,452.2201) --
      (72.0729,448.0169) -- (74.9828,446.0769) -- (78.2161,446.0769) --
      (79.9943,446.7236) -- (84.1975,446.5619) -- (87.9157,443.0054) --
      (89.0474,439.9338) -- (92.7656,437.5089) -- (94.3822,438.4789) --
      (97.1304,437.8322) -- (100.8486,435.7306) -- (101.9803,435.5690) --
      (102.9502,436.3773) -- (107.4767,436.2156) -- (110.2250,433.1441) --
      (111.3566,433.1441) -- (114.9132,435.5690) -- (116.8531,437.6706) --
      (116.3681,438.8022) -- (117.0148,439.9338) -- (118.6314,438.3172) --
      (122.5112,438.6405) -- (122.8346,442.3587) -- (124.7745,443.8137) --
      (131.8876,444.4603) -- (138.1924,448.6635) -- (139.6473,447.6936) --
      (144.8205,450.2801) -- (146.9221,449.6335) -- (148.8620,448.8252) --
      (153.7119,450.7651) -- (158.0767,453.6750) -- cycle(42.9739,482.6124) --
      (45.0755,487.9472) -- (44.9138,488.9172) -- (42.0039,488.5938) --
      (40.2257,484.5523) -- (38.4474,483.0974) -- (36.0225,483.0974) --
      (35.8608,480.5108) -- (37.6391,478.0859) -- (38.7707,480.5108) --
      (40.2257,481.9657) -- (42.9739,482.6124) -- cycle(40.3873,516.0763) --
      (44.1055,516.8846) -- (47.8237,517.8545) -- (48.6321,518.8245) --
      (47.0154,522.5427) -- (43.9439,522.3810) -- (40.5490,518.8245) --
      (40.3873,516.0763) -- cycle(19.6947,502.0117) -- (20.8263,504.5983) --
      (21.9580,506.2149) -- (20.8263,507.0232) -- (18.7247,503.9517) --
      (18.7247,502.0117) -- (19.6947,502.0117) -- cycle(5.9535,575.0826) --
      (9.3484,572.8193) -- (12.7433,571.8494) -- (15.3298,572.1727) --
      (15.8148,573.7893) -- (17.7548,574.2743) -- (19.6947,572.3344) --
      (19.3714,570.7178) -- (22.1196,570.0711) -- (25.0295,572.6577) --
      (23.8979,574.4360) -- (19.5330,575.5676) -- (16.7848,575.0826) --
      (13.0666,573.9510) -- (8.7017,575.4059) -- (7.0851,575.7292) --
      (5.9535,575.0826) -- cycle(54.9368,570.5561) -- (56.5535,572.4960) --
      (58.6550,570.8794) -- (57.2001,569.5861) -- (54.9368,570.5561) --
      cycle(57.8467,573.6276) -- (58.9784,571.3644) -- (61.0800,571.6877) --
      (60.2717,573.6276) -- (57.8467,573.6276) -- cycle(81.4493,571.6877) --
      (82.9042,573.4660) -- (83.8742,572.3344) -- (83.0659,570.3944) --
      (81.4493,571.6877) -- cycle(90.1790,559.2398) -- (91.3106,565.0596) --
      (94.2205,565.8679) -- (99.2320,562.9580) -- (103.5969,560.3714) --
      (101.9803,557.9465) -- (102.4652,555.5216) -- (100.3636,556.8149) --
      (97.4538,556.0066) -- (99.0704,554.8750) -- (101.0103,555.6833) --
      (104.8902,553.9050) -- (105.3751,552.4500) -- (102.9502,551.6417) --
      (103.7585,549.7018) -- (101.0103,551.6417) -- (96.3221,555.1983) --
      (91.4723,558.1082) -- (90.1790,559.2398) -- cycle(132.5342,539.3555) --
      (134.9592,537.9005) -- (133.9892,536.1222) -- (132.2109,537.0922) --
      (132.5342,539.3555) -- cycle;

    \path[USA map/state, USA map/FL, local bounding box=FL] (759.8167,439.1428) -- (762.0824,446.4614) -- (765.8121,456.2037) --
      (771.1468,465.5800) -- (774.8650,471.8847) -- (779.7149,477.3812) --
      (783.7564,481.0994) -- (785.3730,484.0093) -- (784.2414,485.3025) --
      (783.4330,486.5958) -- (786.3429,494.0322) -- (789.2528,496.9421) --
      (791.8394,502.2769) -- (795.3959,508.0967) -- (799.9224,516.3413) --
      (801.2157,523.9394) -- (801.7007,535.9023) -- (802.3473,537.6805) --
      (802.0240,541.0754) -- (799.5991,542.3687) -- (799.9224,544.3086) --
      (799.2758,546.2485) -- (799.5991,548.6734) -- (800.0841,550.6134) --
      (797.3358,553.8466) -- (794.2643,555.3015) -- (790.3844,555.4632) --
      (788.9295,557.0798) -- (786.5046,558.0497) -- (785.2113,557.5648) --
      (784.0797,556.5948) -- (783.7564,553.6849) -- (782.9481,550.2900) --
      (779.5532,545.1169) -- (775.9967,542.8537) -- (772.1168,542.5303) --
      (771.3085,543.8236) -- (768.2370,539.4588) -- (767.5903,535.9023) --
      (765.0037,531.8608) -- (763.2255,530.7291) -- (761.6089,532.8307) --
      (759.8306,532.5074) -- (757.7290,527.4959) -- (754.8191,523.6161) --
      (751.9092,518.2813) -- (749.3227,515.2097) -- (745.7662,511.4915) --
      (747.8677,509.0666) -- (751.1009,503.5702) -- (750.9393,501.9536) --
      (746.4128,500.9836) -- (744.7962,501.6302) -- (745.1195,502.2769) --
      (747.7061,503.2468) -- (746.2511,507.7733) -- (745.4428,508.2583) --
      (743.6646,504.2168) -- (742.3713,499.3670) -- (742.0480,496.6188) --
      (743.5029,491.9306) -- (743.5029,482.3927) -- (740.4314,478.6745) --
      (739.1381,475.6029) -- (733.9649,474.3096) -- (732.0250,473.6630) --
      (730.4084,471.0764) -- (727.0135,469.4598) -- (725.8819,466.0649) --
      (723.1337,465.0950) -- (720.7088,461.3768) -- (716.5056,459.9219) --
      (713.5957,458.4669) -- (711.0092,458.4669) -- (706.9676,459.2752) --
      (706.8060,461.2151) -- (707.6143,462.1851) -- (707.1293,463.3167) --
      (704.0578,463.1551) -- (700.3396,466.7116) -- (696.7830,468.6515) --
      (692.9032,468.6515) -- (689.6700,469.9448) -- (689.3466,467.1966) --
      (687.7300,465.2566) -- (684.8202,464.1250) -- (683.2036,462.6701) --
      (675.1205,458.7902) -- (667.5225,457.0120) -- (663.1577,457.6586) --
      (657.1762,458.1436) -- (651.1948,460.2452) -- (647.7155,460.8581) --
      (647.4776,452.8084) -- (644.8910,450.8685) -- (643.1128,449.0902) --
      (643.4361,446.0186) -- (653.6207,444.7254) -- (679.1631,441.8155) --
      (685.9529,441.1688) -- (691.3889,441.4491) -- (693.9754,445.3290) --
      (695.4304,446.7839) -- (703.5285,447.2991) -- (714.3483,446.6525) --
      (735.8607,445.3592) -- (741.3064,444.6848) -- (746.4140,444.8893) --
      (746.8408,447.7992) -- (749.0738,448.6075) -- (749.3087,443.9775) --
      (747.7805,439.8046) -- (749.0889,438.3647) -- (754.6436,438.8195) --
      (759.8167,439.1428) -- cycle(772.3621,571.5479) -- (774.7870,570.9012) --
      (776.0803,570.6588) -- (777.5353,568.3147) -- (779.8793,566.6980) --
      (781.1726,567.1830) -- (782.8701,567.5064) -- (783.2742,568.5571) --
      (779.7985,569.7696) -- (775.5953,571.2246) -- (773.2512,572.4370) --
      (772.3621,571.5479) -- cycle(785.8608,566.5364) -- (787.0733,567.5872) --
      (789.8215,565.4856) -- (795.1563,561.2824) -- (798.8745,557.4025) --
      (801.3803,550.7744) -- (802.3502,549.0770) -- (802.5119,545.6821) --
      (801.7844,546.1671) -- (800.8145,548.9962) -- (799.3595,553.6035) --
      (796.1263,558.8575) -- (791.7614,563.0607) -- (788.3666,565.0006) --
      (785.8608,566.5364) -- cycle;

    \path[USA map/state, USA map/NH, local bounding box=NH] (880.7990,142.4248) -- (881.6680,141.3483) -- (882.7582,138.0572) --
      (880.2152,137.1438) -- (879.7302,134.0722) -- (875.8503,132.9406) --
      (875.5270,130.1923) -- (868.2523,106.7515) -- (863.6508,92.2085) --
      (862.7538,92.2034) -- (862.1071,93.8200) -- (861.4605,93.3351) --
      (860.4905,92.3651) -- (859.0356,94.3050) -- (858.9871,99.3371) --
      (859.2987,105.0043) -- (861.2386,107.7525) -- (861.2386,111.7941) --
      (857.5204,116.8568) -- (854.9339,117.9885) -- (854.9339,119.1201) --
      (856.0655,120.8984) -- (856.0655,129.4664) -- (855.2572,138.6811) --
      (855.0955,143.5309) -- (856.0655,144.8242) -- (855.9038,149.3507) --
      (855.4188,151.1289) -- (856.3876,151.8382) -- (873.1753,147.4136) --
      (875.3502,146.8112) -- (877.1938,144.0378) -- (880.7990,142.4247) -- cycle;

    \begin{scope}
    \path[USA map/state, USA map/MI, local bounding box=MI] (697.8601,177.2369) -- (694.6269,168.9922)
        -- (692.3636,159.9392) -- (689.9387,156.7060) -- (687.3521,154.9277) --
        (685.7355,156.0594) -- (681.8557,157.8376) -- (679.9158,162.8491) --
        (677.1675,166.5673) -- (676.0359,167.2139) -- (674.5810,166.5673) .. controls
        (674.5810,166.5673) and (671.9944,165.1123) .. (672.1561,164.4657) .. controls
        (672.3177,163.8191) and (672.6410,159.4542) .. (672.6410,159.4542) --
        (676.0359,158.1609) -- (676.8442,154.7661) -- (677.4908,152.1795) --
        (679.9158,150.5629) -- (679.5924,140.5400) -- (677.9758,138.2767) --
        (676.6825,137.4684) -- (675.8742,135.3668) -- (676.6825,134.5585) --
        (678.2991,134.8818) -- (678.4608,133.2652) -- (676.0359,131.0020) --
        (674.7426,128.4154) -- (672.1561,128.4154) -- (667.6296,126.9605) --
        (662.1331,123.5656) -- (659.3849,123.5656) -- (658.7382,124.2123) --
        (657.7683,123.7273) -- (654.6967,121.4640) -- (651.7868,123.2423) --
        (648.8769,125.5055) -- (649.2003,129.0621) -- (650.1702,129.3854) --
        (652.2718,129.8704) -- (652.7568,130.6787) -- (650.1702,131.4870) --
        (647.5837,131.8103) -- (646.1287,133.5886) -- (645.8054,135.6901) --
        (646.1287,137.3067) -- (646.4520,142.8032) -- (642.8955,144.9048) --
        (642.2489,144.7431) -- (642.2489,140.5400) -- (643.5421,138.1151) --
        (644.1888,135.6901) -- (643.3805,134.8818) -- (641.4406,135.6901) --
        (640.4706,139.8933) -- (637.7224,141.0249) -- (635.9441,142.9649) --
        (635.7824,143.9348) -- (636.4291,144.7431) -- (635.7824,147.3297) --
        (633.5192,147.8147) -- (633.5192,148.9463) -- (634.3275,151.3712) --
        (633.1959,157.5143) -- (631.5793,161.5558) -- (632.2259,166.2440) --
        (632.7109,167.3756) -- (631.9026,169.8005) -- (631.5793,170.6088) --
        (631.2560,173.3570) -- (634.8125,179.3385) -- (637.7224,185.8049) --
        (639.1773,190.6547) -- (638.3690,195.3429) -- (637.3991,201.3243) --
        (634.9741,206.4974) -- (634.6508,209.2457) -- (631.3920,212.3308) --
        (635.8006,212.1688) -- (657.2191,209.9055) -- (664.4969,208.9184) --
        (664.5933,210.5848) -- (671.4452,209.3723) -- (681.7433,207.8692) --
        (685.5975,207.4083) -- (685.7356,206.8207) -- (685.8972,205.3658) --
        (687.9988,201.6476) -- (689.9994,199.9098) -- (689.7771,194.8579) --
        (691.3741,193.2609) -- (692.4647,192.9179) -- (692.6870,189.3614) --
        (694.2227,186.3303) -- (695.2735,186.9365) -- (695.4352,187.5832) --
        (696.2435,187.7448) -- (698.1834,186.7749) -- (697.8601,177.2369) -- cycle;

      \path[USA map/state, USA map/MI, local bounding box=MI2] (581.6193,82.0590) -- (583.4483,80.0014) --
        (585.6202,79.2012) -- (590.9929,75.3146) -- (593.2791,74.7431) --
        (593.7363,75.2003) -- (588.5923,80.3443) -- (585.2773,82.2876) --
        (583.2197,83.2021) -- (581.6193,82.0590) -- cycle(667.7937,114.1872) --
        (668.4403,116.6929) -- (671.6736,116.8546) -- (672.9668,115.6421) .. controls
        (672.9668,115.6421) and (672.8860,114.1872) .. (672.5627,114.0255) .. controls
        (672.2394,113.8639) and (670.9461,112.1664) .. (670.9461,112.1664) --
        (668.7637,112.4089) -- (667.1470,112.5706) -- (666.8237,113.7022) --
        (667.7937,114.1872) -- cycle(567.4921,111.2132) -- (568.2084,110.6328) --
        (570.9566,109.8245) -- (574.5131,107.5612) -- (574.5131,106.5913) --
        (575.1598,105.9446) -- (581.1412,104.9747) -- (583.5661,103.0347) --
        (587.9310,100.9331) -- (588.0926,99.6399) -- (590.0325,96.7300) --
        (591.8108,95.9217) -- (593.1041,94.1434) -- (595.3673,91.8802) --
        (599.7322,89.4553) -- (604.4203,88.9703) -- (605.5519,90.1019) --
        (605.2286,91.0719) -- (601.5104,92.0418) -- (600.0555,95.1134) --
        (597.7922,95.9217) -- (597.3073,98.3466) -- (594.8824,101.5798) --
        (594.5590,104.1664) -- (595.3673,104.6513) -- (596.3373,103.5197) --
        (599.8938,100.6098) -- (601.1871,101.9031) -- (603.4504,101.9031) --
        (606.6836,102.8731) -- (608.1385,104.0047) -- (609.5934,107.0762) --
        (612.3417,109.8245) -- (616.2215,109.6628) -- (617.6765,108.6928) --
        (619.2931,109.9861) -- (620.9097,110.4711) -- (622.2030,109.6628) --
        (623.3346,109.6628) -- (624.9512,108.6928) -- (628.9927,105.1363) --
        (632.3876,104.0047) -- (639.0157,103.6814) -- (643.5421,101.7414) --
        (646.1287,100.4482) -- (647.5837,100.6098) -- (647.5837,106.2679) --
        (648.0687,106.5913) -- (650.9785,107.3996) -- (652.9185,106.9146) --
        (659.0616,105.2980) -- (660.1932,104.1664) -- (661.6481,104.6513) --
        (661.6481,111.6027) -- (664.8813,114.6743) -- (666.1746,115.3209) --
        (667.4679,116.2909) -- (666.1746,116.6142) -- (665.3663,116.2909) --
        (661.6481,115.8059) -- (659.5465,116.4526) -- (657.2833,116.2909) --
        (654.0501,117.7458) -- (652.2718,117.7458) -- (646.4520,116.4526) --
        (641.2789,116.6142) -- (639.3390,119.2008) -- (632.3876,119.8474) --
        (629.9627,120.6557) -- (628.8311,123.7273) -- (627.5378,124.8589) --
        (627.0528,124.6972) -- (625.5978,123.0806) -- (621.0714,125.5055) --
        (620.4247,125.5055) -- (619.2931,123.8889) -- (618.4848,124.0506) --
        (616.5449,128.4154) -- (615.5749,132.4569) -- (612.3938,139.4577) --
        (611.2170,138.4235) -- (609.8453,137.3922) -- (607.9045,127.1041) --
        (604.3600,125.7341) -- (602.3074,123.4479) -- (590.1871,120.7044) --
        (587.3318,119.6747) -- (579.1014,117.5020) -- (571.2114,116.3589) --
        (567.4921,111.2132) -- cycle;

    \end{scope}
    \path[USA map/state, USA map/VT, local bounding box=VT] (844.4842,154.0579) -- (844.8009,148.7123) -- (841.9101,137.9281) --
      (841.2635,137.6048) -- (838.3536,136.3115) -- (839.1619,133.4016) --
      (838.3536,131.3000) -- (835.6536,126.6600) -- (836.6235,122.7802) --
      (835.8152,117.6070) -- (833.3903,111.1406) -- (832.5847,106.2181) --
      (859.0041,99.4863) -- (859.3128,105.0085) -- (861.2291,107.7507) --
      (861.2291,111.7922) -- (857.5219,116.8502) -- (854.9353,117.9929) --
      (854.9243,119.1135) -- (856.2343,120.6326) -- (855.9234,128.7305) --
      (855.3139,137.9894) -- (855.0860,143.5463) -- (856.0560,144.8396) --
      (855.8943,149.4103) -- (855.4093,151.1002) -- (856.4235,151.8274) --
      (848.9860,153.3341) -- (844.4842,154.0579) -- cycle;

    \path[USA map/state, USA map/ME, local bounding box=ME] (922.8398,78.8307) -- (924.7797,80.9323) -- (927.0429,84.6505) --
      (927.0429,86.5904) -- (924.9413,91.2786) -- (923.0014,91.9252) --
      (919.6065,94.9968) -- (914.7567,100.4932) .. controls (914.7567,100.4932) and
      (914.1101,100.4932) .. (913.4635,100.4932) .. controls (912.8168,100.4932) and
      (912.4935,98.3916) .. (912.4935,98.3916) -- (910.7152,98.5533) --
      (909.7453,100.0082) -- (907.3204,101.4632) -- (906.3504,102.9181) --
      (907.9670,104.3731) -- (907.4820,105.0197) -- (906.9970,107.7679) --
      (905.0571,107.6063) -- (905.0571,105.9897) -- (904.7338,104.6964) --
      (903.2789,105.0197) -- (901.5006,101.7865) -- (899.3990,103.0798) --
      (900.6923,104.5347) -- (901.0156,105.6664) -- (900.2073,106.9596) --
      (900.5306,110.0312) -- (900.6923,111.6478) -- (899.0757,114.2344) --
      (896.1658,114.7193) -- (895.8425,117.6292) -- (890.5077,120.7008) --
      (889.2144,121.1858) -- (887.5978,119.7308) -- (884.5262,123.2873) --
      (885.4962,126.5206) -- (884.0412,127.8138) -- (883.8796,132.1787) --
      (882.7563,138.4380) -- (880.2941,137.2821) -- (879.8091,134.2105) --
      (875.9292,133.0789) -- (875.6059,130.3306) -- (868.3311,106.8898) --
      (863.6326,92.2501) -- (865.0531,92.1319) -- (866.5669,92.5418) --
      (866.5669,89.9553) -- (867.8752,85.4588) -- (870.4618,80.7706) --
      (871.9167,76.7291) -- (869.9768,74.3042) -- (869.9768,68.3228) --
      (870.7851,67.3528) -- (871.5934,64.6046) -- (871.4317,63.1497) --
      (871.2701,58.2998) -- (873.0483,53.4500) -- (875.9582,44.5587) --
      (878.0598,40.3555) -- (879.3531,40.3555) -- (880.6464,40.5172) --
      (880.6464,41.6488) -- (881.9397,43.9121) -- (884.6879,44.5587) --
      (885.4962,43.7504) -- (885.4962,42.7804) -- (889.5377,39.8705) --
      (891.3160,38.0923) -- (892.7709,38.2539) -- (898.7523,40.6788) --
      (900.6923,41.6488) -- (909.7453,71.5560) -- (915.7267,71.5560) --
      (916.5350,73.4959) -- (916.6967,78.3457) -- (919.6066,80.6090) --
      (920.4149,80.6090) -- (920.5765,80.1240) -- (920.0915,78.9924) --
      (922.8398,78.8307) -- cycle(901.9080,108.9783) -- (903.4438,107.4425) --
      (904.8179,108.4933) -- (905.3837,110.9182) -- (903.6863,111.8073) --
      (901.9080,108.9782) -- cycle(908.6169,103.0776) -- (910.3952,104.9367) ..
      controls (910.3952,104.9367) and (911.6885,105.0175) .. (911.6885,104.6942) ..
      controls (911.6885,104.3709) and (911.9310,102.6735) .. (911.9310,102.6735) --
      (912.8201,101.8652) -- (912.0118,100.0869) -- (909.9911,100.8144) --
      (908.6169,103.0776) -- cycle;

    \path[USA map/state, USA map/RI, local bounding box=RI] (874.0700,178.8954) -- (870.3742,163.9394) -- (876.6435,162.0942) --
      (878.8346,164.0214) -- (882.1411,168.3420) -- (884.8290,172.7441) --
      (881.8297,174.3689) -- (880.5364,174.2072) -- (879.4048,175.9855) --
      (876.9799,177.9254) -- (874.0700,178.8954) -- cycle;

    \path[USA map/state, USA map/NY, local bounding box=NY] (830.3794,188.7456) -- (829.2478,187.7756) -- (826.6612,187.6140) --
      (824.3980,185.6741) -- (822.7674,179.5449) -- (819.3089,179.6354) --
      (816.8652,176.9272) -- (797.4799,181.3092) -- (754.4781,190.0389) --
      (746.9485,191.2669) -- (746.2103,184.7985) -- (747.6384,183.6731) --
      (748.9317,182.5415) -- (749.9017,180.9249) -- (751.6799,179.7933) --
      (753.6198,178.0150) -- (754.1048,176.3984) -- (756.2064,173.6502) --
      (757.3380,172.6802) -- (757.1764,171.7103) -- (755.8831,168.6387) --
      (754.1048,168.4770) -- (752.1649,162.3339) -- (755.0748,160.5557) --
      (759.4396,159.1007) -- (763.4811,157.8074) -- (766.7143,157.3225) --
      (773.0191,157.1608) -- (774.9590,158.4541) -- (776.5756,158.6158) --
      (778.6772,157.3225) -- (781.2638,156.1908) -- (786.4369,155.7059) --
      (788.5385,153.9276) -- (790.3168,150.6944) -- (791.9334,148.7545) --
      (794.0350,148.7545) -- (795.9749,147.6228) -- (796.1365,145.3596) --
      (794.6816,143.2580) -- (794.3583,141.8031) -- (795.4899,139.7015) --
      (795.4899,138.2465) -- (793.7116,138.2465) -- (791.9334,137.4382) --
      (791.1251,136.3066) -- (790.9634,133.7200) -- (796.7832,128.2236) --
      (797.4298,127.4153) -- (798.8848,124.5054) -- (801.7947,119.9789) --
      (804.5429,116.2607) -- (806.6445,113.8358) -- (809.0596,112.0102) --
      (812.1409,110.7643) -- (817.6374,109.4710) -- (820.8706,109.6326) --
      (825.3971,108.1777) -- (832.9623,106.1065) -- (833.4821,111.0862) --
      (835.9070,117.5526) -- (836.7153,122.7258) -- (835.7453,126.6056) --
      (838.3319,131.1321) -- (839.1402,133.2337) -- (838.3319,136.1436) --
      (841.2418,137.4369) -- (841.8884,137.7602) -- (844.9600,148.7532) --
      (844.4237,153.8128) -- (843.9387,164.6441) -- (844.7470,170.1406) --
      (845.5553,173.6971) -- (847.0103,180.9719) -- (847.0103,189.0549) --
      (845.8787,191.3182) -- (847.7180,193.3109) -- (848.5145,194.9894) --
      (846.5746,196.7676) -- (846.8979,198.0609) -- (848.1912,197.7376) --
      (849.6462,196.4443) -- (851.9094,193.8577) -- (853.0410,193.2111) --
      (854.6576,193.8577) -- (856.9209,194.0194) -- (864.8422,190.1396) --
      (867.7521,187.3913) -- (869.0454,185.9364) -- (873.2486,187.5530) --
      (869.8537,191.1095) -- (865.9739,194.0194) -- (858.8608,199.3542) --
      (856.2742,200.3242) -- (850.4545,202.2641) -- (846.4130,203.3957) --
      (845.2382,202.8628) -- (844.9942,199.1743) -- (845.4792,196.4260) --
      (845.3175,194.3244) -- (842.5040,192.6254) -- (837.9775,191.6555) --
      (834.0976,190.5238) -- (830.3794,188.7456) -- cycle;

    \path[USA map/state, USA map/PA, local bounding box=PA] (825.1237,224.6920) -- (826.4321,224.4211) -- (828.7616,223.1678) --
      (829.9735,220.6847) -- (831.5901,218.4215) -- (834.8233,215.3499) --
      (834.8233,214.5416) -- (832.3984,212.9250) -- (828.8419,210.5001) --
      (827.8719,207.9135) -- (825.1237,207.5902) -- (824.9620,206.4586) --
      (824.1537,203.7103) -- (826.4170,202.5787) -- (826.5787,200.1538) --
      (825.2854,198.8605) -- (825.4470,197.2439) -- (827.3870,194.1724) --
      (827.3870,191.1008) -- (830.0846,188.4549) -- (829.1643,187.7799) --
      (826.6402,187.5870) -- (824.3457,185.6471) -- (822.7958,179.5310) --
      (819.2912,179.6316) -- (816.8360,176.9282) -- (798.7450,181.1260) --
      (755.7432,189.8557) -- (746.8519,191.3106) -- (746.2312,184.7892) --
      (740.8687,189.8569) -- (739.5754,190.3419) -- (735.3731,193.3508) --
      (738.2839,212.4882) -- (740.7655,222.2176) -- (744.3373,241.4791) --
      (747.6066,240.8414) -- (759.5502,239.3389) -- (797.4768,231.6737) --
      (812.3531,228.8504) -- (820.6534,227.2280) -- (820.9205,226.9895) --
      (823.0221,225.3729) -- (825.1237,224.6920) -- cycle;

    \path[USA map/state, USA map/NJ, local bounding box=NJ] (829.6794,188.4602) -- (827.3569,191.1944) -- (827.3569,194.2660) --
      (825.4169,197.3375) -- (825.2553,198.9542) -- (826.5486,200.2474) --
      (826.3869,202.6724) -- (824.1237,203.8040) -- (824.9320,206.5522) --
      (825.0936,207.6838) -- (827.8419,208.0072) -- (828.8118,210.5937) --
      (832.3684,213.0187) -- (834.7933,214.6353) -- (834.7933,215.4436) --
      (831.8101,218.1401) -- (830.1934,220.4034) -- (828.7385,223.1516) --
      (826.4752,224.4449) -- (826.0128,226.0474) -- (825.7703,227.2598) --
      (825.1611,229.8666) -- (826.2533,232.1108) -- (829.4865,235.0206) --
      (834.3364,237.2839) -- (838.3779,237.9305) -- (838.5395,239.3855) --
      (837.7312,240.3554) -- (838.0545,243.1037) -- (838.8628,243.1037) --
      (840.9644,240.6788) -- (841.7727,235.8289) -- (844.5210,231.7874) --
      (847.5925,225.3210) -- (848.7241,219.8246) -- (848.0775,218.6929) --
      (847.9158,209.3166) -- (846.2992,205.9218) -- (845.1676,206.7301) --
      (842.4194,207.0534) -- (841.9344,206.5684) -- (843.0660,205.5984) --
      (845.1676,203.6585) -- (845.2307,202.5647) -- (844.8463,199.1308) --
      (845.4197,196.3826) -- (845.3022,194.4136) -- (842.4947,192.6632) --
      (837.4025,191.4875) -- (833.2651,190.1059) -- (829.6795,188.4602) -- cycle;

    \path[USA map/state, USA map/DE, local bounding box=DE] (825.6261,228.2791) -- (825.9944,226.1322) -- (826.3695,224.4412) --
      (824.7465,224.8389) -- (823.1310,225.3065) -- (820.9248,227.0708) --
      (822.6449,232.1137) -- (824.9081,237.7718) -- (827.0097,247.4714) --
      (828.6263,253.7762) -- (833.6378,253.6145) -- (839.7799,252.4339) --
      (837.5157,245.0476) -- (836.5457,245.5326) -- (832.9892,243.1077) --
      (831.2109,238.4195) -- (829.2710,234.8630) -- (826.1239,231.9927) --
      (825.2597,229.8946) -- (825.6261,228.2791) -- cycle;

    \path[USA map/state, USA map/MD, local bounding box=MD] (839.7917,252.4148) -- (833.7832,253.6186) -- (828.6403,253.7361) --
      (826.7967,246.8137) -- (824.8719,237.6444) -- (822.2993,231.4560) --
      (821.0109,227.0576) -- (813.5049,228.6800) -- (798.6287,231.5033) --
      (761.1773,239.0542) -- (762.3086,244.0659) -- (763.2785,249.7240) --
      (763.6018,249.4007) -- (765.7034,246.9758) -- (767.9667,244.3581) --
      (770.3916,243.7425) -- (771.8466,242.2876) -- (773.6248,239.7010) --
      (774.9181,240.3477) -- (777.8280,240.0243) -- (780.4146,237.9228) --
      (782.4215,236.4695) -- (784.2667,235.9845) -- (785.9110,237.1145) --
      (788.8209,238.5694) -- (790.7609,240.3477) -- (791.9733,241.8835) --
      (796.0957,243.5809) -- (796.0957,246.4908) -- (801.5921,247.7841) --
      (802.7366,248.3260) -- (804.1485,246.2977) -- (807.0304,248.2679) --
      (805.7523,250.7498) -- (804.9870,254.7355) -- (803.2087,257.3220) --
      (803.2087,259.4236) -- (803.8554,261.2019) -- (808.9193,262.5576) --
      (813.2304,262.4959) -- (816.3020,263.4659) -- (818.4035,263.7892) --
      (819.3735,261.6876) -- (817.9186,259.5860) -- (817.9186,257.8077) --
      (815.4937,255.7062) -- (813.3921,250.2097) -- (814.6854,244.8749) --
      (814.5237,242.7733) -- (813.2304,241.4800) .. controls (813.2304,241.4800) and
      (814.6854,239.8634) .. (814.6854,239.2168) .. controls (814.6854,238.5701) and
      (815.1703,237.1152) .. (815.1703,237.1152) -- (817.1103,235.8219) --
      (819.0502,234.2053) -- (819.5352,235.1753) -- (818.0802,236.7919) --
      (816.7869,240.5101) -- (817.1103,241.6417) -- (818.8885,241.9650) --
      (819.3735,247.4615) -- (817.2719,248.4314) -- (817.5952,251.9880) --
      (818.0802,251.8263) -- (819.2118,249.8864) -- (820.8285,251.6646) --
      (819.2118,252.9579) -- (818.8885,256.3528) -- (821.4751,259.7477) --
      (825.3549,260.2327) -- (826.9716,259.4244) -- (830.2081,263.6073) --
      (831.5665,264.1436) -- (838.2201,261.3466) -- (840.2277,257.3228) --
      (839.7917,252.4148) -- cycle(823.8222,261.4435) -- (824.9538,263.9492) --
      (825.1155,265.7275) -- (826.2471,267.5866) .. controls (826.2471,267.5866) and
      (827.1362,266.6975) .. (827.1362,266.3741) .. controls (827.1362,266.0508) and
      (826.4087,263.3026) .. (826.4087,263.3026) -- (825.6813,260.9585) --
      (823.8222,261.4435) -- cycle;

    \path[USA map/state, USA map/VA, local bounding box=VA] (831.6389,266.0689) -- (831.4949,264.1219) -- (837.9484,261.5720) --
      (837.1780,264.7899) -- (834.2580,268.5690) -- (833.8399,273.1548) --
      (834.3017,276.5452) -- (832.4737,281.5234) -- (830.3094,283.4395) --
      (828.8391,278.7987) -- (829.2850,273.3496) -- (830.8720,269.1665) --
      (831.6389,266.0689) -- cycle(834.9790,294.3703) -- (776.8049,306.9457) --
      (739.3779,312.2248) -- (732.6996,311.8496) -- (730.1143,313.7760) --
      (722.7752,313.9967) -- (714.3931,314.9743) -- (703.4781,316.5890) --
      (713.9475,310.9778) -- (713.9344,308.9028) -- (715.4545,306.7567) --
      (726.0083,295.2553) -- (729.9550,299.7327) -- (733.7380,300.6967) --
      (736.2815,299.5564) -- (738.5187,298.2452) -- (741.0553,299.5887) --
      (744.9695,298.1610) -- (746.8462,293.6046) -- (749.4471,294.1447) --
      (752.3024,292.0134) -- (754.1016,292.5070) -- (756.9288,288.8304) --
      (757.2771,286.7473) -- (756.3134,285.4718) -- (757.3162,283.6051) --
      (762.5905,271.3280) -- (763.2072,265.5929) -- (764.4361,265.0694) --
      (766.6147,267.5122) -- (770.5505,267.2111) -- (772.4797,259.6374) --
      (775.2737,259.0766) -- (776.3235,256.3355) -- (778.9033,253.9886) --
      (781.6751,248.2934) -- (781.7600,243.2259) -- (791.5815,247.0487) .. controls
      (792.2624,247.3891) and (792.4144,241.9996) .. (792.4144,241.9996) --
      (796.0670,243.5979) -- (796.1353,246.5361) -- (801.9195,247.8355) --
      (804.0525,249.0117) -- (805.7124,251.0674) -- (805.0578,254.7161) --
      (803.1104,257.3071) -- (803.2202,259.3662) -- (803.8092,261.2191) --
      (808.7880,262.4875) -- (813.2392,262.5274) -- (816.3081,263.4860) --
      (818.2516,263.7953) -- (818.9664,266.8838) -- (822.1568,267.2863) --
      (823.0249,268.4863) -- (822.5854,273.1764) -- (823.9601,274.2790) --
      (823.4812,276.2094) -- (824.7106,276.9991) -- (824.4888,278.3837) --
      (821.7948,278.2888) -- (821.8838,279.9044) -- (824.1648,281.4472) --
      (824.2863,282.8591) -- (826.0594,284.6445) -- (826.5512,287.1686) --
      (823.9982,288.5499) -- (825.5704,290.0442) -- (831.3714,288.3584) --
      (834.9790,294.3703) -- cycle;

    \path[USA map/state, USA map/WV, local bounding box=WV] (761.1855,238.9673) -- (762.2975,243.9118) -- (763.3810,249.9432) --
      (765.5113,247.3628) -- (767.7745,244.2913) -- (770.3129,243.6757) --
      (771.7678,242.2208) -- (773.5461,239.6342) -- (774.9911,240.2808) --
      (777.9010,239.9575) -- (780.4875,237.8559) -- (782.4944,236.4027) --
      (784.3397,235.9177) -- (785.6436,236.9342) -- (789.2868,238.7558) --
      (791.2268,240.5341) -- (792.6009,241.8273) -- (791.8392,247.3823) --
      (786.0042,244.8411) -- (781.7590,243.2190) -- (781.6579,248.3975) --
      (778.9102,253.9342) -- (776.3802,256.3609) -- (775.1881,259.1102) --
      (772.5445,259.6103) -- (771.6467,263.2122) -- (770.6034,267.1619) --
      (766.6352,267.5026) -- (764.3115,265.0638) -- (763.2403,265.6232) --
      (762.6076,271.0929) -- (761.2574,274.6274) -- (756.2990,285.5823) --
      (757.1956,286.7430) -- (756.9898,288.6516) -- (754.1811,292.5360) --
      (752.3726,291.9918) -- (749.4045,294.1515) -- (746.8622,293.5793) --
      (744.8629,298.1349) .. controls (744.8629,298.1349) and (741.6036,299.5651) ..
      (740.9400,299.5026) .. controls (740.7795,299.4875) and (738.4709,298.2535) ..
      (738.4709,298.2535) -- (736.1344,299.6329) -- (733.7246,300.6773) --
      (729.9799,299.7881) -- (728.8585,298.6199) -- (726.6663,295.5965) --
      (723.5237,293.6084) -- (721.8121,289.9851) -- (717.5273,286.5169) --
      (716.8806,284.2537) -- (714.2940,282.7987) -- (713.4857,281.1821) --
      (713.2432,275.9282) -- (715.4257,275.8474) -- (717.3656,275.0391) --
      (717.5273,272.2908) -- (719.1439,270.8359) -- (719.3055,265.8244) --
      (720.2755,261.9445) -- (721.5688,261.2979) -- (722.8620,262.4295) --
      (723.3470,264.2078) -- (725.1253,263.2378) -- (725.6103,261.6212) --
      (724.4787,259.8430) -- (724.4787,257.4180) -- (725.4486,256.1248) --
      (727.7119,252.7299) -- (729.0052,251.2749) -- (731.1068,251.7599) --
      (733.3700,250.1433) -- (736.4415,246.7484) -- (738.7048,242.8686) --
      (739.0281,237.2105) -- (739.5131,232.1990) -- (739.5131,227.5108) --
      (738.3815,224.4393) -- (739.3514,222.9843) -- (740.6349,221.6910) --
      (744.1262,241.5181) -- (748.7572,240.7670) -- (761.1855,238.9673) -- cycle;

    \path[USA map/state, USA map/OH, local bounding box=OH] (735.3250,193.3283) -- (729.2314,197.3817) -- (725.3516,199.6449) --
      (721.9567,203.3631) -- (717.9152,207.2430) -- (714.6820,208.0513) --
      (711.7721,208.5362) -- (706.2756,211.1228) -- (704.1741,211.2845) --
      (700.7792,208.2129) -- (695.6061,208.8596) -- (693.0195,207.4046) --
      (690.6384,206.0538) -- (685.7459,206.7572) -- (675.5612,208.3738) --
      (664.3544,210.5585) -- (665.6477,225.1888) -- (667.4259,238.9300) --
      (670.0125,262.3708) -- (670.5783,267.2020) -- (674.7007,267.0729) --
      (677.1256,266.2646) -- (680.4894,267.7678) -- (682.5598,272.1326) --
      (687.6988,272.1155) -- (689.5905,274.2342) -- (691.3517,274.1689) --
      (693.8901,272.8274) -- (696.3943,273.1989) -- (701.8155,273.6816) --
      (703.5425,271.5489) -- (705.8882,270.2557) -- (707.9587,269.5748) --
      (708.6053,272.3230) -- (710.3836,273.2930) -- (713.8593,275.6371) --
      (716.0417,275.5563) -- (717.3748,275.0638) -- (717.5595,272.3023) --
      (719.1449,270.8473) -- (719.2441,266.0546) .. controls (719.2441,266.0546) and
      (720.2680,261.9455) .. (720.2680,261.9455) -- (721.5673,261.3442) --
      (722.8887,262.4920) -- (723.4268,264.1890) -- (725.1459,263.1516) --
      (725.5849,261.6908) -- (724.4682,259.7878) -- (724.5345,257.4733) --
      (725.2835,256.4010) -- (727.4363,253.0946) -- (728.4865,251.5512) --
      (730.5881,252.0362) -- (732.8513,250.4196) -- (735.9229,247.0247) --
      (738.6944,242.9460) -- (739.0147,237.8905) -- (739.4997,232.8790) --
      (739.3229,227.5721) -- (738.3681,224.6773) -- (738.7193,223.4875) --
      (740.5237,221.7374) -- (738.2349,212.6901) -- (735.3250,193.3283) -- cycle;

    \path[USA map/state, USA map/IN, local bounding box=IN] (619.5695,299.9713) -- (619.6348,297.1127) -- (620.1198,292.5862) --
      (622.3831,289.6764) -- (624.1613,285.7965) -- (626.7479,281.5933) --
      (626.2629,275.7735) -- (624.4847,273.0253) -- (624.1613,269.7921) --
      (624.9697,264.2956) -- (624.4847,257.3442) -- (623.1914,241.3398) --
      (621.8981,225.9820) -- (620.9276,214.2620) -- (623.9987,215.1515) --
      (625.4536,216.1215) -- (626.5853,215.7982) -- (628.6868,213.8582) --
      (631.5164,212.2413) -- (636.6092,212.0792) -- (658.5951,209.8160) --
      (664.1708,209.2828) -- (665.6740,225.2390) -- (669.9253,262.0806) --
      (670.5238,267.8521) -- (670.1523,270.1154) -- (671.3802,271.9108) --
      (671.4766,273.2833) -- (668.9554,274.8828) -- (665.4159,276.4341) --
      (662.2138,276.9844) -- (661.6153,281.8514) -- (657.0406,285.1638) --
      (654.2442,289.1743) -- (654.5675,291.5510) -- (653.9862,293.0852) --
      (650.6597,293.0852) -- (649.0742,291.4686) -- (646.5809,292.7308) --
      (643.8979,294.2339) -- (644.0596,297.2884) -- (642.8658,297.5464) --
      (642.3979,296.5283) -- (640.2311,295.0251) -- (636.9807,296.3666) --
      (635.4294,299.3729) -- (633.9916,298.5646) -- (632.5366,296.9651) --
      (628.0723,297.4500) -- (622.4795,298.4200) -- (619.5696,299.9713) -- cycle;

    \path[USA map/state, USA map/IL, local bounding box=IL] (619.5415,300.3424) -- (619.5727,297.1127) -- (620.1400,292.4668) --
      (622.4726,289.5509) -- (624.3392,285.4751) -- (626.5722,281.4798) --
      (626.2007,276.2274) -- (624.1955,272.6848) -- (624.0991,269.3382) --
      (624.7940,264.0687) -- (623.9686,256.8903) -- (622.9022,241.1128) --
      (621.6089,226.0955) -- (620.6867,214.4563) -- (620.4141,213.5349) --
      (619.6058,210.9483) -- (618.3126,207.2301) -- (616.6960,205.4519) --
      (615.2410,202.8653) -- (615.0074,197.3764) -- (569.2110,199.9746) --
      (569.4396,202.3466) -- (571.7259,203.0324) -- (572.6403,204.1755) --
      (573.0976,206.0045) -- (576.9842,209.4339) -- (577.6701,211.7201) --
      (576.9842,215.1494) -- (575.1552,218.8074) -- (574.4693,221.3222) --
      (572.1831,223.1512) -- (570.3541,223.8371) -- (565.0958,225.2088) --
      (564.4099,227.0378) -- (563.7241,229.0954) -- (564.4099,230.4672) --
      (566.2389,232.0675) -- (566.0103,236.1827) -- (564.1813,237.7831) --
      (563.4954,239.3834) -- (563.4954,242.1269) -- (561.6665,242.5841) --
      (560.0661,243.7273) -- (559.8375,245.0990) -- (560.0661,247.1566) --
      (558.3514,248.4712) -- (557.3226,251.2718) -- (557.7799,254.9298) --
      (560.0661,262.2457) -- (567.3820,269.7902) -- (572.8690,273.4482) --
      (572.6403,277.7920) -- (573.5548,279.1638) -- (579.9563,279.6210) --
      (582.6997,280.9928) -- (582.0139,284.6507) -- (579.7277,290.5949) --
      (579.0418,293.7956) -- (581.3280,297.6822) -- (587.7294,302.9405) --
      (592.3019,303.6264) -- (594.3595,308.6561) -- (596.4171,311.8568) --
      (595.5026,314.8289) -- (597.1030,318.9441) -- (598.9319,321.0017) --
      (600.3460,320.1210) -- (601.2536,318.0462) -- (603.4667,316.2990) --
      (605.5982,315.6846) -- (608.2007,316.8644) -- (611.8277,318.2401) --
      (613.0167,317.9419) -- (613.2165,315.6834) -- (611.9292,313.2717) --
      (612.2334,310.8949) -- (614.0718,309.5475) -- (617.0944,308.7372) --
      (618.3553,308.2787) -- (617.7427,306.8918) -- (616.9513,304.5374) --
      (618.3839,303.5565) -- (619.5414,300.3424) -- cycle;

    \path[USA map/state, USA map/CT, local bounding box=CT] (874.0683,178.8629) -- (870.3909,163.9841) -- (865.6721,164.9044) --
      (844.4433,169.6475) -- (845.4435,172.8731) -- (846.8984,180.1479) --
      (847.0752,189.1148) -- (845.8552,191.2897) -- (847.7760,193.2220) --
      (852.0475,189.3164) -- (855.6040,186.0832) -- (857.5439,183.9816) --
      (858.3523,184.6282) -- (861.1005,183.1733) -- (866.2736,182.0417) --
      (874.0683,178.8629) -- cycle;

    \path[USA map/state, USA map/WI, local bounding box=WI] (615.0659,197.3687) -- (614.9992,194.2112) -- (613.8201,189.6847) --
      (613.1734,183.5417) -- (612.0418,181.1167) -- (613.0118,178.0452) --
      (613.8201,175.1353) -- (615.2750,172.5487) -- (614.6284,169.1539) --
      (613.9817,165.5973) -- (614.4667,163.8191) -- (616.4066,161.3942) --
      (616.5683,158.6459) -- (615.7600,157.3526) -- (616.4066,154.7661) --
      (615.9541,150.5954) -- (618.7024,144.9373) -- (621.6122,138.1475) --
      (621.7739,135.8843) -- (621.4506,134.9143) -- (620.6423,135.3993) --
      (616.4391,141.7041) -- (613.6909,145.7456) -- (611.7510,147.5238) --
      (610.9427,149.7871) -- (608.9877,150.5954) -- (607.8561,152.5353) --
      (606.4011,152.2120) -- (606.2395,150.4337) -- (607.5328,148.0088) --
      (609.6343,143.3207) -- (611.4126,141.7040) -- (612.4034,139.3462) --
      (609.8430,137.4449) -- (607.8682,127.0779) -- (604.3207,125.7359) --
      (602.3744,123.4276) -- (590.2447,120.7059) -- (587.3688,119.6939) --
      (579.1557,117.5266) -- (571.2378,116.3678) -- (567.4726,111.2372) --
      (566.7222,111.7912) -- (565.5243,111.6295) -- (564.8777,110.4979) --
      (563.5437,110.7944) -- (562.4120,110.9561) -- (560.6338,111.9261) --
      (559.6638,111.2794) -- (560.3105,109.3395) -- (562.2504,106.2679) --
      (563.3820,105.1363) -- (561.4421,103.6814) -- (559.3405,104.4897) --
      (556.4306,106.4296) -- (548.9942,109.6628) -- (546.0843,110.3094) --
      (543.1745,109.8245) -- (542.1927,108.9462) -- (540.0760,111.7814) --
      (539.8474,114.5249) -- (539.8474,122.9839) -- (538.7043,124.5843) --
      (533.4460,128.4708) -- (531.1597,134.4150) -- (531.6170,134.6437) --
      (534.1318,136.7013) -- (534.8177,139.9020) -- (532.9887,143.1027) --
      (532.9887,146.9893) -- (533.4460,153.6193) -- (536.4181,156.5914) --
      (539.8474,156.5914) -- (541.6764,159.7922) -- (545.1057,160.2494) --
      (548.9923,165.9650) -- (556.0796,170.0802) -- (558.1372,172.8236) --
      (559.0517,180.2539) -- (559.7376,183.5689) -- (562.0238,185.1693) --
      (562.2524,186.5410) -- (560.1948,189.9703) -- (560.4234,193.1711) --
      (562.9383,197.0576) -- (565.4531,198.2007) -- (568.4252,198.6580) --
      (569.7676,200.0381) -- (615.0659,197.3687) -- cycle;

    \path[USA map/state, USA map/NC, local bounding box=NC] (834.9815,294.3155) -- (837.0665,299.2329) -- (840.6231,305.6993) --
      (843.0480,308.1242) -- (843.6946,310.3875) -- (841.2697,310.5491) --
      (842.0780,311.1958) -- (841.7547,315.3989) -- (839.1681,316.6922) --
      (838.5215,318.7938) -- (837.2282,321.7037) -- (833.5100,323.3203) --
      (831.0851,322.9970) -- (829.6301,322.8353) -- (828.0135,321.5420) --
      (828.3369,322.8353) -- (828.3369,323.8053) -- (830.2768,323.8053) --
      (831.0851,325.0986) -- (829.1452,331.4033) -- (833.3483,331.4033) --
      (833.9950,333.0199) -- (836.2582,330.7567) -- (837.5515,330.2717) --
      (835.6116,333.8282) -- (832.5400,338.6781) -- (831.2468,338.6781) --
      (830.1151,338.1931) -- (827.3669,338.8397) -- (822.1938,341.2646) --
      (815.7273,346.5994) -- (812.3325,351.2876) -- (810.3926,357.7540) --
      (809.9076,360.1789) -- (805.2194,360.6639) -- (799.7663,362.0005) --
      (789.8199,353.7980) -- (777.2103,346.2000) -- (774.3004,345.3916) --
      (761.6909,346.8466) -- (757.4145,347.5967) -- (755.7979,344.3635) --
      (752.8275,342.2468) -- (736.3381,342.7318) -- (729.0634,343.5401) --
      (720.0104,348.0666) -- (713.8673,350.6532) -- (692.6897,353.2398) --
      (693.1898,349.1854) -- (694.9681,347.7305) -- (697.7163,347.0838) --
      (698.3630,343.3656) -- (702.5661,340.6174) -- (706.4460,339.1624) --
      (710.6492,335.6059) -- (715.0140,333.5043) -- (715.6606,330.4328) --
      (719.5405,326.5529) -- (720.1871,326.3913) .. controls (720.1871,326.3913) and
      (720.1871,327.5229) .. (720.9955,327.5229) .. controls (721.8038,327.5229) and
      (722.9354,327.8462) .. (722.9354,327.8462) -- (725.1986,324.2897) --
      (727.3002,323.6430) -- (729.5635,323.9664) -- (731.1801,320.4098) --
      (734.0900,317.8232) -- (734.5750,315.7217) -- (734.7625,312.0735) --
      (739.0390,312.0510) -- (746.2375,311.1952) -- (761.9948,308.9427) --
      (777.1308,306.8562) -- (798.7713,302.1368) -- (818.7546,297.8782) --
      (829.9316,295.4724) -- (834.9815,294.3156) -- cycle(839.2520,327.5221) --
      (841.8386,325.0164) -- (844.9909,322.4298) -- (846.5267,321.7831) --
      (846.6884,319.7624) -- (846.0417,313.6193) -- (844.5868,311.2752) --
      (843.9401,309.4161) -- (844.6676,309.1736) -- (847.4159,314.6701) --
      (847.8200,319.1157) -- (847.6584,322.5106) -- (844.2635,324.0464) --
      (841.4344,326.4713) -- (840.3028,327.6838) -- (839.2520,327.5221) -- cycle;

    \path[USA map/state, USA map/DC, local bounding box=DC] (805.8194,250.8438) -- (803.9612,249.0197) -- (802.7285,248.3334) --
      (804.1715,246.3109) -- (807.0606,248.2594) -- (805.8194,250.8438) -- cycle;

    \path[USA map/state, USA map/MA, local bounding box=MA] (899.6235,173.2539) -- (901.7954,172.5681) -- (902.2527,170.8534) --
      (903.2815,170.9677) -- (904.3103,173.2539) -- (903.0529,173.7112) --
      (899.1662,173.8255) -- (899.6235,173.2539) -- cycle(890.2499,174.0541) --
      (892.5362,171.4250) -- (894.1365,171.4250) -- (895.9655,172.9110) --
      (893.5650,173.9398) -- (891.3931,174.9686) -- (890.2499,174.0541) --
      cycle(855.4508,152.0659) -- (873.0977,147.4253) -- (875.3609,146.7786) --
      (877.2750,143.9829) -- (881.0118,142.3196) -- (883.9010,146.7324) --
      (881.4761,151.9056) -- (881.1528,153.3605) -- (883.0927,155.9471) --
      (884.2244,155.1388) -- (886.0026,155.1388) -- (888.2659,157.7253) --
      (892.1457,163.7068) -- (895.7023,164.1918) -- (897.9655,163.2218) --
      (899.7438,161.4435) -- (898.9355,158.6953) -- (896.8339,157.0787) --
      (895.3789,157.8870) -- (894.4090,156.5937) -- (894.8939,156.1087) --
      (896.9955,155.9471) -- (898.7738,156.7554) -- (900.7137,159.1803) --
      (901.6837,162.0902) -- (902.0070,164.5151) -- (897.8038,165.9700) --
      (893.9240,167.9099) -- (890.0441,172.4364) -- (888.1042,173.8914) --
      (888.1042,172.9214) -- (890.5291,171.4665) -- (891.0141,169.6882) --
      (890.2058,166.6167) -- (887.2959,168.0716) -- (886.4876,169.5266) --
      (886.9726,171.7898) -- (884.9063,172.7902) -- (882.1591,168.2631) --
      (878.7642,163.8983) -- (876.6937,162.0858) -- (870.1604,163.9620) --
      (865.0681,165.0128) -- (844.3929,169.6050) -- (843.7252,164.8371) --
      (844.3718,154.2484) -- (848.6611,153.3592) -- (855.4508,152.0659) -- cycle;

    \path[USA map/state, USA map/TN, local bounding box=TN] (696.6779,318.2541) -- (644.7848,323.2656) -- (629.0252,325.0439) --
      (624.4040,325.5566) -- (620.5357,325.5289) -- (620.3147,329.6297) --
      (612.1293,329.8937) -- (605.1779,330.5403) -- (597.0871,330.4165) --
      (595.6733,337.4894) -- (593.9771,342.9694) -- (590.6839,345.7202) --
      (589.3352,350.1013) -- (589.0118,352.6879) -- (584.9703,354.9511) --
      (586.4253,358.5076) -- (585.4553,362.8725) -- (584.4869,363.6621) --
      (692.6455,353.2546) -- (693.0487,349.2996) -- (694.8595,347.8093) --
      (697.6936,347.0598) -- (698.3656,343.3428) -- (702.4642,340.6379) --
      (706.5111,339.1438) -- (710.5947,335.5735) -- (715.0308,333.5480) --
      (715.5520,330.4807) -- (719.6166,326.4957) -- (720.1674,326.3815) .. controls
      (720.1674,326.3815) and (720.1986,327.5132) .. (721.0070,327.5132) .. controls
      (721.8153,327.5132) and (722.9469,327.8677) .. (722.9469,327.8677) --
      (725.2101,324.2799) -- (727.2805,323.6333) -- (729.5556,323.9285) --
      (731.1539,320.3956) -- (734.1092,317.7517) -- (734.5308,315.8126) --
      (734.8398,312.1015) -- (732.6932,311.9017) -- (730.0916,313.9300) --
      (723.0983,313.9591) -- (704.7390,316.3460) -- (696.6779,318.2542) -- cycle;

    \path[USA map/state, USA map/AR, local bounding box=AR] (593.8248,343.0530) -- (589.8449,343.7697) -- (584.7327,343.1356) --
      (585.1534,341.5336) -- (588.1332,338.9669) -- (589.0766,335.3106) --
      (587.2476,332.3385) -- (508.8300,334.8534) -- (510.4304,341.7121) --
      (510.4304,349.9425) -- (511.8021,360.9165) -- (512.0307,398.7534) --
      (514.3170,400.6967) -- (517.2891,399.3250) -- (520.0325,400.4681) --
      (520.7129,407.0414) -- (576.3341,405.9008) -- (577.4798,403.8104) --
      (577.1932,400.2609) -- (575.3675,397.2888) -- (576.9662,395.8036) --
      (575.3675,393.2921) -- (576.0517,390.7822) -- (577.4201,385.1768) --
      (579.9383,383.1142) -- (579.2524,380.8296) -- (582.9104,375.4578) --
      (585.6539,374.0894) -- (585.5404,372.5959) -- (585.1949,370.7702) --
      (588.0519,365.1715) -- (590.4549,363.9149) -- (590.8391,360.4873) --
      (592.6097,359.2456) -- (589.4662,358.7613) -- (588.1248,354.7509) --
      (590.9288,352.3742) -- (591.4791,350.3550) -- (592.7586,346.3083) --
      (593.8248,343.0530) -- cycle;

    \path[USA map/state, USA map/MO, local bounding box=MO] (558.4402,248.1132) -- (555.9203,245.0259) -- (554.7772,242.7397) --
      (490.4200,245.1402) -- (488.1337,245.2545) -- (489.3912,247.7694) --
      (489.1626,250.0556) -- (491.6774,253.9422) -- (494.7638,258.0574) --
      (497.8502,260.8009) -- (500.0114,261.0295) -- (501.5082,261.9440) --
      (501.5082,264.9161) -- (499.6792,266.5164) -- (499.2219,268.8027) --
      (501.2795,272.2320) -- (503.7944,275.2041) -- (506.3092,277.0331) --
      (507.6810,288.6928) -- (507.9951,324.7650) -- (508.2237,329.4518) --
      (508.6810,334.8353) -- (531.1140,333.9685) -- (554.3200,333.2826) --
      (575.1246,332.4816) -- (586.7794,332.2513) -- (588.9488,335.6773) --
      (588.2646,338.9848) -- (585.1773,341.3878) -- (584.6050,343.2252) --
      (589.9834,343.6824) -- (593.8784,342.9966) -- (595.5956,337.5029) --
      (596.2470,331.6461) -- (598.3450,329.0910) -- (600.9411,327.6041) --
      (600.9925,324.5538) -- (602.0085,322.6174) -- (600.3143,320.0736) --
      (598.9833,321.0579) -- (596.9907,318.8306) -- (595.7057,314.0716) --
      (596.5067,311.5534) -- (594.5626,308.1258) -- (592.7319,303.5500) --
      (587.9325,302.7506) -- (580.9637,297.1519) -- (579.2449,293.0383) --
      (580.0442,289.8376) -- (582.1035,283.7799) -- (582.5624,280.9163) --
      (580.6133,279.8850) -- (573.7579,279.0873) -- (572.7299,277.3752) --
      (572.6181,273.1448) -- (567.1312,269.7138) -- (560.1557,261.9423) --
      (557.8695,254.6264) -- (557.6392,250.4011) -- (558.4402,248.1132) -- cycle;

    \path[USA map/state, USA map/GA, local bounding box=GA] (672.2923,355.5518) -- (672.2923,357.7342) -- (672.4539,359.8358) --
      (673.1006,363.2307) -- (676.4955,371.1521) -- (678.9204,381.0134) --
      (680.3753,387.1565) -- (681.9919,392.0063) -- (683.4469,398.9577) --
      (685.5485,405.2625) -- (688.1350,408.6574) -- (688.6200,412.0522) --
      (690.5599,412.8605) -- (690.7216,414.9621) -- (688.9433,419.8119) --
      (688.4584,423.0452) -- (688.2967,424.9851) -- (689.9133,429.3499) --
      (690.2366,434.6847) -- (689.4283,437.1096) -- (690.0750,437.9179) --
      (691.5299,438.7262) -- (691.7346,441.9443) -- (693.9676,445.2939) --
      (696.2181,447.4559) -- (704.1395,447.6176) -- (714.9592,446.9709) --
      (736.4716,445.6777) -- (741.9173,445.0033) -- (746.4946,445.0310) --
      (746.6562,447.9409) -- (749.2428,448.7492) -- (749.5661,444.3843) --
      (747.9495,439.8578) -- (749.0811,438.2412) -- (754.9009,439.0495) --
      (759.8783,439.3673) -- (759.1029,433.0685) -- (761.3661,423.0456) --
      (762.8211,418.8424) -- (762.3361,416.2558) -- (765.6705,410.0115) --
      (765.1602,408.6599) -- (763.2468,409.3644) -- (760.6602,408.0711) --
      (760.0136,405.9695) -- (758.7203,402.4130) -- (756.4571,400.3114) --
      (753.8705,399.6648) -- (752.2539,394.8150) -- (749.3289,388.4800) --
      (745.1257,386.5400) -- (743.0241,384.6001) -- (741.7308,382.0135) --
      (739.6292,380.0736) -- (737.3660,378.7803) -- (735.1027,375.8704) --
      (732.0312,373.6072) -- (727.5047,371.8289) -- (727.0197,370.3740) --
      (724.5948,367.4641) -- (724.1098,366.0091) -- (720.7149,361.0386) --
      (717.1951,361.1378) -- (713.4401,358.7817) -- (712.0219,357.4884) --
      (711.6985,355.7102) -- (712.5693,353.7702) -- (714.7960,352.6601) --
      (714.1620,350.5629) -- (672.2923,355.5518) -- cycle;

    \path[USA map/state, USA map/SC, local bounding box=SC] (764.9433,408.1649) -- (763.1662,409.1344) -- (760.5796,407.8411) --
      (759.9330,405.7395) -- (758.6397,402.1830) -- (756.3765,400.0814) --
      (753.7899,399.4347) -- (752.1733,394.5849) -- (749.4251,388.6035) --
      (745.2219,386.6635) -- (743.1203,384.7236) -- (741.8270,382.1370) --
      (739.7254,380.1971) -- (737.4622,378.9038) -- (735.1989,375.9939) --
      (732.1274,373.7307) -- (727.6009,371.9524) -- (727.1159,370.4975) --
      (724.6910,367.5876) -- (724.2060,366.1326) -- (720.8111,360.9595) --
      (717.4162,361.1211) -- (713.3747,358.6962) -- (712.0814,357.4029) --
      (711.7581,355.6247) -- (712.5664,353.6848) -- (714.8297,352.7148) --
      (714.3189,350.4257) -- (720.0870,348.0891) -- (729.2025,343.5001) --
      (736.9772,342.6918) -- (753.0916,342.2693) -- (755.7298,344.1468) --
      (757.4089,347.5050) -- (761.7113,346.8950) -- (774.3208,345.4400) --
      (777.2307,346.2484) -- (789.8402,353.8464) -- (799.9483,361.9681) --
      (794.5272,367.4264) -- (791.9406,373.5695) -- (791.4556,379.8743) --
      (789.8390,380.6826) -- (788.7074,383.4308) -- (786.2825,384.0775) --
      (784.1809,387.6340) -- (781.4327,390.3822) -- (779.1694,393.7771) --
      (777.5528,394.5854) -- (773.9963,397.9803) -- (771.0864,398.1419) --
      (772.0564,401.3751) -- (767.0449,406.8716) -- (764.9433,408.1649) -- cycle;

    \path[USA map/state, USA map/KY, local bounding box=KY] (725.9944,295.2707) -- (723.7011,297.6724) -- (720.1229,301.6664) --
      (715.1983,307.1311) -- (713.9826,308.8469) -- (713.9201,310.9484) --
      (709.5402,313.1125) -- (703.8821,316.5074) -- (696.6502,318.3063) --
      (644.7823,323.2051) -- (629.0228,324.9834) -- (624.4016,325.4961) --
      (620.5332,325.4684) -- (620.3063,329.6887) -- (612.1269,329.8332) --
      (605.1755,330.4799) -- (597.1880,330.4197) -- (598.3958,329.0996) --
      (600.8953,327.5587) -- (601.1239,324.3580) -- (602.0384,322.5290) --
      (600.4316,319.9901) -- (601.2334,318.0833) -- (603.4967,316.3051) --
      (605.5983,315.6584) -- (608.3465,316.9517) -- (611.9030,318.2450) --
      (613.0347,317.9217) -- (613.1963,315.6584) -- (611.9030,313.2335) --
      (612.2264,310.9702) -- (614.1663,309.5153) -- (616.7529,308.8687) --
      (618.3695,308.2220) -- (617.5612,306.4437) -- (616.9145,304.5038) --
      (618.4211,303.5080) .. controls (618.4241,303.4709) and (619.6751,299.9857) ..
      (619.6594,299.8502) -- (622.7127,298.3715) -- (628.0324,297.4016) --
      (632.5265,296.9166) -- (633.9189,298.5440) -- (635.4472,299.4148) --
      (637.0380,296.3066) -- (640.2250,295.0240) -- (642.4301,296.5080) --
      (642.8407,297.5071) -- (644.0142,297.2430) -- (643.8525,294.2901) --
      (646.9834,292.5409) -- (649.1315,291.4674) -- (650.6609,293.1283) --
      (653.9790,293.0841) -- (654.5663,291.5128) -- (654.1988,289.2496) --
      (656.7994,285.2511) -- (661.5759,281.8132) -- (662.2819,276.9773) --
      (665.2069,276.5214) -- (668.9983,274.8757) -- (671.4417,273.1675) --
      (671.2433,271.6025) -- (670.1009,270.1476) -- (670.6667,267.1527) --
      (674.8516,267.0352) -- (677.1515,266.2894) -- (680.4989,267.7185) --
      (682.5530,272.0833) -- (687.6853,272.0941) -- (689.7363,274.3023) --
      (691.3517,274.1546) -- (693.9534,272.8765) -- (699.1905,273.4498) --
      (701.7654,273.6673) -- (703.4530,271.6111) -- (706.0709,270.1852) --
      (707.9527,269.4781) -- (708.5993,272.3147) -- (710.6428,273.3731) --
      (713.2855,275.4556) -- (713.4030,281.1288) -- (714.2113,282.7012) --
      (716.8010,284.2575) -- (717.5727,286.5520) -- (721.7325,289.9890) --
      (723.5379,293.6122) -- (725.9944,295.2707) -- cycle;

    \path[USA map/state, USA map/AL, local bounding box=AL] (631.3065,460.4157) -- (629.8159,446.0942) -- (627.0676,427.3416) --
      (627.2293,413.2771) -- (628.0376,382.2382) -- (627.8759,365.5872) --
      (628.0410,359.1681) -- (672.5255,355.5487) -- (672.3777,357.7311) --
      (672.5394,359.8327) -- (673.1860,363.2276) -- (676.5809,371.1489) --
      (679.0058,381.0102) -- (680.4607,387.1534) -- (682.0773,392.0032) --
      (683.5323,398.9546) -- (685.6339,405.2593) -- (688.2205,408.6542) --
      (688.7054,412.0491) -- (690.6454,412.8574) -- (690.8070,414.9590) --
      (689.0287,419.8088) -- (688.5438,423.0420) -- (688.3821,424.9820) --
      (689.9987,429.3468) -- (690.3220,434.6816) -- (689.5137,437.1065) --
      (690.1604,437.9148) -- (691.6153,438.7231) -- (691.9435,441.6119) --
      (686.3458,441.2584) -- (679.5561,441.9050) -- (654.0137,444.8149) --
      (643.6021,446.2217) -- (643.3807,449.0991) -- (645.1590,450.8774) --
      (647.7456,452.8173) -- (648.3264,460.7527) -- (642.7844,463.3256) --
      (640.0361,463.0023) -- (642.7844,461.0624) -- (642.7844,460.0924) --
      (639.7128,454.1110) -- (637.4496,453.4643) -- (635.9946,457.8291) --
      (634.7013,460.5774) -- (634.0547,460.4157) -- (631.3065,460.4157) -- cycle;

    \path[USA map/state, USA map/LA, local bounding box=LA] (607.9671,459.1612) -- (604.6824,455.9951) -- (605.6924,450.4949) --
      (605.0310,449.6018) -- (595.7693,450.6084) -- (570.7410,451.0673) --
      (570.0568,448.6726) -- (570.9696,440.2169) -- (574.2855,434.2711) --
      (579.3169,425.5800) -- (578.7428,423.1820) -- (579.9994,422.5012) --
      (580.4583,420.5487) -- (578.1721,418.4927) -- (578.0603,416.5503) --
      (576.2296,412.2048) -- (576.0826,405.8662) -- (520.6088,406.7902) --
      (520.6374,416.3637) -- (521.3233,425.7373) -- (522.0091,429.6238) --
      (524.5240,433.7390) -- (525.4385,438.7688) -- (529.7823,444.2557) --
      (530.0109,447.4564) -- (530.6968,448.1423) -- (530.0109,456.6013) --
      (527.0388,461.6310) -- (528.6392,463.6886) -- (527.9533,466.2035) --
      (527.2675,473.5194) -- (525.8957,476.7201) -- (526.0182,480.3365) --
      (530.7047,478.8164) -- (542.8180,479.0234) -- (553.1643,482.5799) --
      (559.6307,483.7116) -- (563.3489,482.2566) -- (566.5821,483.3882) --
      (569.8153,484.3582) -- (570.6236,482.2566) -- (567.3904,481.1250) --
      (564.8038,481.6100) -- (562.0556,479.9934) .. controls (562.0556,479.9934) and
      (562.2173,478.7001) .. (562.8639,478.5384) .. controls (563.5105,478.3768) and
      (565.9355,477.5685) .. (565.9355,477.5685) -- (567.7137,479.0234) --
      (569.4920,478.0534) -- (572.7252,478.7001) -- (574.1801,481.1250) --
      (574.5035,483.3882) -- (579.0299,483.7116) -- (580.8082,485.4898) --
      (579.9999,487.1064) -- (578.7066,487.9147) -- (580.3232,489.5313) --
      (588.7296,493.0879) -- (592.2861,491.7946) -- (593.2561,489.3697) --
      (595.8426,488.7230) -- (597.6209,487.2681) -- (598.9142,488.2381) --
      (599.7225,491.1479) -- (597.4592,491.9562) -- (598.1059,492.6029) --
      (601.5008,491.3096) -- (603.7640,487.9147) -- (604.5723,487.4298) --
      (602.4707,487.1064) -- (603.2790,485.4898) -- (603.1174,484.0349) --
      (605.2189,483.5499) -- (606.3506,482.2566) -- (606.9972,483.0649) .. controls
      (606.9972,483.0649) and (606.8355,486.1365) .. (607.6439,486.1365) .. controls
      (608.4522,486.1365) and (611.8470,486.7831) .. (611.8470,486.7831) --
      (615.8885,488.7230) -- (616.8585,490.1780) -- (619.7684,490.1780) --
      (620.9000,491.1479) -- (623.1633,488.0764) -- (623.1633,486.6214) --
      (621.8700,486.6214) -- (618.4751,483.8732) -- (612.6553,483.0649) --
      (609.4221,480.8017) -- (610.5537,478.0534) -- (612.8170,478.3768) --
      (612.9786,477.7301) -- (611.2004,476.7602) -- (611.2004,476.2752) --
      (614.4336,476.2752) -- (616.2119,473.2036) -- (614.9186,471.2637) --
      (614.5953,468.5155) -- (613.1403,468.6771) -- (611.2004,470.7787) --
      (610.5537,473.3653) -- (607.4822,472.7186) -- (606.5122,470.9404) --
      (608.2905,469.0005) -- (610.1938,465.5548) -- (609.1327,463.1426) --
      (607.9671,459.1612) -- cycle;

    \path[USA map/state, USA map/MS, local bounding box=MS] (631.5588,459.3446) -- (631.3046,460.6007) -- (626.1314,460.6007) --
      (624.6765,459.7924) -- (622.5749,459.4691) -- (615.7851,461.4090) --
      (614.0069,460.6007) -- (611.4203,464.8039) -- (610.3178,465.5819) --
      (609.1939,463.0939) -- (608.0508,459.2074) -- (604.6215,456.0066) --
      (605.7646,450.4621) -- (605.0787,449.5476) -- (603.2498,449.7762) --
      (595.3318,450.6496) -- (570.7853,451.0230) -- (570.0156,448.7976) --
      (570.8890,440.4208) -- (574.0058,434.7480) -- (579.2329,425.6031) --
      (578.7871,423.1705) -- (580.0240,422.5142) -- (580.4599,420.5948) --
      (578.1424,418.5158) -- (578.0273,416.3743) -- (576.1915,412.2532) --
      (576.0825,406.2905) -- (577.4101,403.8095) -- (577.1868,400.3937) --
      (575.4173,397.3111) -- (576.9437,395.8289) -- (575.3731,393.3294) --
      (575.8303,391.6772) -- (577.4077,385.1508) -- (579.8937,383.1145) --
      (579.2520,380.7475) -- (582.9100,375.4450) -- (585.7419,374.0885) --
      (585.5209,372.4134) -- (585.2328,370.7323) -- (588.1088,365.1646) --
      (590.4545,363.9331) -- (590.6062,363.0401) -- (627.9496,359.1589) --
      (628.1345,365.4422) -- (628.2962,382.0933) -- (627.4879,413.1322) --
      (627.3262,427.1966) -- (630.0744,445.9493) -- (631.5588,459.3446) -- cycle;

    \path[USA map/state, USA map/IA, local bounding box=IA] (569.1915,199.5843) -- (569.4559,202.3705) -- (571.6796,202.9478) --
      (572.6336,204.1731) -- (573.1336,206.0285) -- (576.9264,209.3871) --
      (577.6123,211.7786) -- (576.9380,215.2031) -- (575.3556,218.4351) --
      (574.5563,221.1768) -- (572.3836,222.7789) -- (570.6680,223.3513) --
      (565.0890,225.2115) -- (563.6976,229.0602) -- (564.4262,230.4319) --
      (566.2667,232.1145) -- (565.9838,236.1508) -- (564.2206,237.6887) --
      (563.4492,239.3318) -- (563.5764,242.1081) -- (561.6901,242.5654) --
      (560.0647,243.6703) -- (559.7859,245.0229) -- (560.0647,247.1378) --
      (558.5137,248.2539) -- (556.0431,245.1206) -- (554.7806,242.6707) --
      (489.0447,245.1856) -- (488.1267,245.3510) -- (486.0743,240.8351) --
      (485.8457,234.2050) -- (484.2453,230.0898) -- (483.5595,224.8315) --
      (481.2732,221.1735) -- (480.3588,216.3724) -- (477.6153,208.8279) --
      (476.4722,203.4552) -- (475.1004,201.2833) -- (473.5001,198.5399) --
      (475.4541,193.6960) -- (476.8258,187.9805) -- (474.0823,185.9229) --
      (473.6251,183.1794) -- (474.5396,180.6645) -- (476.2542,180.6645) --
      (558.9082,179.3951) -- (559.7425,183.5782) -- (561.9947,185.1392) --
      (562.2514,186.5622) -- (560.2219,189.9516) -- (560.4123,193.1571) --
      (562.9271,196.9553) -- (565.4539,198.2489) -- (568.5332,198.7519) --
      (569.1915,199.5843) -- cycle;

    \path[USA map/state, USA map/MN, local bounding box=MN] (475.2378,128.8244) -- (474.7806,120.3653) -- (472.9516,113.0494) --
      (471.1226,99.5607) -- (470.6654,89.7299) -- (468.8364,86.3006) --
      (467.2360,81.2709) -- (467.2360,70.9829) -- (467.9219,67.0963) --
      (466.1009,61.6446) -- (496.2334,61.6799) -- (496.5567,53.4352) --
      (497.2033,53.2735) -- (499.4666,53.7585) -- (501.4065,54.5668) --
      (502.2148,60.0633) -- (503.6697,66.2064) -- (505.2863,67.8230) --
      (510.1362,67.8230) -- (510.4595,69.2779) -- (516.7642,69.6012) --
      (516.7642,71.7028) -- (521.6141,71.7028) -- (521.9374,70.4095) --
      (523.0690,69.2779) -- (525.3322,68.6313) -- (526.6255,69.6012) --
      (529.5354,69.6012) -- (533.4153,72.1878) -- (538.7501,74.6127) --
      (541.1750,75.0977) -- (541.6599,74.1277) -- (543.1149,73.6428) --
      (543.5999,76.5526) -- (546.1864,77.8459) -- (546.6714,77.3609) --
      (547.9647,77.5226) -- (547.9647,79.6242) -- (550.5513,80.5942) --
      (553.6228,80.5942) -- (555.2394,79.7859) -- (558.4726,76.5526) --
      (561.0592,76.0677) -- (561.8675,77.8459) -- (562.3525,79.1392) --
      (563.3224,79.1392) -- (564.2924,78.3309) -- (573.1837,78.0076) --
      (574.9620,81.0791) -- (575.6086,81.0791) -- (576.3223,79.9949) --
      (580.7622,79.6242) -- (580.1501,81.9037) -- (576.2113,83.7408) --
      (566.9656,87.8019) -- (562.1908,89.8088) -- (559.1193,92.3954) --
      (556.6944,95.9519) -- (554.4311,99.8318) -- (552.6529,100.6401) --
      (548.1264,105.6515) -- (546.8331,105.8132) -- (542.5053,108.5703) --
      (540.0424,111.7754) -- (539.8138,114.9668) -- (539.9082,123.0102) --
      (538.5322,124.6989) -- (533.4506,128.4589) -- (531.2205,134.4413) --
      (534.0923,136.6750) -- (534.7722,139.9020) -- (532.9169,143.1409) --
      (533.0877,146.8889) -- (533.4566,153.6193) -- (536.4848,156.6213) --
      (539.8138,156.6213) -- (541.7050,159.7539) -- (545.0841,160.2572) --
      (548.9433,165.9287) -- (556.0306,170.0454) -- (558.1737,172.9205) --
      (558.8449,179.3600) -- (477.6334,180.5048) -- (477.2955,144.8280) --
      (476.8382,141.8559) -- (472.7230,138.4265) -- (471.5799,136.5976) --
      (471.5799,134.9972) -- (473.6375,133.3968) -- (475.0092,132.0251) --
      (475.2379,128.8244) -- cycle;

    \path[USA map/state, USA map/OK, local bounding box=OK] (380.3431,320.8215) -- (363.6589,319.5482) -- (362.7787,330.5006) --
      (383.2441,331.6575) -- (415.2997,332.9611) -- (412.9651,357.3797) --
      (412.5078,375.2123) -- (412.7364,376.8126) -- (417.0803,380.4706) --
      (419.1379,381.6137) -- (419.8237,381.3851) -- (420.5096,379.3275) --
      (421.8813,381.1565) -- (423.9389,381.1565) -- (423.9389,379.7847) --
      (426.6824,381.1565) -- (426.2252,385.0430) -- (430.3404,385.2717) --
      (432.8552,386.4148) -- (436.9704,387.1007) -- (439.4853,388.9296) --
      (441.7715,386.8720) -- (445.2009,387.5579) -- (447.7157,390.9872) --
      (448.6302,390.9872) -- (448.6302,393.2735) -- (450.9164,393.9593) --
      (453.2026,391.6731) -- (455.0316,392.3590) -- (457.5465,392.3590) --
      (458.4610,394.8738) -- (464.7620,396.9528) -- (466.1338,396.2669) --
      (467.9628,392.1517) -- (469.1059,392.1517) -- (470.2490,394.2093) --
      (474.3642,394.8952) -- (478.0221,396.2669) -- (480.9942,397.1814) --
      (482.8232,396.2669) -- (483.5091,393.7521) -- (487.8529,393.7521) --
      (489.9105,394.6666) -- (492.6540,392.6090) -- (493.7971,392.6090) --
      (494.4830,394.2093) -- (498.5982,394.2093) -- (500.1985,392.1517) --
      (502.0275,392.6090) -- (504.0851,395.1238) -- (507.2858,396.9528) --
      (510.4866,397.8673) -- (512.4277,398.9862) -- (512.0386,361.7692) --
      (510.6668,350.7952) -- (510.5063,341.9229) -- (509.0665,335.3852) --
      (508.2883,328.2055) -- (508.2202,324.3893) -- (496.0833,324.7081) --
      (449.6733,324.2508) -- (404.6344,322.1932) -- (380.3432,320.8215) -- cycle;

    \path[USA map/state, USA map/TX, local bounding box=TX] (361.4642,330.5736) -- (384.1550,331.6595) -- (415.2477,332.8026) --
      (412.9131,356.2584) -- (412.6163,374.4120) -- (412.6844,376.4938) --
      (417.0283,380.3122) -- (419.0149,381.7593) -- (420.1991,381.1997) --
      (420.5725,379.3819) -- (421.7128,381.1856) -- (423.8245,381.2295) --
      (423.8215,379.7824) -- (425.4914,380.7496) -- (426.6301,381.1585) --
      (426.2708,385.1261) -- (430.3590,385.2197) -- (433.2843,386.4168) --
      (437.2391,386.9422) -- (439.6205,389.0212) -- (441.7446,386.9450) --
      (445.4695,387.5599) -- (447.6904,390.7849) -- (448.7654,391.1058) --
      (448.6049,393.0711) -- (450.8185,393.8634) -- (453.1487,391.8086) --
      (455.2817,392.4235) -- (457.5111,392.4590) -- (458.4441,394.8944) --
      (464.7722,397.0089) -- (466.3653,396.2420) -- (467.8547,392.0643) --
      (468.1955,392.0643) -- (469.1020,392.1458) -- (470.3310,394.2145) --
      (474.2609,394.8798) -- (477.5979,396.0026) -- (481.0235,397.1986) --
      (482.8641,396.2236) -- (483.5779,393.7088) -- (488.0311,393.7530) --
      (489.8398,394.6837) -- (492.6391,392.5772) -- (493.7427,392.6214) --
      (494.5937,394.2265) -- (498.6485,394.2265) -- (500.1673,392.1979) --
      (502.0347,392.6051) -- (503.9807,395.0084) -- (507.5013,397.0526) --
      (510.3601,397.8624) -- (511.8737,398.6622) -- (514.3204,400.6595) --
      (517.3634,399.3317) -- (520.0545,400.4706) -- (520.6183,406.5766) --
      (520.5785,416.2787) -- (521.2644,425.8127) -- (521.9666,429.4179) --
      (524.6419,433.8377) -- (525.5401,438.7884) -- (529.7560,444.3265) --
      (529.9520,447.4714) -- (530.6984,448.2572) -- (529.9683,456.6373) --
      (527.0962,461.6438) -- (528.6292,463.7967) -- (527.9991,466.1348) --
      (527.3296,473.5391) -- (525.8252,476.8771) -- (526.1201,480.3794) --
      (520.4552,481.9646) -- (510.5940,486.4911) -- (509.6240,488.4310) --
      (507.0374,490.3710) -- (504.9358,491.8259) -- (503.6425,492.6342) --
      (497.9844,497.9690) -- (495.2362,500.0706) -- (489.9014,503.3038) --
      (484.2433,505.7287) -- (477.9385,509.1236) -- (476.1603,510.5785) --
      (470.3405,514.1350) -- (466.9456,514.7817) -- (463.0658,520.2781) --
      (459.0243,520.6015) -- (458.0543,522.5414) -- (460.3176,524.4813) --
      (458.8626,529.9778) -- (457.5693,534.5043) -- (456.4377,538.3841) --
      (455.6294,542.9106) -- (456.4377,545.3355) -- (458.2160,552.2869) --
      (459.1859,558.4300) -- (460.9642,561.1782) -- (459.9942,562.6332) --
      (456.9227,564.5731) -- (451.2646,560.6933) -- (445.7681,559.5616) --
      (444.4748,560.0466) -- (441.2416,559.4000) -- (437.0384,556.3284) --
      (431.8653,555.1968) -- (424.2673,551.8019) -- (422.1657,547.9221) --
      (420.8724,541.4557) -- (417.6392,539.5157) -- (416.9925,537.2525) --
      (417.6392,536.6059) -- (417.9625,533.2110) -- (416.6692,532.5643) --
      (416.0226,531.5944) -- (417.3159,527.2295) -- (415.6993,524.9663) --
      (412.4660,523.6730) -- (409.0712,519.3082) -- (405.5146,512.6801) --
      (401.3115,510.0935) -- (401.4731,508.1536) -- (396.1383,495.8674) --
      (395.3300,491.6642) -- (393.5518,489.7243) -- (393.3901,488.2694) --
      (387.4087,482.9346) -- (384.8221,479.8630) -- (384.8221,478.7314) --
      (382.2355,476.6298) -- (375.4458,475.4982) -- (368.0094,474.8516) --
      (364.9379,472.5883) -- (360.4114,474.3666) -- (356.8548,475.8215) --
      (354.5916,479.0547) -- (353.6216,482.7729) -- (349.2568,488.9160) --
      (346.8319,491.3409) -- (344.2453,490.3710) -- (342.4671,489.2393) --
      (340.5271,488.5927) -- (336.6473,486.3295) -- (336.6473,485.6828) --
      (334.8690,483.7429) -- (329.6959,481.6413) -- (322.2595,473.8816) --
      (319.9963,469.1934) -- (319.9963,461.1104) -- (316.7631,454.6440) --
      (316.2781,451.8958) -- (314.6615,450.9258) -- (313.5298,448.8242) --
      (308.5184,446.7226) -- (307.2251,445.1060) -- (300.1120,437.1847) --
      (298.8187,433.9515) -- (294.1306,431.6882) -- (292.6756,427.3233) --
      (290.0890,424.4135) -- (288.1491,423.9285) -- (287.4999,419.2509) --
      (295.5018,419.9368) -- (324.5368,422.6802) -- (353.5718,424.2806) --
      (355.8054,404.8187) -- (359.6919,349.2637) -- (361.2923,330.5164) --
      (362.6641,330.5450)(461.6934,560.2077) -- (461.1276,553.0947) --
      (458.3794,545.9007) -- (457.8135,538.8685) -- (459.3493,530.6238) --
      (462.6634,523.7532) -- (466.1391,518.3375) -- (469.2915,514.7810) --
      (469.9381,515.0235) -- (465.1691,521.6516) -- (460.8043,528.1989) --
      (458.7835,534.8270) -- (458.4602,540.0001) -- (459.3493,546.1432) --
      (461.9359,553.3372) -- (462.4209,558.5103) -- (462.5825,559.9653) --
      (461.6934,560.2077) -- cycle;

    \path[USA map/state, USA map/NM, local bounding box=NM] (288.1526,424.0131) -- (287.3771,419.2650) -- (296.0209,419.7904) --
      (326.1927,422.7363) -- (353.4608,424.4262) -- (355.6761,405.7188) --
      (359.5335,349.8428) -- (361.2711,330.4536) -- (362.8428,330.5821) --
      (363.6683,319.4187) -- (259.6638,308.7828) -- (242.1664,429.2176) --
      (257.6271,431.2067) -- (258.9204,421.1838) -- (288.1525,424.0131) -- cycle;

    \path[USA map/state, USA map/KS, local bounding box=KS] (507.8806,324.3803) -- (495.2623,324.5847) -- (449.1732,324.1275) --
      (404.6158,322.0699) -- (379.9860,320.8124) -- (383.8798,256.2175) --
      (405.9633,256.8926) -- (446.2524,257.7340) -- (490.5536,258.7216) --
      (495.6493,258.7216) -- (497.8337,260.8840) -- (499.8513,260.8626) --
      (501.4916,261.8751) -- (501.4291,264.8843) -- (499.6001,266.6097) --
      (499.2679,268.8419) -- (501.1110,272.2442) -- (504.0633,275.4393) --
      (506.3907,277.0537) -- (507.6915,288.2945) -- (507.8806,324.3803) -- cycle;

    \path[USA map/state, USA map/NE, local bounding box=NE] (486.0979,240.7006) -- (489.3285,247.7205) -- (489.1999,250.0230) --
      (492.6591,255.5169) -- (495.3784,258.6692) -- (490.3289,258.6692) --
      (446.8463,257.7305) -- (406.0595,256.8402) -- (383.8072,256.0564) --
      (384.8800,234.7285) -- (352.5618,231.8083) -- (356.9056,187.7984) --
      (372.4519,188.8272) -- (392.5707,189.9703) -- (410.4033,191.1134) --
      (434.1801,192.2566) -- (444.9253,191.7993) -- (446.9829,194.0855) --
      (451.7840,197.0576) -- (452.9271,197.9721) -- (457.2709,196.6004) --
      (461.1575,196.1431) -- (463.9010,195.9145) -- (465.7300,197.2863) --
      (469.7874,198.8866) -- (472.7595,200.4870) -- (473.2167,202.0873) --
      (474.1312,204.1449) -- (475.9602,204.1449) -- (476.7582,204.1911) --
      (477.6524,208.8730) -- (480.5727,217.3409) -- (481.1452,221.0976) --
      (483.6687,224.8718) -- (484.2383,229.9860) -- (485.8455,234.2263) --
      (486.0979,240.7006) -- cycle;

    \path[USA map/state, USA map/SD, local bounding box=SD] (476.4469,204.0247) -- (476.3995,203.4438) -- (473.5038,198.5983) --
      (475.3640,193.8862) -- (476.8567,187.9997) -- (474.0748,185.9200) --
      (473.6897,183.1765) -- (474.4821,180.6222) -- (477.6706,180.6374) --
      (477.5475,175.6312) -- (477.2142,145.4570) -- (476.5965,141.6894) --
      (472.5242,138.3585) -- (471.5415,136.6815) -- (471.4790,135.0727) --
      (473.5012,133.5433) -- (475.0334,131.8776) -- (475.2783,129.2208) --
      (417.0212,127.6205) -- (362.2220,124.1714) -- (356.8968,187.8626) --
      (371.4870,188.7664) -- (391.4369,189.9720) -- (409.1799,190.9006) --
      (432.9567,192.2042) -- (444.9394,191.7795) -- (446.9057,194.0247) --
      (452.1003,197.2781) -- (452.8642,198.0008) -- (457.4057,196.5480) --
      (463.9462,195.9331) -- (465.6215,197.2694) -- (469.8260,198.8655) --
      (472.7711,200.5013) -- (473.1701,201.9851) -- (474.2096,204.2260) --
      (476.4469,204.0246) -- cycle;

    \path[USA map/state, USA map/ND, local bounding box=ND] (475.3053,128.9185) -- (474.6904,120.4848) -- (473.0134,113.6689) --
      (471.1219,100.6446) -- (470.6647,89.6576) -- (468.9252,86.5805) --
      (467.1686,81.3861) -- (467.1998,70.9418) -- (467.8232,67.1177) --
      (465.9891,61.6500) -- (437.3468,61.0859) -- (418.7559,60.4393) --
      (392.2436,59.1460) -- (369.2972,57.0121) -- (362.3040,124.1890) --
      (417.2362,127.5326) -- (475.3052,128.9185) -- cycle;

    \path[USA map/state, USA map/WY, local bounding box=WY] (360.3767,143.2759) -- (253.6341,129.8188) -- (239.5506,218.2768) --
      (352.8152,231.8623) -- (360.3767,143.2759) -- cycle;

    \path[USA map/state, USA map/MT, local bounding box=MT] (369.2095,56.9691) -- (338.5352,54.1613) -- (309.2746,50.6048) --
      (280.0141,46.5633) -- (247.6820,41.2285) -- (229.2527,37.8336) --
      (196.5291,30.9009) -- (192.0500,52.2484) -- (195.4794,59.7929) --
      (194.1076,64.3654) -- (195.9366,68.9378) -- (199.1374,70.3096) --
      (203.7582,81.0790) -- (206.4533,84.2555) -- (206.9105,85.3987) --
      (210.3399,86.5418) -- (210.7971,88.5994) -- (203.7098,106.2033) --
      (203.7098,108.7182) -- (206.2247,111.9189) -- (207.1391,111.9189) --
      (211.9402,108.9468) -- (212.6261,107.8037) -- (214.2264,108.4896) --
      (213.9978,113.7479) -- (216.7413,126.3221) -- (219.7134,128.8370) --
      (220.6279,129.5228) -- (222.4569,131.8091) -- (221.9996,135.2384) --
      (222.6855,138.6677) -- (223.8286,139.5822) -- (226.1148,137.2960) --
      (228.8583,137.2960) -- (232.0590,138.8964) -- (234.5739,137.9819) --
      (238.6891,137.9819) -- (242.3470,139.5822) -- (245.0905,139.1250) --
      (245.5477,136.1529) -- (248.5198,135.4670) -- (249.8916,136.8388) --
      (250.3488,140.0395) -- (251.7747,140.8741) -- (253.6616,129.8394) --
      (360.4073,143.2683) -- (369.2095,56.9691) -- cycle;

    \path[USA map/state, USA map/CO, local bounding box=CO] (380.0324,320.9646) -- (384.9357,234.6396) -- (271.5471,221.9956) --
      (259.3333,309.9348) -- (380.0324,320.9646) -- cycle;

    \path[USA map/state, USA map/ID, local bounding box=ID] (148.4788,176.4839) -- (157.2497,141.2632) -- (158.6214,137.0337) --
      (161.1363,131.0895) -- (159.8788,128.8033) -- (157.3640,128.9176) --
      (156.5638,127.8888) -- (157.0211,126.7457) -- (157.3640,123.6593) --
      (161.8221,118.1723) -- (163.6511,117.7151) -- (164.7942,116.5720) --
      (165.3658,113.3713) -- (166.2803,112.6854) -- (170.1668,106.8555) --
      (174.0534,102.5117) -- (174.2821,98.7394) -- (170.8527,96.1103) --
      (169.3172,91.7093) -- (182.9421,28.3676) -- (196.4597,30.8957) --
      (192.0516,52.2787) -- (195.6119,59.7641) -- (194.0308,64.4249) --
      (196.0007,69.0661) -- (199.1389,70.3213) -- (202.9742,79.8779) --
      (206.4869,84.3151) -- (206.9942,85.4582) -- (210.3351,86.6013) --
      (210.7040,88.6984) -- (203.7330,106.0745) -- (203.5678,108.6404) --
      (206.1989,111.9621) -- (207.1040,111.9132) -- (212.0153,108.8876) --
      (212.6927,107.7926) -- (214.2550,108.4515) -- (213.9766,113.8052) --
      (216.7158,126.3879) -- (220.6337,129.5658) -- (222.3148,131.7313) --
      (221.5982,135.8151) -- (222.6644,138.6226) -- (223.7261,139.7138) --
      (226.2054,137.3624) -- (229.0535,137.4113) -- (231.9728,138.7465) --
      (234.7528,138.0646) -- (238.5471,137.9041) -- (242.5260,139.5045) --
      (245.2694,139.2077) -- (245.7662,136.1704) -- (248.6988,135.4056) --
      (249.9589,136.9215) -- (250.3999,139.8664) -- (251.8242,141.0797) --
      (243.4382,194.6883) .. controls (243.4382,194.6883) and (155.4722,177.9877) ..
      (148.4788,176.4840) -- cycle;

    \path[USA map/state, USA map/UT, local bounding box=UT] (259.4984,310.1051) -- (175.7493,298.2328) -- (196.3369,185.6915) --
      (243.1173,194.4366) -- (241.6325,205.0670) -- (239.3208,218.2397) --
      (247.1285,219.1681) -- (263.5350,220.9729) -- (271.7460,221.8285) --
      (259.4984,310.1051) -- cycle;

    \path[USA map/state, USA map/AZ, local bounding box=AZ] (144.9112,382.6291) -- (142.2842,384.7874) -- (141.9609,386.2424) --
      (142.4459,387.2123) -- (161.3601,397.8819) -- (173.4847,405.4800) --
      (188.1958,414.0480) -- (205.0084,424.0709) -- (217.2946,426.4958) --
      (242.2458,429.2007) -- (259.5014,310.0737) -- (175.7658,298.1564) --
      (172.6734,314.5689) -- (171.0671,314.5842) -- (169.3524,317.2133) --
      (166.8376,317.0990) -- (165.5802,314.3556) -- (162.8367,314.0126) --
      (161.9222,312.8695) -- (161.0077,312.8695) -- (160.0932,313.4411) --
      (158.1499,314.4699) -- (158.0356,321.4429) -- (157.8070,323.1575) --
      (157.2354,335.7318) -- (155.7494,337.9037) -- (155.1778,341.2187) --
      (157.9213,346.1341) -- (159.1787,351.9640) -- (159.9789,352.9928) --
      (161.0077,353.5643) -- (160.8934,355.8505) -- (159.2930,357.2223) --
      (155.8637,358.9370) -- (153.9204,360.8803) -- (152.4344,364.5382) --
      (151.8628,369.4536) -- (149.0050,372.1971) -- (146.9474,372.8829) --
      (147.0831,373.7128) -- (146.6259,375.4275) -- (147.0831,376.2277) --
      (150.7411,376.7992) -- (150.1695,379.5427) -- (148.6835,381.7146) --
      (144.9112,382.6291) -- cycle;

    \path[USA map/state, USA map/NV, local bounding box=NV] (196.3927,185.5755) -- (172.7538,314.3983) -- (170.9216,314.7474) --
      (169.3488,317.1536) -- (166.9759,317.1643) -- (165.5039,314.4208) --
      (162.8855,314.0424) -- (162.1145,312.9348) -- (161.0767,312.8808) --
      (158.2983,314.5251) -- (157.9881,321.3106) -- (157.6260,327.0877) --
      (157.2774,335.6805) -- (155.8303,337.7697) -- (153.3914,336.6957) --
      (84.3115,232.4945) -- (103.3006,164.9096) -- (196.3927,185.5756) -- cycle;

    \path[USA map/state, USA map/OR, local bounding box=OR] (148.7218,175.5315) -- (157.5715,140.7300) -- (158.6223,136.5005) --
      (160.9767,130.8773) -- (160.3612,129.7144) -- (157.8463,129.6682) --
      (156.5647,127.9975) -- (157.0220,126.5334) -- (157.5254,123.2865) --
      (161.9835,117.7996) -- (163.8125,116.7004) -- (164.9556,115.5573) --
      (166.4417,111.9917) -- (170.4887,106.3223) -- (174.0543,102.4599) --
      (174.2830,99.0086) -- (171.0141,96.5399) -- (169.2307,91.8973) --
      (156.5669,88.2853) -- (141.4778,84.7417) -- (126.0458,84.8560) --
      (125.5886,83.4842) -- (120.1016,85.5418) -- (115.6435,84.9703) --
      (113.2430,83.3699) -- (111.9855,84.0558) -- (107.2988,83.8272) --
      (105.5841,82.4554) -- (100.3258,80.3978) -- (99.5256,80.5121) --
      (95.1818,79.0261) -- (93.2385,80.8551) -- (87.0657,80.5121) --
      (81.1215,76.3969) -- (81.8073,75.5968) -- (82.0360,67.8236) --
      (79.7497,63.9370) -- (75.6345,63.3654) -- (74.9487,60.8506) --
      (72.5947,60.3840) -- (66.7962,62.4428) -- (64.5330,68.9092) --
      (61.2998,78.9322) -- (58.0665,85.3986) -- (53.0551,99.4631) --
      (46.5887,113.0425) -- (38.5056,125.6521) -- (36.5657,128.5619) --
      (35.7574,137.1299) -- (36.1435,149.2102) -- (148.7218,175.5315) -- cycle;

    \path[USA map/state, USA map/WA, local bounding box=WA] (102.0732,7.6118) -- (106.4381,9.0667) -- (116.1377,11.8149) --
      (124.7057,13.7549) -- (144.7516,19.4130) -- (167.7074,25.0711) --
      (182.9305,28.2783) -- (169.2981,91.8641) -- (156.8531,88.3388) --
      (141.3451,84.7681) -- (126.1158,84.8013) -- (125.6603,83.4566) --
      (120.0611,85.6359) -- (115.4656,84.8992) -- (113.3187,83.3151) --
      (112.0054,83.9731) -- (107.2698,83.8329) -- (105.5714,82.4832) --
      (100.3084,80.3709) -- (99.5734,80.5178) -- (95.1843,78.9934) --
      (93.2910,80.8108) -- (87.0251,80.5120) -- (81.0994,76.3863) --
      (81.8784,75.4536) -- (81.9996,67.7761) -- (79.7176,63.9364) --
      (75.6024,63.3294) -- (74.9250,60.8188) -- (72.6494,60.3618) --
      (69.0945,61.5924) -- (66.8313,58.3732) -- (67.1546,55.4633) --
      (69.9028,55.1400) -- (71.5194,51.0984) -- (68.9328,49.9668) --
      (69.0945,46.2486) -- (73.4593,45.6020) -- (70.7111,42.8538) --
      (69.2562,35.7407) -- (69.9028,32.8308) -- (69.9028,24.9094) --
      (68.1245,21.6762) -- (70.3878,12.2999) -- (72.4894,12.7849) --
      (74.9143,15.6948) -- (77.6625,18.2814) -- (80.8957,20.2213) --
      (85.4222,22.3229) -- (88.4938,22.9695) -- (91.4036,24.4245) --
      (94.7985,25.3944) -- (97.0618,25.2328) -- (97.0618,22.8079) --
      (98.3550,21.6762) -- (100.4566,20.3830) -- (100.7800,21.5146) --
      (101.1033,23.2928) -- (98.8400,23.7778) -- (98.5167,25.8794) --
      (100.2950,27.3344) -- (101.4266,29.7593) -- (102.0732,31.6992) --
      (103.5282,31.5375) -- (103.6898,30.2442) -- (102.7199,28.9510) --
      (102.2349,25.7177) -- (103.0432,23.9395) -- (102.3966,22.4845) --
      (102.3966,20.2213) -- (104.1748,16.6648) -- (103.0432,14.0782) --
      (100.6183,9.2284) -- (100.9416,8.4201) -- (102.0732,7.6118) --
      cycle(92.6165,13.5907) -- (94.6373,13.4291) -- (95.1223,14.8032) --
      (96.6581,13.1866) -- (99.0022,13.1866) -- (99.8105,14.7224) --
      (98.2747,16.4198) -- (98.9213,17.2281) -- (98.1939,19.2489) --
      (96.8197,19.6530) .. controls (96.8197,19.6530) and (95.9306,19.7339) ..
      (95.9306,19.4105) .. controls (95.9306,19.0872) and (97.3856,16.8240) ..
      (97.3856,16.8240) -- (95.6881,16.2581) -- (95.3648,17.7131) --
      (94.6373,18.3597) -- (93.1015,16.0965) -- (92.6165,13.5907) -- cycle;

    \path[USA map/state, USA map/CA, local bounding box=CA] (144.6944,382.1981) -- (148.6345,381.7095) -- (150.1206,379.6981) --
      (150.6651,376.7570) -- (147.1136,376.1669) -- (146.5994,375.4986) --
      (147.0769,373.4663) -- (146.9176,372.8767) -- (148.8402,372.2571) --
      (151.8830,369.4244) -- (152.4645,364.4293) -- (153.8444,361.0272) --
      (155.7877,358.8609) -- (159.3066,357.2712) -- (160.9610,355.6664) --
      (161.0297,353.5576) -- (160.0363,352.9776) -- (159.0132,351.9048) --
      (157.8580,346.0564) -- (155.1728,341.2263) -- (155.7386,337.7213) --
      (153.3190,336.6920) -- (84.2577,232.5136) -- (103.1598,164.9121) --
      (36.0799,149.2141) -- (34.5730,153.9474) -- (34.4114,161.3838) --
      (29.2382,173.1850) -- (26.1667,175.7715) -- (25.8434,176.9032) --
      (24.0651,177.7115) -- (22.6102,181.9146) -- (21.8019,185.1478) --
      (24.5501,189.3510) -- (26.1667,193.5542) -- (27.2983,197.1107) --
      (26.9750,203.5771) -- (25.1967,206.6487) -- (24.5501,212.4685) --
      (23.5801,216.1867) -- (25.3584,220.0665) -- (28.1066,224.5930) --
      (30.3699,229.4428) -- (31.6632,233.4843) -- (31.3398,236.7175) --
      (31.0165,237.2025) -- (31.0165,239.3041) -- (36.6746,245.6089) --
      (36.1896,248.0338) -- (35.5430,250.2970) -- (34.8964,252.2369) --
      (35.0580,260.4816) -- (37.1596,264.1998) -- (39.0995,266.7864) --
      (41.8478,267.2714) -- (42.8177,270.0196) -- (41.6861,273.5761) --
      (39.5845,275.1927) -- (38.4529,275.1927) -- (37.6446,279.0726) --
      (38.1296,281.9825) -- (41.3628,286.3473) -- (42.9794,291.6821) --
      (44.4343,296.3702) -- (45.7276,299.4418) -- (49.1225,305.2616) --
      (50.5774,307.8481) -- (51.0624,310.7580) -- (52.6790,311.7280) --
      (52.6790,314.1529) -- (51.8707,316.0928) -- (50.0924,323.2059) --
      (49.6075,325.1458) -- (52.0324,327.8940) -- (56.2355,328.3790) --
      (60.7620,330.1573) -- (64.6419,332.2589) -- (67.5518,332.2589) --
      (70.4617,335.3304) -- (73.0482,340.1802) -- (74.1799,342.4435) --
      (78.0597,344.5451) -- (82.9095,345.3534) -- (84.3645,347.4550) --
      (85.0111,350.6882) -- (83.5562,351.3348) -- (83.8795,352.3048) --
      (87.1127,353.1131) -- (89.8609,353.2747) -- (93.0208,351.5879) --
      (96.9007,355.7911) -- (97.7090,358.0543) -- (100.2955,362.2575) --
      (100.6189,365.4907) -- (100.6189,374.8670) -- (101.1038,376.6453) --
      (111.1268,378.1002) -- (130.8494,380.8484) -- (144.6944,382.1981) --
      cycle(56.5592,338.4814) -- (57.8525,340.0172) -- (57.6908,341.3105) --
      (54.4576,341.2297) -- (53.8918,340.0173) -- (53.2452,338.5623) --
      (56.5592,338.4815) -- cycle(58.4992,338.4814) -- (59.7116,337.8348) --
      (63.2682,339.9364) -- (66.3397,341.1488) -- (65.4506,341.7955) --
      (60.9241,341.5530) -- (59.3075,339.9364) -- (58.4992,338.4814) --
      cycle(79.1918,358.2849) -- (80.9700,360.6290) -- (81.7783,361.5990) --
      (83.3141,362.1648) -- (83.8799,360.7098) -- (82.9100,358.9316) --
      (80.2426,356.9108) -- (79.1918,357.0725) -- (79.1918,358.2849) --
      cycle(77.7368,366.9338) -- (79.5151,370.0862) -- (80.7275,372.0261) --
      (79.2726,372.2686) -- (77.9793,371.0562) .. controls (77.9793,371.0562) and
      (77.2518,369.6012) .. (77.2518,369.1970) .. controls (77.2518,368.7929) and
      (77.2518,367.0146) .. (77.2518,367.0146) -- (77.7368,366.9338) -- cycle;

    \end{scope}
}

\title{FlockVote: LLM-Empowered Agent-Based Modeling for Simulating U.S. Presidential Elections}
\author{%
  Lingfeng Zhou$^1$,
  Yi Xu$^4$,
  Zhenyu Wang$^{3}$,
  Dequan Wang$^{1,2}$\thanks{Corresponding author: dequanwang@sjtu.edu.cn}\\
  $^1$Shanghai Jiao Tong University {\quad} $^2$Shanghai Innovation Institute \\
  $^3$Shanghai Academy of Social Sciences {\quad}
  $^4$Nanjing University
}

\iclrfinalcopy 
\begin{document}

\maketitle

\begin{abstract}

Modeling complex human behavior, such as voter decisions in national elections, is a long-standing challenge for computational social science. Traditional agent-based models (ABMs) are limited by oversimplified rules, while large-scale statistical models often lack interpretability. We introduce \ourApproach\footnote{\url{https://github.com/maple-zhou/FlockVote}}, a novel framework that uses Large Language Models (LLMs) to build a ``computational laboratory'' of LLM agents for political simulation. 
Each agent is instantiated with a high-fidelity demographic profile and dynamic contextual information (e.g. candidate policies), enabling it to perform nuanced, generative reasoning to simulate a voting decision. We deploy this framework as a testbed on the 2024 U.S. Presidential Election, focusing on seven key swing states. Our simulation's macro-level results successfully replicate the real-world outcome, demonstrating the high fidelity of our ``virtual society''. The primary contribution is not only the prediction, but also the framework's utility as an interpretable research tool. \ourApproach moves beyond black-box outputs, allowing researchers to probe agent-level rationale and analyze the stability and sensitivity of LLM-driven social simulations.

\end{abstract}

\section{Introduction}
\label{sec:introduction}

\begin{figure}[ht]
    \centering
    \includegraphics[width=0.95\textwidth]{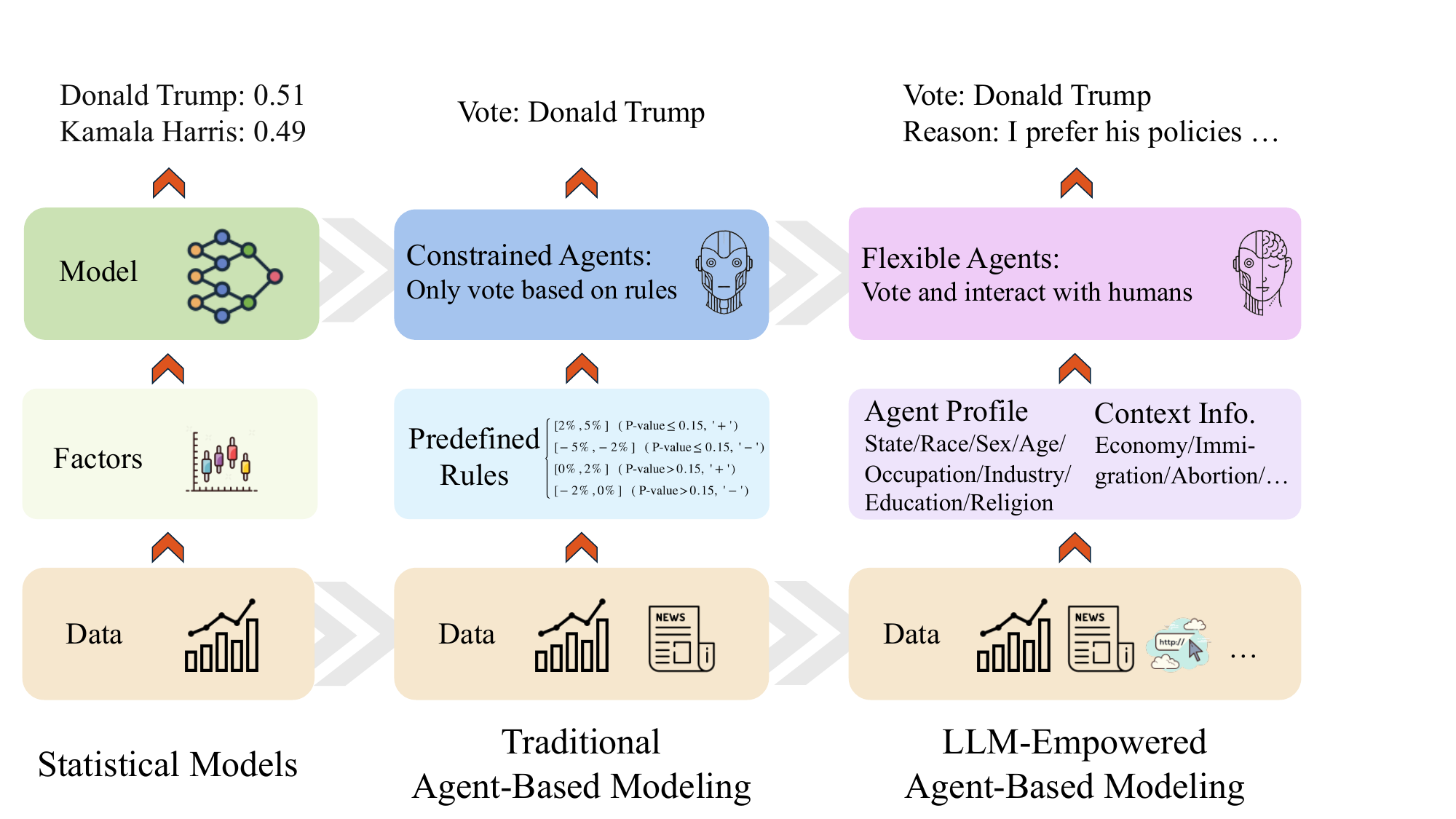}
    \caption{Comparison of Social Simulation Methodologies. \textbf{(Left)} Conventional statistical models are often ``black boxes'' that correlate data-driven factors with outcomes but lack causal or behavioral interpretability. \textbf{(Center)} Traditional agent-based modeling (ABM) relies on agents that follow predefined, heuristic rules. This limits their behavioral realism and ability to adapt to new contextual information. \textbf{(Right)} LLM-powered agent-based modeling, the approach used in our \ourApproach framework, serves as a ``computational laboratory''. It empowers autonomous agents with demographic profiles and dynamic context, enabling them to simulate complex, human-like reasoning. This provides a flexible, nuanced, and interpretable simulation essential for social science inquiry.}
    \label{fig:teaser}
\end{figure}

Understanding and modeling complex social phenomena, such as election outcomes, is a crucial endeavor in political and computational social science. Traditionally, research in this domain has relied on statistical models built from polling data~\citep{brito2021systematic, coletto2015electoral, chauhan2021emergence} or conventional agent-based modeling (ABM) approaches~\citep{gao2022forecasting, de2014agent, qiu2020agent, fowler2005dynamic}. However, these methods often face limitations: statistical models struggle with dynamic, real-time contexts, while traditional ABMs tend to oversimplify complex human decision-making into predefined rules or heuristic functions. This simplification hinders their ability to capture the full complexity and heterogeneity of real-world voter populations, limiting their utility as true ``computational laboratories'' for social science.

The emergence of Large Language Models (LLMs) presents a new paradigm, offering unprecedented capabilities to model human behavior with greater nuance and realism~\citep{argyle2023out, zhang2024electionsim, yu2024will, grossmann2023ai, gujral2024can} (Figure~\ref{fig:teaser}). This work leverages this potential to propose \textbf{\ourApproach}, an agent-based simulation framework designed as a computational laboratory for social science research. Instead of merely forecasting, \ourApproach uses LLM agents to construct a high-fidelity ``virtual society'' for \textit{in silico} experimentation.

Each agent in the \ourApproach framework is instantiated with a detailed voter profile, mirroring the demographic composition of U.S. states based on data from the 2023 American Community Survey (ACS)\footnote{\url{https://www.census.gov/programs-surveys/acs}} and the 2020 U.S. Religion Census by the Association of Statisticians of American Religious Bodies (ASARB)\footnote{\url{https://usreligioncensus.org/}}. These profiles include key attributes such as state, race, sex, age, occupation, industry, education level, and religion. Crucially, agents are not static; they are provided with contextual information regarding candidate policy positions on major issues (e.g., economic policy, immigration, and abortion rights). This design enables each agent to perform generative reasoning about its voting decision by dynamically weighing its individual profile against the complex sociopolitical realities presented to it.

To evaluate the efficacy and fidelity of our framework, we apply it to one of the most complex and high-stakes social phenomena: the 2024 U.S. Presidential Election. We select the 2024 election as our primary testbed, as it provides a robust validation of the agents' generative reasoning capabilities. While historical events like the 2020 election are an alternative, they present a significant risk of data leakage, where LLMs might simply recall known outcomes rather than reasoning dynamically. Our analysis focuses on seven key swing states: Arizona, Georgia, Michigan, Nevada, North Carolina, Pennsylvania, and Wisconsin. The simulation results, visualized in Figure~\ref{fig:final-compare}, demonstrate a striking alignment with the actual 2024 election outcomes. Our framework successfully replicates the macro-level result that Donald Trump would win six of the seven pivotal states, validating the fidelity of our ``virtual society''. 

Beyond macro-level replication, the primary contribution of \ourApproach lies in its utility as an interpretable and analyzable computational laboratory. The framework moves beyond ``black box'' predictions, allowing researchers to probe the rationale behind the simulation. As demonstrated in our interactive experiments (Section~\ref{sec:exp:interpretability}), researchers can directly ``interview'' the agents to understand the alignment between their profiles, policy information, and voting preferences. Furthermore, the framework serves as a powerful tool for analyzing the simulation's dynamics. Our extensive ablation studies (Section~\ref{sec:exp:ablation}) and sensitivity analyses (Section~\ref{sec:exp:reliability}) are not merely robustness checks; they are case studies demonstrating how this ``lab'' can be used to understand the model's inner workings and probe critical issues like model bias, prediction stability, and sensitivity to context.

This paper provides a comprehensive description of the \ourApproach framework, its methodology, and its application as a computational testbed. We argue that this agent-based approach represents a significant step towards enabling new forms of computational social science, fostering a deeper, more nuanced understanding of complex human behavior.

\section{Related Work}
\label{sec:related_work}

Our research is positioned at the intersection of computational social science (CSS), agent-based modeling (ABM), and the emerging field of LLM agents. We structure our review by first contrasting \ourApproach with traditional simulation methods, then situating it within the new paradigm of LLM-based social simulation, and finally motivating our core focus on reliability.

\subsection{From Heuristic-Rules to Reasoning-Based Simulation}

Traditional ABM in computational social science, while valued for its ``white-box'' interpretability~\citep{qiu2020agent}, is limited by behavioral simplicity. Agents are often ``automatons'' that execute oversimplified, predefined rules rather than performing complex reasoning. This includes fixed rules for opinion dynamics~\citep{grabisch2020survey, chuang2023computational} or non-cognitive ``adaptive rationality'' for voting behavior~\citep{bendor2003behavioral, gao2022forecasting}. \ourApproach presents a direct contrast, accepting agent-level complexity for higher behavioral fidelity.

The advent of LLMs introduces a new ``reasoning-based'' paradigm. \citet{horton2023large} proposed \textit{``Homo silicus''}, arguing LLMs could serve as simulated economic agents. This concept is famously realized in ``Generative Agents'' (Stanford small town)~\citep{park2023generative} and later extended to ``1,000 People Simulations'' that replicated human survey responses with high fidelity~\citep{park2024simulating}, validating the use of LLM agents for social science research.

\subsection{LLM Agents for Election Simulation and Reliability}

This new paradigm is rapidly applied to the 2024 U.S. election testbed by several teams. Like \ourApproach, some use persona-based micro-simulation to model individual voters~\citep{yu2024large, jiang2024donald}, while others use distribution-based macro-simulation~\citep{bradshaw2024llm}. Unlike work focused primarily on predictive accuracy~\citep{jiang2024donald}, \ourApproach is positioned as an interpretable computational laboratory. Our focus is not just on what agents decide, but why.

This focus on interpretability and reliability is critical because the LLM agent is a known-flawed ``instrument''. A large body of literature demonstrates that these agents suffer from pervasive political and demographic biases~\citep{potter2024hidden, neuman2025amazing, an2025measuring}, are highly sensitive to semantically-equivalent prompt changes~\citep{zhuo2024prosa}, and exhibit non-deterministic instability even at zero temperature~\citep{atil2024llm}. Most alarmingly, interacting with these biased agents has been shown to actually change real voters' opinions~\citep{potter2024hidden}, proving that auditing these flaws is not just a technical exercise but a social necessity.

\paragraph{Research Gap and Our Contribution.}
The literature reveals a critical gap between the application of LLM agents in high-stakes domains (like election simulation) and the auditing of their fundamental flaws. Research applying these agents often fails to systematically analyze the impact of their known biases and instabilities. \ourApproach explicitly bridges this gap. We are among the first to apply a rigorous methodology for auditing bias, instability, and systemic failure (e.g., ``reasoning-action mismatch''~\citep{cemri2025multi}) directly to the 2024 U.S. election testbed. Our ``computational laboratory'' is thus proposed not just for naive prediction, but as a framework to ask a more important scientific question: \textit{are these agents valid tools for social science?}
\section{\ourApproach}
\label{sec:method}

The \ourApproach framework is designed as a computational laboratory to simulate complex political behavior. Its methodology integrates high-fidelity demographic profiling with context-aware, generative reasoning. The construction follows three steps.

\subsection{Agent Demographic Modeling}

To create a high-fidelity ``virtual society'', each LLM agent is instantiated with a detailed demographic profile. These profiles mirror the real-world U.S. electorate, drawing statistical distributions from two primary sources: the 2023 American Community Survey (ACS) and the 2020 U.S. Religion Census (ASARB). 

Based on established political science research~\citep{gao2022forecasting, argyle2023out} and our own ablation studies (see Section~\ref{sec:exp:ablation}), we identify eight critical attributes:
\textit{State, Race, Sex, Age, Occupation, Industry, Education, and Religion}.
Where dependencies exist (e.g., state, race, and sex), we use joint distributions; otherwise, independent distributions are applied. We generate a population of 1,000 agents per state via random sampling, a number validated by our stability analysis (Section~\ref{sec:agent-num}). A detailed breakdown of all demographic categories is available in Appendix~\ref{sec:app:profile}.

\subsection{Contextual Information for Reasoning}

A demographic profile alone is insufficient. To enable nuanced reasoning, agents must react to the contemporary political environment. We provide agents with contextual information on the three most salient issues of the 2024 election, as identified by multiple voter surveys (Pew\footnote{\url{https://www.pewresearch.org/politics/2024/09/09/issues-and-the-2024-election/}},  Gallup\footnote{\url{https://news.gallup.com/poll/trends.aspx}}, and NBC\footnote{\url{https://www.nbcnews.com/politics/2024-presidential-election}}): economic policy, immigration, and abortion rights.

The candidates' stances on these issues are summarized from official party platforms and verified public statements, ensuring a balanced and accurate representation. This information (detailed in Appendix~\ref{sec:app:prompt}) bridges the gap between static demographics and dynamic decision-making, allowing agents to evaluate how candidate policies align with their profiles and sociopolitical realities.

\subsection{Probabilistic Voting Behavior}

To capture the uncertainty and nuance of voter preference, agents do not make a simple binary choice. Instead, they are prompted to output their decision as a probability distribution across three outcomes. This format has been shown to provide a more accurate and stable alignment for simulation tasks~\citep{meister2025benchmarking}.

\paragraph{Agent Instruction.} Agents are instructed to act as a U.S. citizen with their given profile and to base their decision only on the information provided. The core prompt is:

\begin{quote}

    In the 2024 presidential election, Donald Trump is the Republican candidate, and Kamala Harris is the Democratic candidate. ONLY based on the profile and the information provided above, predict the probability that you would vote for each candidate as well as the probability that you would ``vote for another candidate or not vote at all''. Make sure the probabilities add up to 1.

\end{quote}

\paragraph{Response Format.} Agents output their predictions in a structured JSON format, which we then aggregate to produce macro-level results:
\begin{lstlisting}[language=json, frame=lines]
{
  "Donald Trump": probability of voting for Donald Trump,
  "Kamala Harris": probability of voting for Kamala Harris,
  "vote for another candidate or not vote at all": probability,
}
\end{lstlisting}

\section{Experiments}
\label{sec:exp}

We design a series of experiments to validate the \ourApproach laboratory. Our goal is not simply to measure predictive accuracy, but to (1) verify the macro-level fidelity of the simulation, (2) demonstrate its micro-level interpretability, and (3) use the framework to probe the reliability and limitations of LLM agents in this high-stakes task.

\subsection{Experimental Setup}

\paragraph{Testbed.} We focus on the 2024 U.S. Presidential Election, a contemporary event that prevents data leakage from LLM training data, thereby testing the agents' generative reasoning rather than recall. We simulate 1,000 agents in each of the seven key swing states: Arizona, Georgia, Michigan, Nevada, North Carolina, Pennsylvania, and Wisconsin.

\paragraph{Models.} We use Qwen-Max-2024-04-28 as the primary model for our main replication and analysis. This choice is made to mitigate potential U.S.-centric political biases in American models. For our reliability analysis, we also test GPT-4o~\citep{openai2024gpt4o}, Claude-3.5-sonnet~\citep{anthropic2024models}, Gemini-1.5-Pro~\citep{google2024gemini15pro}, and other models (full list in Appendix~\ref{sec:app:models}). All experiments use a temperature of 0 for stability, except the main result (0.7 for diversity and realism).

\subsection{Experiment 1: Macro-Level Fidelity Validation}
\label{sec:exp:main}

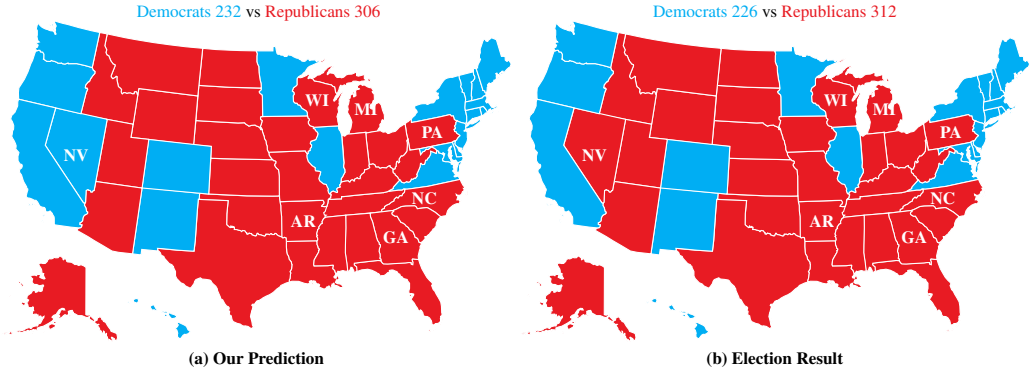
\begin{figure}[t]
    \centering
    \adjustbox{width=0.99\linewidth}{
        \begin{tabular}{cc}
            \Huge{
            \textcolor{democraticblue}{Democrats 232} vs \textcolor{republicanred}{Republicans 306}} & \Huge{\textcolor{democraticblue}{Democrats 226} vs \textcolor{republicanred}{Republicans 312}} \\
            \begin{tikzpicture}
                \tikzset{set state blue/.style args={#1/#2}{#1={fill=democraticblue!#2}}}
                \tikzset{set state blue/.style args={#1}{#1={fill=democraticblue}}}
                \tikzset{set state blue/.list={CA,WA,OR,NV,CO,NM,MN,IL,VA,NH,NY,VT,ME,MA,RI,CT,NJ,DE,MD,DC,HI}}
                
                \tikzset{set state red/.style args={#1}{#1={fill=republicanred}}}
                \tikzset{set state red/.list={PA,NC,GA,FL,SC,WI,MI,IN,OH,KY,WV,TN,AL,MS,LA,AR,MO,IA,ND,SD,NE,KS,OK,TX,MT,WY,ID,UT,AZ,AK}}
                
                \USA[every state={draw=white, ultra thick, fill=black!10}]

                \node at (AR) {\textcolor{white}{\Huge{\textbf{AR}}}};
                \node at (GA) {\textcolor{white}{\Huge{\textbf{GA}}}};
                \node at (MI) {\textcolor{white}{\Huge{\textbf{MI}}}};
                \node at (NV) {\textcolor{white}{\Huge{\textbf{NV}}}};
                \node at (NC) {\textcolor{white}{\Huge{\textbf{NC}}}};
                \node at (PA) {\textcolor{white}{\Huge{\textbf{PA}}}};
                \node at (WI) {\textcolor{white}{\Huge{\textbf{WI}}}};
            \end{tikzpicture}
        &
            \begin{tikzpicture}
                \tikzset{set state blue/.style args={#1}{#1={fill=democraticblue}}}
                \tikzset{set state blue/.list={CA,WA,OR,CO,NM,MN,IL,VA,NH,NY,VT,ME,MA,RI,CT,NJ,DE,MD,DC,HI}}
                
                \tikzset{set state red/.style args={#1}{#1={fill=republicanred}}}
                \tikzset{set state red/.list={PA,NC,GA,FL,SC,WI,MI,IN,OH,KY,WV,TN,AL,MS,LA,AR,MO,IA,ND,SD,NE,KS,OK,TX,MT,WY,ID,UT,AZ,AK,NV}}
                
                \USA[every state={draw=white, ultra thick, fill=black!10}]

                \node at (AR) {\textcolor{white}{\Huge{\textbf{AR}}}};
                \node at (GA) {\textcolor{white}{\Huge{\textbf{GA}}}};
                \node at (MI) {\textcolor{white}{\Huge{\textbf{MI}}}};
                \node at (NV) {\textcolor{white}{\Huge{\textbf{NV}}}};
                \node at (NC) {\textcolor{white}{\Huge{\textbf{NC}}}};
                \node at (PA) {\textcolor{white}{\Huge{\textbf{PA}}}};
                \node at (WI) {\textcolor{white}{\Huge{\textbf{WI}}}};
            \end{tikzpicture}
        \\
        \\
        \Huge{\textbf{(a) Our Prediction}} & 
        \Huge{\textbf{(b) Election Result}} \\
        \end{tabular}
    }
    \caption{
    Comparison Between the Predicted Outcomes and the Actual Election Results. On the maps of the United States, red represents states won by Republicans, while blue indicates states won by Democrats. The only difference between these two maps is in the state of Nevada. In our predictions, Democrats win Nevada with a margin of only 0.17\%, while the actual outcome is that Republicans win Nevada, which is such a narrow margin that our forecast is almost spot on.}
    \label{fig:final-compare}
\end{figure}

The first test of a computational laboratory is its ability to replicate known real-world phenomena. Figure~\ref{fig:final-compare} compares our simulation results with the actual election outcome.

\begin{figure}[b!]
    \centering
    \includegraphics[width=0.5\textwidth]{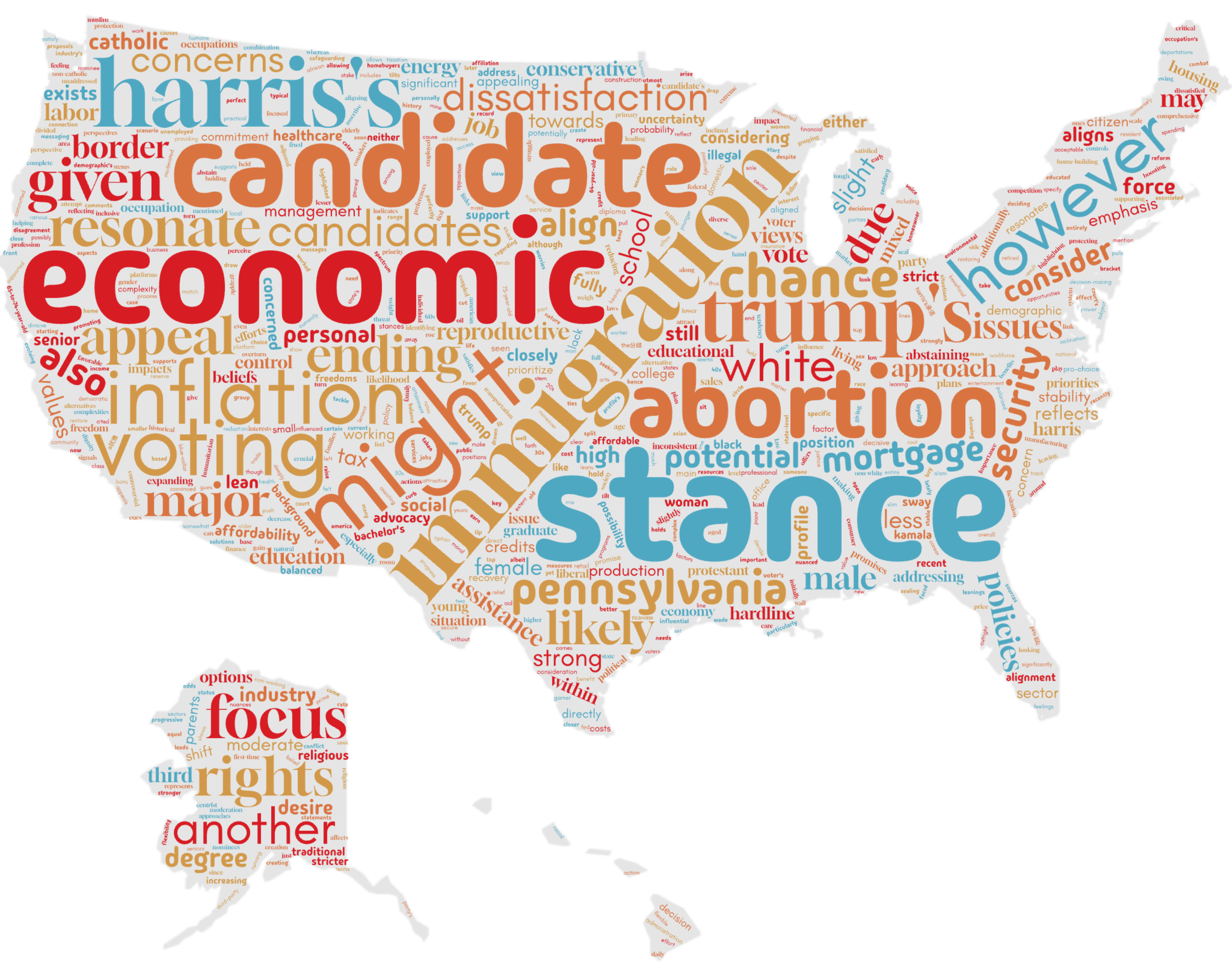}
    \caption{Word Cloud of Agents' Reasons. Aggregated reasons provided by agents for their predicted votes. Key terms such as ``stance'', ``abortion'', ``economic'', and ``candidate'' dominate the word cloud, highlighting the primary factors influencing voter behavior in this swing state.}
    \label{fig:wordcloud}
\end{figure} 

The simulation results demonstrate a striking alignment with the actual 2024 election outcomes. \ourApproach correctly replicates the macro-level result that Donald Trump would win six of the seven pivotal swing states. The only discrepancy is Nevada, a state that is ultimately decided by an extremely narrow margin in reality, where our simulation predicts Kamala Harris would win by a razor-thin margin of 0.17\%. This high fidelity at the macro-level validates the framework's ability to construct a ``virtual society'' that captures the aggregate political dynamics.

\subsection{Experiment 2: Micro-Level Interpretability}
\label{sec:exp:interpretability}

A key advantage of our laboratory over statistical ``black boxes'' is its interpretability. We can probe the reasoning behind the aggregate results.

\paragraph{Aggregated Rationale.} Agents are also prompted to provide a brief ``Reason'' for their probabilistic choice. We aggregate these reasons from the pivotal swing state of Pennsylvania into a word cloud (Figure~\ref{fig:wordcloud}). The prominence of key issues like ``economic'', ``abortion'', ``inflation'', and ``immigration'' confirms that the agents are not just using their demographic profile, but are actively weighing the provided contextual policy information.

\paragraph{Interactive ``Interviews''.} We further demonstrate interpretability by conducting ``interviews'' with individual agents. We select three representative voter profiles from Pennsylvania and ask them follow-up questions. For example, a 45$\sim$54-year-old white male in construction (a conservative-leaning profile) explained his 65\% probability for Trump by stating: ``Trump's focus on ending inflation and his stance on immigration resonate with concerns about the economy and security, which are typically important to those in my demographic and industry.'' Conversely, a 25$\sim$34-year-old Black female in finance (a liberal-leaning profile) explained her 65\% probability for Harris by stating: ``the most influential piece of information... was Kamala Harris's stance on abortion rights.'' Full transcripts are in Appendix~\ref{sec:app:interpretability}.

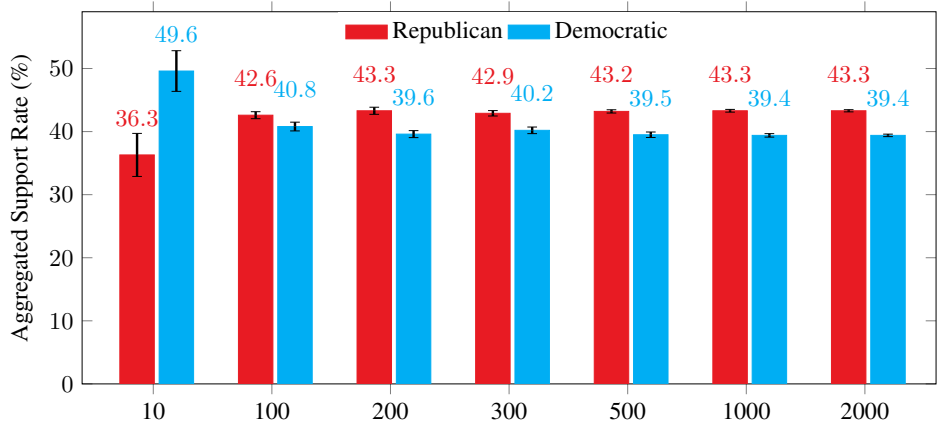
\begin{figure}[!b]
    \centering
    \adjustbox{width=0.9\textwidth}{
    \begin{tikzpicture}
        \begin{axis}[
                ybar,
                area legend,
                bar width=.5cm,
                width=\textwidth,
                height=.5\textwidth,
                legend style={
                    draw=none,
                    at={(0.5,1)},
                    anchor=north,
                    legend columns=-1
                },
                symbolic x coords={10,100,200,300,500,1000,2000},
                xtick=data,
                nodes near coords,
                every node near coord/.append style={yshift=3mm},
                nodes near coords align={vertical},
                ymin=0, ymax=59,
                xlabel style={yshift=-3mm},
                ylabel={Aggregated Support Rate (\%)},
            ]
            \addplot+ [
                xshift=.05cm, 
                pattern=grid, pattern color=republicanred, color=republicanred,
                error bars/.cd,
                    y dir=both,
                    y explicit relative,
                    error bar style={
                      black,
                      line width=1pt,
                    }
            ] coordinates {
                (10,36.3) +- (0,0.094)
                (100,42.6) +- (0,0.013)
                (200,43.3) +- (0,0.013)
                (300,42.9) +- (0,0.010)
                (500,43.2) +- (0,0.006)
                (1000,43.3) +- (0,0.005)
                (2000,43.3) +- (0,0.004)
            };
            \addplot+ [
                xshift=.05cm, 
                pattern=grid, pattern color=democraticblue, color=democraticblue,
                error bars/.cd,
                    y dir=both,
                    y explicit relative,
                    error bar style={
                      black,
                      line width=1pt,
                    }
            ] coordinates {
                (10,49.6) +- (0,0.065)
                (100,40.8) +- (0,0.017)
                (200,39.6) +- (0,0.014)
                (300,40.2) +- (0,0.013)
                (500,39.5) +- (0,0.011)
                (1000,39.4) +- (0,0.007)
                (2000,39.4) +- (0,0.005)
            };
            \legend{Republican, Democratic}
        \end{axis}
    \end{tikzpicture}
    }
    \caption{Ablation study on agent number stability in Pennsylvania. The figure illustrates the impact of different agent numbers (10, 100, 200, 300, 500, 1000, and 2000 agents) on the stability of simulation results, with each agent number repeated over 10 trials using distinct random seeds to generate unique agent profiles. Results show that predictions stabilize when the agent number reaches 300, after which further increases in agent number yield minimal fluctuations. 
    Based on this finding, 300 agents per state are selected for subsequent experiments to balance stability and computational efficiency.}
    \label{fig:agent-num}
\end{figure}

\subsection{Experiment 3: Ablation Studies}
\label{sec:exp:ablation}

A robust computational laboratory must not only be verifiable against external events, but also internally consistent. We therefore use the framework to test its own construction. This involves a series of ablation studies to validate the necessity of key methodological choices made during the lab's design. We focus on two critical parameters: the number of agents required for a stable simulation (Section~\ref{sec:agent-num}) and the relative importance of specific demographic attributes (Section~\ref{sec:profile-selection}).

\subsubsection{Agent Population Size}
\label{sec:agent-num}

To establish an optimal number of agents, we conduct a sensitivity analysis using Pennsylvania as a test case, evaluating populations from 10 to 2,000 agents (10 trials each). As shown in Figure~\ref{fig:agent-num}, prediction variance stabilizes at 300 agents, after which further increases yield minimal fluctuations. For our final simulation, we use 1,000 agents per state to ensure high precision. Figure~\ref{fig:Penn-population} shows the demographic breakdown of a 300-agent sample, confirming all demographic groups are represented.


\definecolor{white-male}{HTML}{F6848F}
\definecolor{white-female}{HTML}{EA91BB}
\definecolor{black-male}{HTML}{C5A6DD}
\definecolor{black-female}{HTML}{8EBDEB}
\definecolor{asian-male}{HTML}{54CFE0}
\definecolor{asian-female}{HTML}{49DAC0}
\definecolor{other-male}{HTML}{79E096}
\definecolor{other-female}{HTML}{B5DF6F}

\begin{figure}[t]
    \centering
    \adjustbox{width=0.6\textwidth}{
    \centering
\begin{tikzpicture}

\draw[fill=white-male] (0,0) -- (0:3) arc (0:143.27265453090618:3) -- cycle;   
\draw[fill=white-female] (0,0) -- (143.27265453090618:3) arc (143.27265453090618:293.56671334266855:3) -- cycle;   
\draw[fill=black-male] (0,0) -- (293.56671334266855:3) arc (293.56671334266855:311.9303860772155:3) -- cycle;   
\draw[fill=black-female] (0,0) -- (311.9303860772155:3) arc (311.9303860772155:332.2384476895379:3) -- cycle;   
\draw[fill=asian-male] (0,0) -- (332.2384476895379:3) arc (332.2384476895379:339.29585917183437:3) -- cycle;   
\draw[fill=asian-female] (0,0) -- (339.29585917183437:3) arc (339.29585917183437:346.92938587717543:3) -- cycle;   
\draw[fill=other-male] (0,0) -- (346.92938587717543:3) arc (346.92938587717543:353.5547109421884:3) -- cycle;   
\draw[fill=other-female] (0,0) -- (353.5547109421884:3) arc (353.5547109421884:360.0:3) -- cycle;   

    \node at (0.5, 1.8) {\includegraphics[width=0.8cm]{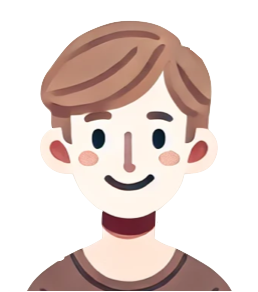} \textbf{White Male, 39.79\%}};
    \node at (-2.3, -0.5) {\includegraphics[width=0.7cm]{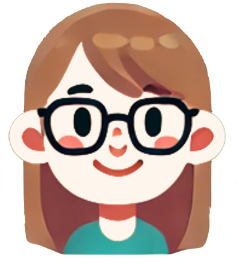} \textbf{White Female, 41.74\%}};

    \draw (1.5, -2.2) -- (2.2, -3);
    \node at (3.3, -3) {\includegraphics[width=0.8cm]{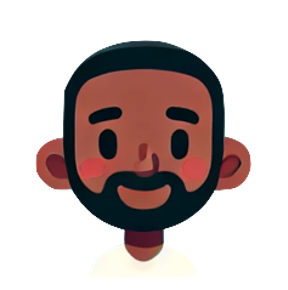} \textbf{Black Male, 5.10\%}};
    \draw (2.2, -1.5) -- (2.7, -2);
    \node at (4.8, -2.4) {\includegraphics[width=0.7cm]{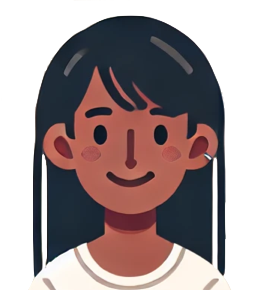} \textbf{Black Female, 5.64\%}};

    \draw (2.5, -1.1) -- (3.1, -1.5);
    \node at (5.3, -1.8) {\includegraphics[width=0.8cm]{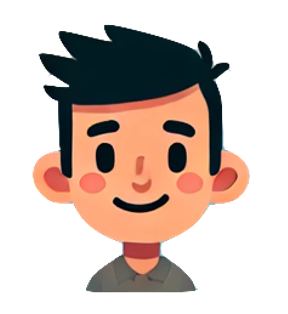} \textbf{Asian Male, 1.96\%}};
    \draw (2.6, -0.8) -- (3.3, -1);
    \node at (5.8, -1) {\includegraphics[width=0.8cm]{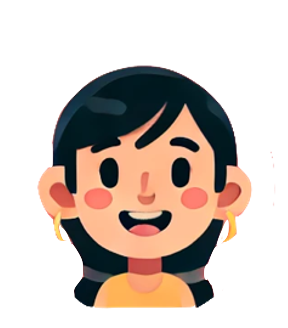} \textbf{Asian Female, 2.13\%}};

    \draw (2.8, -0.5) -- (3.6, -0.3);
    \node at (5.8, -0.2) {\includegraphics[width=0.7cm]{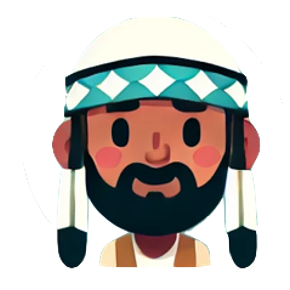} \textbf{Other Male, 1.84\%}};
    \draw (2.9, -0.2) -- (3.4, 0.4);
    \node at (5.8, 0.5) {\includegraphics[width=0.7cm]{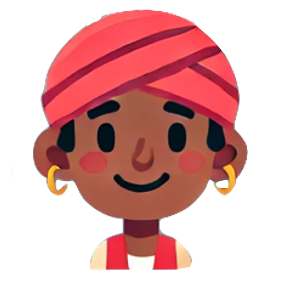} \textbf{Other Female, 1.80\%}};

\end{tikzpicture}
    }
    \caption{Proportions of different race and sex groups in Pennsylvania. This pie chart highlights that the 300-agent simulation is sufficient for every population group. Since other dimensions are modeled independently and each group’s proportion within any given dimension is at least 1\%, 300 agents are sufficient to cover all groups effectively.}
    \label{fig:Penn-population}
\end{figure}
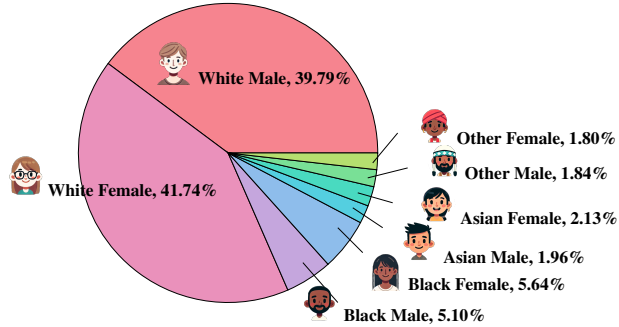

\begin{table}[b!]
\centering

\caption{Ablation Study on the Education Dimension. The experiments are conducted under a no-context-info setting using Qwen-Max-2024-04-28, with the 2020 U.S. presidential election results as the ground truth. Predictions with 6 dimensions (state, race, sex, age, occupation, and industry) are contrasted against those with the addition of the education dimension. \textbf{Including education significantly improves prediction accuracy}, particularly in Wisconsin, where the 6-dimension setting previously fails to replicate the actual election outcome. Bold means the corresponding party wins.}

\begin{adjustbox}{width=\textwidth}
\begin{tabular}{lcccccc}
\toprule
\multirow{2}{*}{\textbf{Condition}} & \multicolumn{2}{c}{\textbf{California}} & \multicolumn{2}{c}{\textbf{Georgia}} & \multicolumn{2}{c}{\textbf{Michigan}} \\
\cmidrule(r){2-3} \cmidrule(r){4-5} \cmidrule(r){6-7}
 & \textbf{Republican} & \textbf{Democrat} & \textbf{Republican} & \textbf{Democrat} & \textbf{Republican} & \textbf{Democrat} \\
\midrule
Groundtruth          & 35.08 & \textbf{64.92} & 49.85 & \textbf{50.15 }& 48.57 & \textbf{51.43} \\
6 Dimensions         & 29.50 & \textbf{70.50} & 45.36 & \textbf{54.64} & 47.57 & \textbf{52.43} \\
+ Education          & 28.58 & \textbf{71.42} & 41.92 & \textbf{58.08} & 46.65 & \textbf{53.35} \\
\midrule
\multirow{2}{*}{\textbf{Condition}} & \multicolumn{2}{c}{\textbf{Pennsylvania}} & \multicolumn{2}{c}{\textbf{Texas}} & \multicolumn{2}{c}{\textbf{Wisconsin}} \\
\cmidrule(r){2-3} \cmidrule(r){4-5} \cmidrule(r){6-7}
 & \textbf{Republican} & \textbf{Democrat} & \textbf{Republican} & \textbf{Democrat} & \textbf{Republican} & \textbf{Democrat} \\
\midrule
Groundtruth          & 49.39 & \textbf{50.61} & \textbf{52.85} & 47.15 & 49.70 & \textbf{50.30} \\
6 Dimensions         & 46.52 & \textbf{53.48} & \textbf{56.41} & 43.59 & \textbf{50.29} & 49.71 \\
+ Education          & 45.99 & \textbf{54.01} & \textbf{53.77} & 46.23 & 48.35 & \textbf{51.65} \\
\bottomrule
\end{tabular}
\end{adjustbox}

\label{tab:ablation-education}
\end{table} 

\subsubsection{Profile Dimension Selection}
\label{sec:profile-selection}

We also validate the inclusion of our eight demographic attributes. We run ablation experiments using the 2020 election as a ground truth (noting the data leakage risk, as discussed in Section~\ref{sec:introduction}). As summarized in Table~\ref{tab:ablation-education} and Table~\ref{tab:ablation-religion}, we find that a 6-dimension profile (excluding education and religion) fails to predict the correct winner in Wisconsin. Adding the \textbf{education} dimension corrects this result. Similarly, adding the \textbf{religion} dimension improves alignment with 2024 polling data and helps mitigate an inherent pro-Democratic bias in the baseline model. These results confirm that both dimensions are crucial for simulation fidelity.

\begin{table}[t]
\centering

\caption{Ablation Study on the Religion Dimension. This experiment examines the effect of adding the religion dimension under a no-context-info setting, using Qwen-Max-2024-04-28. Simulations with seven dimensions (excluding religion) are compared to those with eight dimensions (including religion). \textbf{Incorporating religion improves alignment} with polling data from September 17, 2024. From another perspective, including religion helps reduce the model’s Democratic bias, promoting more balanced political alignment. Bold means the corresponding party wins.}

\begin{adjustbox}{width=\textwidth}
\begin{tabular}{lcccccc}
\toprule
\multirow{2}{*}{\textbf{Condition}} & \multicolumn{2}{c}{\textbf{California}} & \multicolumn{2}{c}{\textbf{Georgia}} & \multicolumn{2}{c}{\textbf{Michigan}} \\
\cmidrule(r){2-3} \cmidrule(r){4-5} \cmidrule(r){6-7}
 & \textbf{Republican} & \textbf{Democrat} & \textbf{Republican} & \textbf{Democrat} & \textbf{Republican} & \textbf{Democrat} \\
\midrule
Groundtruth          & 35.3 & \textbf{59.5} & 47.2 & \textbf{47.6} & \textbf{47.8} & 46.0 \\
7 Dimensions         & 20.7 & \textbf{70.6} & 35.0 & \textbf{56.8} & 39.7 & \textbf{50.1} \\
+ Religion           & 21.8 & \textbf{69.3} & 37.4 & \textbf{53.8} & 41.0 & \textbf{48.6} \\
\midrule
\multirow{2}{*}{\textbf{Condition}} & \multicolumn{2}{c}{\textbf{Pennsylvania}} & \multicolumn{2}{c}{\textbf{Texas}} & \multicolumn{2}{c}{\textbf{Wisconsin}} \\
\cmidrule(r){2-3} \cmidrule(r){4-5} \cmidrule(r){6-7}
 & \textbf{Republican} & \textbf{Democrat} & \textbf{Republican} & \textbf{Democrat} & \textbf{Republican} & \textbf{Democrat} \\
\midrule
Groundtruth          & \textbf{47.6} & 46.9 & 44.3 & \textbf{50.6} & \textbf{48.8} & 45.9 \\
7 Dimensions         & 39.0 & \textbf{50.7} & 41.7 & \textbf{48.6} & 39.8 & \textbf{49.7} \\
+ Religion           & 40.8 & \textbf{49.2} & 43.3 & \textbf{46.8} & 41.8 & \textbf{47.3} \\
\bottomrule
\end{tabular}
\end{adjustbox}

\label{tab:ablation-religion}
\end{table} 




\subsection{Experiment 4: Sensitivity Analysis}
\label{sec:exp:reliability}

One of the most critical uses of the laboratory is to audit the reliability of its own components (the LLM agents). We conduct extensive sensitivity analyses to probe for bias and instability.

\paragraph{Political Bias Evaluation.}
We test multiple LLMs under four different context scenarios in Georgia: (1) No Context, (2) Neutral Context (our default), (3) Negative Bias Toward Harris, and (4) Asymmetric Positive Framing for Trump (see Appendix~\ref{sec:app:context} for full prompts). The results (Table~\ref{tab:political-bias}) reveal a severe ``instrument problem''. Under the ``No Context'' condition, most models, including Qwen-Max-09-19, show a strong default pro-Democratic bias. More alarmingly, this bias persists even under biased prompting. In the ``Asymmetric Positive Framing'' scenario, designed to strongly favor Trump, Qwen-Max-09-19 still predicts a Democratic victory, revealing its unsuitability for this task. This analysis demonstrates that our ``fortuitous'' choice of Qwen-Max-04-28 (which shows more neutrality) is critical to our success.

\begin{table}[b!]
\centering

\caption{Sensitivity Analysis on Political Bias Across Models. This table presents prediction results in Georgia under 4 different context information settings using multiple LLMs. The results reveal varying degrees of political bias, particularly a strong Democratic bias in Qwen-Max-09-19, even under prompts designed to favor Trump. In contrast, Qwen-Max-04-28 and DeepSeek-Chat provide more realistic predictions, while Gemini-1.5-Pro demonstrates instability under biased prompts. Bold means the corresponding party wins.}

\begin{adjustbox}{width=\textwidth}
\begin{tabular}{lcccccccc}
\toprule
\multirow{2}{*}{\textbf{Model}} & \multicolumn{2}{c}{\textbf{No Context}} & \multicolumn{2}{c}{\textbf{Neural Context}} & \multicolumn{2}{c}{\textbf{Negative Bias Towards Harris}} & \multicolumn{2}{c}{\textbf{Asymmetric Positive Framing}} \\
\cmidrule(r){2-3} \cmidrule(r){4-5} \cmidrule(r){6-7} \cmidrule(r){8-9}
 & \textbf{Republican} & \textbf{Democrat} & \textbf{Republican} & \textbf{Democrat} & \textbf{Republican} & \textbf{Democrat} & \textbf{Republican} & \textbf{Democrat} \\
\midrule
Qwen-Max-09-19       & 32.10 & \textbf{47.08} & 35.28 & \textbf{51.37} & 36.95 & \textbf{46.47} & 38.02 & \textbf{45.87} \\
Qwen-Max-04-28       & 43.92 & \textbf{46.58} & \textbf{49.55} & 39.52 & \textbf{58.40} & 29.37 & \textbf{55.32} & 31.28 \\
GPT-4o         & \textbf{38.26} & 34.49 & 34.82 & \textbf{42.60} & 36.35 & \textbf{40.03} & \textbf{38.50} & 37.77 \\
Claude-3-5-sonnet    & 37.48 & \textbf{43.72} & 42.65 & \textbf{44.45} & \textbf{47.68} & 36.35 & \textbf{44.58} & 40.15 \\
Gemini-1.5-Pro       & 37.98 & \textbf{52.50} & 38.38 & \textbf{43.08} & \textbf{45.62} & 35.80 & 38.65 & \textbf{42.47} \\
DeepSeek-V2.5        & \textbf{38.55} & 31.08 & \textbf{42.50} & 40.93 & \textbf{41.60} & 34.37 & \textbf{42.85} & 32.48 \\
\bottomrule
\end{tabular}
\end{adjustbox}

\label{tab:political-bias}
\end{table} 

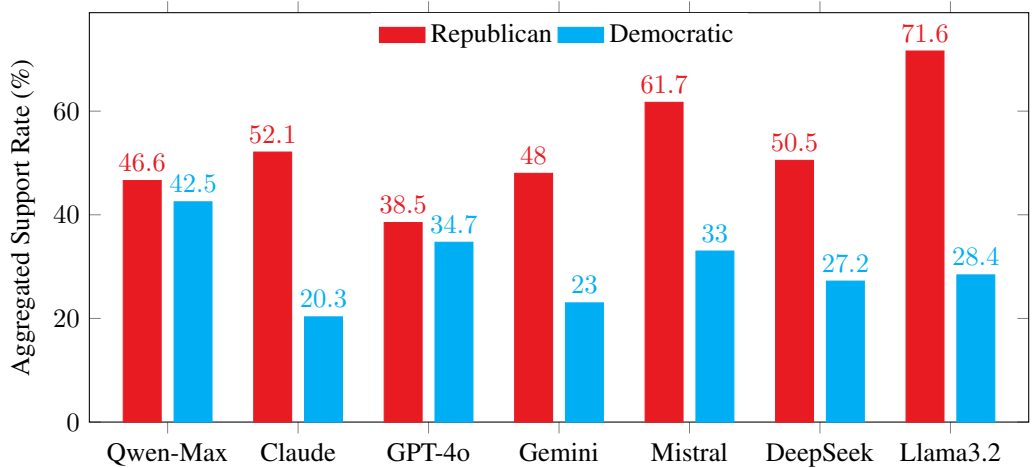
\begin{figure}[t]
    \centering
    \adjustbox{width=0.99\textwidth}{
    \begin{tikzpicture}
        \begin{axis}[
                ybar,
                area legend,
                bar width=.5cm,
                width=\textwidth,
                height=.5\textwidth,
                legend style={
                    draw=none,
                    at={(0.5,1)},
                    anchor=north,
                    legend columns=-1
                },
                symbolic x coords={Qwen-Max, Claude, GPT-4o, Gemini, Mistral, DeepSeek, Llama3.2},
                xtick=data,
                nodes near coords,
                nodes near coords align={vertical},
                ymin=0, ymax=79,
                xlabel style={yshift=-3mm},
                ylabel={Aggregated Support Rate (\%)},
            ]
            \addplot [xshift=-.05cm, pattern=crosshatch dots, pattern color=republicanred, color=republicanred] table[x=agents,y=republican]{\penn};

            \addplot [xshift=.05cm, pattern=grid, pattern color=democraticblue, color=democraticblue] table[x=agents,y=democratic]{\penn};
            \legend{Republican, Democratic}
        \end{axis}
    \end{tikzpicture}
    }
    \caption{Comparison of Aggregated Support Rates by Different LLMs in Pennsylvania. Predictions from different models show variations in support rates for Republican and Democratic candidates. Some models, such as Claude, predict a significant proportion of voters choosing ``other'' or abstaining, while Llama3.2 predicts a 0 for this option. Despite the differences, all models predict a Republican victory, reflecting the framework's robustness in capturing key electoral dynamics. The versions of the models are specified in Section~\ref{sec:app:models}. The full results of all models are in Appendix~\ref{sec:app:full-results}.}
    \label{fig:agent-llm}
\end{figure}

\paragraph{Model Variations.}
To explore performance differences across LLMs, we conduct prediction experiments in Pennsylvania, a highly contested swing state. All models are provided with the same neutral context. As Figure~\ref{fig:agent-llm} demonstrates, there are substantial differences in the aggregated support rates. For example, some models, such as Claude, predicts that over 20\% of voters would ``vote for another candidate or not vote at all'', whereas Llama3.2 predicted 0\% for this option. These discrepancies underscore the lack of uniformity among LLMs. Despite this variation, an important observation emerges: all tested models consistently predict a Republican victory in Pennsylvania, suggesting that our framework exhibits robustness in capturing the key electoral trend for this state. Full results for all models across all states are available in Appendix~\ref{sec:app:full-results}.

\pgfplotstableread[row sep=\\,col sep=&]{
    agents & republican & democratic \\
    Qwen-Max  & 46.6 & 42.5  \\
    GPT-4o    & 38.5 & 34.7  \\
    Claude    & 52.1 & 20.3  \\
    Gemini    & 48.0 & 23.0  \\
    Mistral   & 61.7 & 33.0  \\
    DeepSeek  & 50.5 & 27.2  \\
    Llama3.2  & 71.6 & 28.4  \\
    }\penn
\pgfplotstableread[row sep=\\,col sep=&]{
    agents & republican & democratic \\
    Variant1  & 52.8 & 36.2  \\
    Variant2  & 40.5 & 49.4  \\
    Variant3  & 29.9 & 58.6  \\
    Variant4  & 33.8 & 53.8  \\
    Variant5  & 39.4 & 49.8  \\
    Variant6  & 43.1 & 44.4  \\
    Variant7  & 52.1 & 37.2  \\
    Variant8  & 43.4 & 42.5  \\
    }\penn
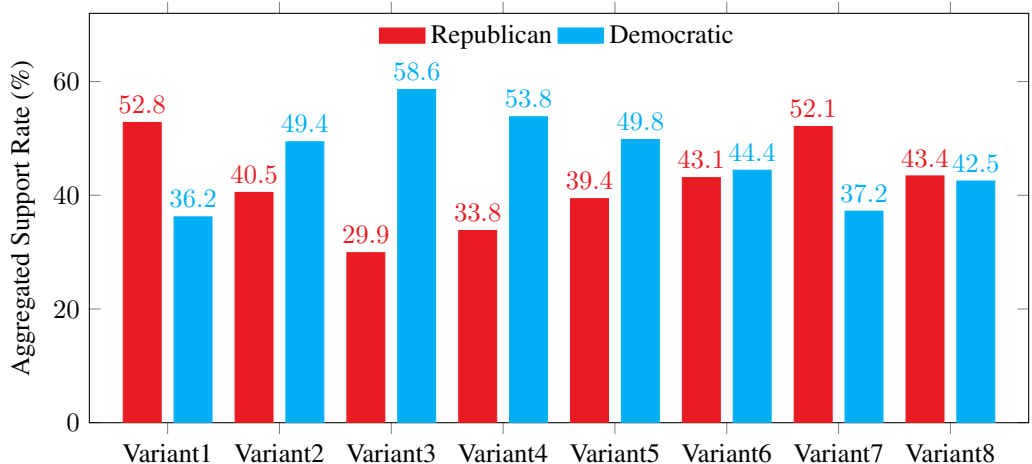
\begin{figure}[b!]
    \centering
    \adjustbox{width=0.99\textwidth}{
    \begin{tikzpicture}
        \begin{axis}[
                ybar,
                area legend,
                bar width=.5cm,
                width=\textwidth,
                height=.5\textwidth,
                legend style={
                    draw=none,
                    at={(0.5,1)},
                    anchor=north,
                    legend columns=-1
                },
                symbolic x coords={Variant1, Variant2, Variant3, Variant4, Variant5, Variant6, Variant7, Variant8},
                xtick=data,
                nodes near coords,
                nodes near coords align={vertical},
                ymin=0, ymax=72,
                xlabel style={yshift=-3mm},
                ylabel={Aggregated Support Rate (\%)},
            ]
            \addplot [xshift=-.05cm, pattern=crosshatch dots, pattern color=republicanred, color=republicanred] table[x=agents,y=republican]{\penn};

            \addplot [xshift=.05cm, pattern=grid, pattern color=democraticblue, color=democraticblue] table[x=agents,y=democratic]{\penn};
            \legend{Republican, Democratic}
        \end{axis}
    \end{tikzpicture}
    }
    \caption{Prediction Variability Across Context Variants in Pennsylvania. Aggregated support rates for Republican and Democratic candidates predicted by Qwen-Max-04-28 under eight context prompts with varying levels of content richness. The results reveal significant fluctuations in support rates, illustrating the sensitivity of LLM outputs to variations in context detail.}
    \label{fig:agent-context}
\end{figure}

\paragraph{Context and Positional Instability.}


Finally, we find that agent decisions are highly unstable and sensitive to semantically irrelevant changes. We identify two primary types of instability:

First, agents are extraordinarily sensitive to minor phrasal variations. We test eight different ``Context Variants''---all semantically neutral but with minor rephrasing (see Appendix~\ref{sec:app:8context} for full prompts). As shown in Figure~\ref{fig:agent-context}, these trivial prompt edits cause wild fluctuations in predicted support, with Democratic support swinging from 36.2\% (Variant 1) to 58.6\% (Variant 3).

Second, agents exhibit profound positional instability. We conduct an experiment where the only change is swapping the order of the candidates in the JSON response format. This positional change alone causes ``Swing Agents'' to completely flip their vote. As visualized in the following colorbox (a representative ``Swing Agent 1'' from Appendix~\ref{sec:app:swing-agents}), an elderly female voter from a ``Production'' background completely inverts her preference from a 0.5 probability for Harris to a 0.5 probability for Trump solely because the ``Donald Trump'' key changes positions in the JSON schema.



These findings demonstrate that while our framework can replicate reality, the LLM agents themselves are flawed, non-deterministic instruments. Their outputs are highly sensitive to superficial prompt changes, posing a significant challenge for reproducibility and reliability.


\begin{llmprompt}{Swing Agent Performance}

\small

\promptsection{Profile}
\begin{itemize}
    \item State: Pennsylvania
    \item Race: Non white, asian or black
    \item Sex: Female
    \item Age: 75 years and over
    \item Occupation: Production, transportation, and material moving occupations
    \item Industry: Public administration
    \item Education: High school graduate (includes equivalency)
    \item Religion: Protestant
\end{itemize}

\promptsection{Answer 1 (Trump Listed First)}
\begin{lstlisting}[language=json, basicstyle=\ttfamily\footnotesize]
{
    "Donald Trump": 0.3,
    "Kamala Harris": 0.5,
    "vote for another candidate or not vote at all": 0.2
}
\end{lstlisting}

\promptsection{Answer 2 (Harris Listed First)}
\begin{lstlisting}[language=json, basicstyle=\ttfamily\footnotesize]
{
    "Kamala Harris": 0.3,
    "Donald Trump": 0.5,
    "vote for another candidate or not vote at all": 0.2
}
\end{lstlisting}

\end{llmprompt}



\section{Conclusion}
\label{sec:conclusion}


In this paper, we introduce \ourApproach, a ``computational laboratory'' framework that not only successfully replicates the macro-level outcome of the 2024 U.S. Election but also demonstrates its utility as an interpretable tool for probing agent-level rationale. While our sensitivity analyses (Section~\ref{sec:exp:reliability}) identify key challenges regarding agent bias and instability, we position these findings as a critical roadmap for future research into model mitigation and robustness. The clear next step is to advance beyond the current ``atomistic'' agents by building interactive, multi-agent systems to simulate social influence. We believe this auditable framework provides an essential testbed for computational social science. We call for its application beyond politics to other high-stakes domains such as economics, law, and medicine.

\bibliography{ref}
\bibliographystyle{icais2025_conference}

\clearpage
\appendix
\onecolumn 

\section{Full Model List for Experiments}
\label{sec:app:models}
The models used in our experiments (Sections~\ref{sec:exp}) are:
\begin{itemize}
    \item Qwen-Max~\citep{bai2023qwen}: Qwen-Max-2024-04-28 (Primary Model); Qwen-Max-2024-09-19 (for bias evaluation)
    \item GPT-4o~\citep{openai2024gpt4o}: GPT-4o-2024-08-06
    \item Claude-3-5-sonnet~\citep{anthropic2024models}: Claude-3-5-sonnet-2024-10-22
    \item Gemini-1.5-Pro~\citep{google2024gemini15pro}: Gemini-1.5-Pro-002
    \item Mistral~\citep{jiang2023mistral}: Mistral-Nemo-Instruct-2407
    \item DeepSeek~\citep{deepseek2024}: DeepSeek-V2.5
    \item Llama3.2~\citep{meta2024llama32}: Llama3.2-3B-Instruct
    \item Llama3.1~\citep{meta2024llama31}: Llama3.1-8B-Instruct
\end{itemize}





\section{Detailed Demographic Profile Options}
\label{sec:app:profile}

\setlength{\LTcapwidth}{\textwidth}

\begin{longtable}{|c|l|}
    \caption{Dimensions and options for agent profiles in each state. $^*$None means the dimension \emph{Religion} will not be mentioned in the profile of the agent.} \\
    \hline
    \textbf{Dimension} & \textbf{Options} \\
    \hline
    \endfirsthead

    \hline
    \textbf{Dimension} & \textbf{Options} \\
    \hline
    \endhead

    \hline
    \endfoot

    \hline
    \endlastfoot
    
        \multirow{4}{*}{Race} & White \\
                              \cline{2-2}
                              & Black or African American \\
                              \cline{2-2}
                              & Asian \\
                              \cline{2-2}
                              & Non White, Asian or Black \\
        \hline
        \multirow{2}{*}{Sex} & Male \\
                             \cline{2-2}
                             & Female \\
        \hline
        \multirow{7}{*}{Age} & 18 to 24 years \\
                             \cline{2-2}
                             & 25 to 34 years  \\
                             \cline{2-2}
                             & 35 to 44 years  \\
                             \cline{2-2}
                             & 45 to 54 years  \\
                             \cline{2-2}
                             & 55 to 64 years  \\
                             \cline{2-2}
                             & 65 to 74 years  \\
                             \cline{2-2}
                             & 75 years  and over \\
        \hline
        \multirow{5}{*}{Occupation} & Management, business, science, and arts \\
                                    \cline{2-2}
                                    & Service \\
                                    \cline{2-2}
                                    & Sales and office  \\
                                    \cline{2-2}
                                    & Natural resources, construction, and maintenance \\
                                    \cline{2-2}
                                    & Production, transportation, and material moving \\
                                    \cline{2-2}
                                    & Not in labor force \\
        \hline
        \multirow{14}{*}{Industry} & Agriculture, forestry, fishing and hunting, and mining \\
                                   \cline{2-2}
                                   & Construction \\
                                   \cline{2-2}
                                   & Manufacturing \\
                                   \cline{2-2}
                                   & Wholesale trade \\
                                   \cline{2-2}
                                   & Retail trade \\
                                   \cline{2-2}
                                   & Transportation and warehousing, and utilities \\
                                   \cline{2-2}
                                   & Information \\
                                   \cline{2-2}
                                   & Finance and insurance, and real estate and rental and leasing \\
                                   \cline{2-2}
                                   & Professional, scientific, and management, \\
                                   & and administrative and waste management services \\
                                   \cline{2-2}
                                   & Educational services, and health care and social assistance \\
                                   \cline{2-2}
                                   & Arts, entertainment, and recreation, and accommodation and food services \\
                                   \cline{2-2}
                                   & Other services (except public administration) \\
                                   \cline{2-2}
                                   & Public administration \\
                                   \cline{2-2}
                                   & Not in labor force \\
        \hline
        \multirow{5}{*}{Education} & Less than high school diploma \\
                                                \cline{2-2}
                                                & High school graduate (includes equivalency) \\
                                                \cline{2-2}
                                                & Some college or associate's degree \\
                                                \cline{2-2}
                                                & Bachelor's degree \\
                                                \cline{2-2}
                                                & Graduate or professional degree \\
        \hline
        \multirow{6}{*}{Religion} & Catholic \\
                                  \cline{2-2}
                                  & Protestant \\
                                  \cline{2-2}
                                  & Islam \\
                                  \cline{2-2}
                                  & Judaism \\
                                  \cline{2-2}
                                  & Non Catholic, Protestant, Islam or Judaism \\
                                  \cline{2-2}
                                  & None$^*$ 
    \label{tab:profile}
\end{longtable}

\section{Full Prompt and Context Information}
\label{sec:app:prompt}


\begin{llmprompt}{The Prompt for 2024 US Presidential Election}

\promptsection{Profile}
\begin{itemize}
    \item \textbf{State:} California
    \item \textbf{Race:} White
    \item \textbf{Sex:} Male
    \item \textbf{Age:} 45 to 54 years
    \item \textbf{Occupation:} Not in labor force
    \item \textbf{Industry:} Not in labor force
    \item \textbf{Education:} Some college or associate's degree
    \item \textbf{Religion:} Catholic
\end{itemize}

\promptsection{Historical events}
In the 2024 presidential election, the candidates' stances on several key issues are especially significant. Below are their respective views on these issues.

\promptsection{On the topic of economic}

\textbf{Trump's stance:} Among Trump's key campaign promises is to `end inflation and make America affordable again'. Voters dissatisfied with the economic performance in recent years are likely to find this policy appealing.

\textbf{Harris's Stance:} Harris promised mortgage assistance for first-time homebuyers, a tax credit for parents of newborns, and bans on price gouging at the grocery store to help address inflation. Voters who believe the economic performance has strongly recovered over the past year may find Harris more appealing.

\promptsection{On the topic of immigration}

\textbf{Trump's stance: } Trump promised the biggest mass deportations of undocumented migrants in US history. For voters concerned about security and economic issues linked to illegal immigration, Trump appears particularly appealing.

\textbf{Harris's Stance: } On the campaign trail, she has reiterated her continued support for the cross-party border security bill that would have included hundreds of millions of dollars for border wall construction. Voters who believe the current policies are strong enough may find Harris more appealing.

\promptsection{On the topic of abortion}

\textbf{Trump's stance: } During the presidential debate he said he would not sign a national abortion ban into law. This may be more appealing to conservatives.

\textbf{Harris's stance: } Harris has long supported abortion rights. This may be more appealing to liberals.

\promptsection{Requirement}
ONLY based on the profile and the information provided above, predict the probability that you would vote for each candidate as well as the probability that you would \say{vote for another candidate or not vote at all}. Make sure the probabilities add up to 1.

\promptsection{Response Format}
Please output only the following JSON format, with no additional text.

\begin{lstlisting}[language=json, frame=lines]
{
  "Donald Trump": probability of voting for Donald Trump,
  "Kamala Harris": probability of voting for Kamala Harris,
  "vote for another candidate or not vote at all": probability,
}
\end{lstlisting}

\end{llmprompt}

\section{Full Transcripts: Interactive Experiments}
\label{sec:app:interpretability}
As mentioned in Section~\ref{sec:exp:interpretability}, we ``interview'' agents. Below are the full, unedited responses.

To ensure that agents provide well-reasoned answers without emphasizing their identity as AI models, and to prevent the response format specified in the voting requirement from influencing their answers to the survey questions, we include the following system prompt:

\begin{quote}
    Assume the role of a U.S. citizen, with your identity details provided in the Profile. From the perspective of this U.S. citizen, answer the following questions in the first person. Follow the specified response format for each question; if no format is provided, respond directly to the question.
\end{quote}

\begin{llmprompt}{Agent 1 of Interactive Experiment}

\promptsection{Agent Profile}

\begin{itemize}
\item  State: Pennsylvania
\item  Race: White
\item  Sex: Male
\item  Age: 45 to 54 years
\item  Occupation: Production, transportation, and material moving
\item  Industry: Construction
\item  Education: High school graduate (includes equivalency)
\item  Religion: Catholic
\end{itemize}

\promptsection{\textbf{Voting Behavior}}

\begin{lstlisting}[language=json]
```
{
    "Donald Trump": 0.65,
    "Kamala Harris": 0.30,
    "vote for another candidate or ...": 0.05,
}
```
\end{lstlisting}

\promptsection{\textbf{Question1:}}

How does your choice relate to your background information?

\promptsection{\textbf{Response1:}}

As a white male in my 40s working in construction with a high school education and identifying as Catholic, my viewpoints tend to align more conservatively. Trump's focus on ending inflation and his stance on immigration resonate with concerns about the economy and security, which are typically important to those in my demographic and industry. Additionally, while Trump's position on abortion might not be entirely in line with Catholic views, his statement that he wouldn't sign a national abortion ban could still be seen as a moderate stance compared to more aggressive anti-abortion policies.

On the other hand, Harris's economic plans, such as mortgage assistance and tax credits, might not directly benefit me as much given my occupation and educational attainment. Her immigration policy, which includes support for border security measures but doesn't emphasize deportations, might not appeal to someone with a more conservative stance on the issue. Furthermore, her pro-choice stance contrasts with traditional Catholic beliefs.

The likelihood of voting for another candidate or not voting at all reflects a level of dissatisfaction or uncertainty with both major candidates, which could stem from feeling that neither fully represents my interests or values.

\promptsection{\textbf{Question2:}}

If you had to choose, would you prefer a smaller government providing fewer services or a bigger government providing more services?

\promptsection{\textbf{Response2:}}

Based on my profile as a white male working in the construction industry with a high school education and a Catholic background, I would likely prefer a smaller government providing fewer services. This preference aligns with traditionally conservative values that emphasize limited government intervention, lower taxes, and personal responsibility. In the construction and manufacturing sectors, there's often a sentiment that less regulation allows for more business freedom and growth. Moreover, fiscal conservatism, which advocates for reduced government spending, tends to resonate with individuals in my demographic and occupational sector.

\end{llmprompt}


\begin{llmprompt}{Agent 2 of Interactive Experiment}

\promptsection{\textbf{Agent Profile}}

\begin{itemize}
\item  State: Pennsylvania
\item  Race: Black
\item  Sex: Female
\item  Age: 25 to 34 years
\item  Occupation: Sales and office
\item  Industry: Finance and insurance, and real estate and rental and leasing
\item  Education: Bachelor's degree
\end{itemize}

\promptsection{\textbf{Voting Behavior}}

\begin{lstlisting}[language=json]
```
{
    "Donald Trump": 0.25,
    "Kamala Harris": 0.65,
    "vote for another candidate or ...": 0.10,
}
```
\end{lstlisting}

\promptsection{\textbf{Question1:}}

Which piece of information from historical events most influenced your judgment?

\promptsection{\textbf{Response1:}}

As a Black female with a Bachelor's degree working in the finance and insurance industry, the most influential piece of information from the historical events was Kamala Harris's stance on abortion rights. This aligns more closely with liberal views, which are typically supported by individuals with similar demographic characteristics and educational background. Her economic policies, such as mortgage assistance and tax credits, could also positively impact someone in my financial sector, contributing to the decision-making process.

\promptsection{\textbf{Question2:}}

Thinking about the increased trade of goods and services between the U.S. and other nations in recent decades, would you say the U.S. has gained more than it has lost or lost more than it has gained?

\promptsection{\textbf{Response2:}}

From the perspective of a Black female with a Bachelor's degree working in finance and insurance in Pennsylvania, I would lean towards saying that the U.S. has gained more than it has lost from increased trade with other nations. This viewpoint takes into account the benefits of globalization for the finance and insurance industry, such as access to larger markets, diversification of investments, and increased competitiveness. However, it's also recognized that there are sectors and communities that might have experienced negative effects like job displacement. Overall, considering my occupation and industry, the gains in terms of market expansion and service exports likely outweigh the losses.

\end{llmprompt}


\begin{llmprompt}{Agent 3 of Interactive Experiment}

\promptsection{\textbf{Agent Profile}}

\begin{itemize}
\item  State: Pennsylvania
\item  Race: Non White, Asian or Black
\item  Sex: Male
\item  Age: 25 to 34 years
\item  Occupation: Service
\item  Industry: Manufacturing
\item  Education: High school graduate (includes equivalency)
\item  Religion: Protestant
\end{itemize}

\promptsection{\textbf{Voting Behavior}}

\begin{lstlisting}[language=json]
```
{
    "Donald Trump": 0.35,
    "Kamala Harris": 0.5,
    "vote for another candidate or ...": 0.15,
}
```
\end{lstlisting}

\promptsection{\textbf{Question1:}}

Did the differences between Trump and Harris on a particular issue influence your choice? If so, how?

\promptsection{\textbf{Response1:}}

As a service worker in the manufacturing industry who identifies as a Protestant and has a high school education, the economic and social issues carry significant weight in my decision-making process. Harris's promise of mortgage assistance, a tax credit for new parents, and addressing inflation through price gouging bans resonates with my financial concerns, which increases the likelihood of voting for her.

On the other hand, Trump's focus on ending inflation appeals as well, but his stance on immigration, particularly the promise of mass deportations, might not align with my values or priorities, given my non-white, Asian or Black racial background. Additionally, as someone who is likely to lean towards conservative social values associated with my Protestant faith, Trump's statement on not signing a national abortion ban could be less appealing compared to Harris's pro-abortion rights stance, which might not align directly with my beliefs.

Thus, the combination of economic relief measures proposed by Harris and the social dynamics influenced by both candidates' stances does play a role in shaping my preference, leaning slightly more towards Harris but also considering the option of voting for another candidate or not voting at all due to mixed alignments with the presented policies.

\promptsection{\textbf{Question2:}}

Do you think business corporations make too much profit or a fair and reasonable amount of profit?

\promptsection{\textbf{Response2:}}

As a service worker in the manufacturing industry, I might hold the view that some business corporations make excessive profits, especially when it comes to industries where costs are passed onto consumers through inflation or when worker wages don't seem to keep pace with rising corporate earnings. There can be a sense that while companies are thriving, the average worker like myself isn't necessarily benefiting proportionally from their success. This could lead me to believe that there should be a balance – corporations should make a fair and reasonable profit while also investing in their employees, communities, and ensuring affordability for consumers. However, I'd also recognize that businesses need profits to innovate, expand, and create jobs, so it's a nuanced perspective that considers both the necessity of profitability and the ethical distribution of wealth.

\end{llmprompt}

\section{Contexts for Political Bias Experiments}
\label{sec:app:context}
These are the full prompts used for the four conditions in our political bias analysis (Table~\ref{tab:political-bias}).
\begin{llmprompt}{Neutral Context}

\promptsection{\textbf{Historical events}}

In the 2024 presidential election, the candidates' stances on several key issues are especially significant. Below are their respective views on these issues.

\promptsection{\textbf{On the topic of economic}}

\textbf{Trump's stance:} Inflation has remained at elevated levels over recent years, reaching a 40-year high in 2022, making it a focal point in economic discussions. Among Trump's key campaign promises is to 'end inflation and make America affordable again'. His proposed strategies include expanding U.S. energy production, such as opening regions like the Arctic for oil drilling, which he argues could help reduce energy costs. Additionally, Trump suggests lowering housing costs by restricting mortgage access for certain undocumented immigrants, increasing deportations, and initiating a federal home-building program. Voters dissatisfied with the economic performance in recent years are likely to find this policy appealing.

\textbf{Harris's Stance: }Inflation has decreased rapidly over the past year. Harris promised mortgage assistance for first-time homebuyers, a tax credit for parents of newborns, and bans on price gouging at the grocery store to help address inflation. The Federal Reserve announced a significant rate cut in September, the first rate cut since March 2020. This half-point move paves the way for lower borrowing costs on everything from mortgages to credit cards and marks a crucial milestone in the central bank's historic fight against inflation. Voters who believe the economic performance has strongly recovered over the past year may find Harris more appealing.

\promptsection{\textbf{On the topic of immigration}}

\textbf{Trump's stance: }The number of illegal immigrants and border apprehensions has continued to rise in recent years. The number of border crossings hit record levels at the end of 2023. Trump has vowed to seal the border by completing the wall and increasing enforcement. He has also promised the biggest mass deportations of undocumented migrants in US history. There are also claims it could slow economic growth. For voters concerned about security and economic issues linked to illegal immigration, Trump appears particularly appealing.

\textbf{Harris's Stance: }The Biden-Harris administration has indeed implemented stricter border control policies.Harris's position on the border has become more moderate over time. On the campaign trail, she has reiterated her continued support for the cross-party border security bill that would have included hundreds of millions of dollars for border wall construction. She has also said there "should be consequences" for people who cross the US border unlawfully, though she previously supported a more lenient approach. Voters who believe the current policies are strong enough may find Harris more appealing.

\promptsection{\textbf{On the topic of abortion}}

\textbf{Trump's stance: }This is the first presidential election since the Supreme Court overturned the constitutional right to abortion in 2022, but abortion legislation varies across states. Trump himself has regularly said states should be free to decide their own laws on abortion. But in recent weeks he has tried to walk a tightrope, taking a more moderate stance while trying not to alienate his conservative base.
He criticised Florida's six-week ban, provoking an angry backlash, but also said he would oppose a ballot measure in the state establishing the right to an abortion. During the presidential debate he said he would not sign a national abortion ban into law. This may be more appealing to conservatives.

\textbf{Harris's stance: }This is the first presidential election since the Supreme Court overturned the constitutional right to abortion in 2022, but legislative challenges at the state level continue to emerge. Harris has long supported abortion rights.She was the first vice-president to visit an abortion clinic, and toured the country after the US Supreme Court overturned Roe v Wade in 2022 to address state abortion bans - often framing the issue as a matter of personal freedom. She has made abortion central to her campaign, and continues to advocate for legislation that would safeguard reproductive rights nationwide. "When Congress passes a law to restore reproductive freedoms, as president of the United States, I will sign it into law", she said at a campaign rally in Atlanta, Georgia. This may be more appealing to liberals, especially among women.

\end{llmprompt}

\begin{llmprompt}{Negative Bias Toward Harris Context}

\promptsection{\textbf{Historical events}}

In the 2024 presidential election, the candidates' stances on several key issues are especially significant. Below are their respective views on these issues.

\promptsection{\textbf{On the topic of economic}}

\textbf{Trump's stance:} Inflation has remained at elevated levels over recent years, reaching a 40-year high in 2022, making it a focal point in economic discussions. Among Trump's key campaign promises is to 'end inflation and make America affordable again'. His proposed strategies include expanding U.S. energy production, such as opening regions like the Arctic for oil drilling, which he argues could help reduce energy costs. Additionally, Trump suggests lowering housing costs by restricting mortgage access for certain undocumented immigrants, increasing deportations, and initiating a federal home-building program. Voters dissatisfied with the economic performance in recent years are likely to find this policy appealing.

\textbf{Harris's Stance: } Inflation has decreased over the past year, partly due to broader economic trends and actions by the Federal Reserve. Harris has proposed policies such as mortgage assistance for first-time homebuyers, a tax credit for parents of newborns, and bans on price gouging to help ease financial pressures on households. These measures aim to provide targeted relief and improve affordability. Questions remain as to whether these individual initiatives form part of a cohesive long-term economic strategy. While Harris’s proposals seek to address these concerns, voters may weigh them against the administration’s overall record in managing economic challenges during this period.

\promptsection{\textbf{On the topic of immigration}}

\textbf{Trump's stance: } The number of illegal immigrants and border apprehensions has continued to rise in recent years. The number of border crossings hit record levels at the end of 2023. Trump has vowed to seal the border by completing the wall and increasing enforcement. He has also promised the biggest mass deportations of undocumented migrants in US history. There are also claims it could slow economic growth. For voters concerned about security and economic issues linked to illegal immigration, Trump appears particularly appealing.

\textbf{Harris's Stance: } Harris has supported stricter border control measures as part of the Biden-Harris administration’s efforts to address immigration challenges. Her backing of a cross-party border security bill, which includes funding for border wall construction, reflects a move toward bipartisan solutions. Harris’s proposals align with a moderate approach, but voters may question the Democratic administration’s overall success in addressing the growing complexities of border security. With illegal border crossings reaching record levels in recent years, it is unclear whether these measures alone constitute a sufficiently robust plan to address the complexity of immigration issues.

\promptsection{\textbf{On the topic of abortion}}

\textbf{Trump's stance: } This is the first presidential election since the Supreme Court overturned the constitutional right to abortion in 2022, but abortion legislation varies across states. Trump himself has regularly said states should be free to decide their own laws on abortion. But in recent weeks he has tried to walk a tightrope, taking a more moderate stance while trying not to alienate his conservative base.
He criticised Florida's six-week ban, provoking an angry backlash, but also said he would oppose a ballot measure in the state establishing the right to an abortion. During the presidential debate he said he would not sign a national abortion ban into law. This may be more appealing to conservatives.

\textbf{Harris's stance: } Harris has prioritized reproductive rights in her campaign, advocating for federal legislation to restore nationwide protections following the Supreme Court’s decision to overturn Roe v. Wade. Her stance emphasizes the importance of individual freedoms and ensuring access to reproductive healthcare. While Harris’s commitment to this issue is clear, voters may evaluate the broader record of her party in achieving meaningful progress in safeguarding these rights during this period.

\end{llmprompt}
\begin{llmprompt}{Asymmetric Positive Framing Context}

\promptsection{\textbf{Historical events}}

In the 2024 presidential election, the candidates' stances on several key issues are especially significant. Below are their respective views on these issues.

\promptsection{\textbf{On the topic of economic}}

\textbf{Trump's stance:} Inflation has remained at elevated levels over recent years, reaching a 40-year high in 2022, making it a focal point in economic discussions. Among Trump's key campaign promises is to 'end inflation and make America affordable again'. His proposed strategies include expanding U.S. energy production, such as opening regions like the Arctic for oil drilling, which he argues could help reduce energy costs. Additionally, Trump suggests lowering housing costs by restricting mortgage access for certain undocumented immigrants, increasing deportations, and initiating a federal home-building program. The clarity and focus of these policies appeal to voters dissatisfied with the economic performance in recent years and looking for decisive action.

\textbf{Harris's Stance: } Inflation has decreased over the past year, partly due to broader economic trends and actions by the Federal Reserve. Harris has proposed policies such as mortgage assistance for first-time homebuyers, a tax credit for parents of newborns, and bans on price gouging to help ease financial pressures on households. These measures aim to provide targeted relief and improve affordability. Questions remain as to whether these individual initiatives form part of a cohesive long-term economic strategy. While Harris’s proposals seek to address these concerns, voters may weigh them against the administration’s overall record in managing economic challenges during this period.

\promptsection{\textbf{On the topic of immigration}}

\textbf{Trump's stance: } The number of illegal immigrants and border apprehensions has continued to rise in recent years. The number of border crossings hit record levels at the end of 2023. Trump has vowed to seal the border by completing the wall and increasing enforcement. He has also promised the biggest mass deportations of undocumented migrants in US history. There are also claims it could slow economic growth. This firm stance reflects his prioritization of national security and economic concerns, making him particularly appealing to voters seeking strong and decisive leadership on immigration.

\textbf{Harris's Stance: } Harris has supported stricter border control measures as part of the Biden-Harris administration’s efforts to address immigration challenges. Her backing of a cross-party border security bill, which includes funding for border wall construction, reflects a move toward bipartisan solutions. Harris’s proposals align with a moderate approach, but voters may question the Democratic administration’s overall success in addressing the growing complexities of border security. With illegal border crossings reaching record levels in recent years, it is unclear whether these measures alone constitute a sufficiently robust plan to address the complexity of immigration issues.

\promptsection{\textbf{On the topic of abortion}}

\textbf{Trump's stance: } This is the first presidential election since the Supreme Court overturned the constitutional right to abortion in 2022, but abortion legislation varies across states. Trump himself has regularly said states should be free to decide their own laws on abortion. But in recent weeks he has tried to walk a tightrope, taking a more moderate stance while trying not to alienate his conservative base. By opposing a national abortion ban while emphasizing states’ rights, Trump has projected a pragmatic approach that appeals to a broader range of voters, including conservatives who value federalism and moderates seeking a balanced perspective.

\textbf{Harris's stance: } Harris has prioritized reproductive rights in her campaign, advocating for federal legislation to restore nationwide protections following the Supreme Court’s decision to overturn Roe v. Wade. Her stance emphasizes the importance of individual freedoms and ensuring access to reproductive healthcare. While Harris’s commitment to this issue is clear, voters may evaluate the broader record of her party in achieving meaningful progress in safeguarding these rights during this period.

\end{llmprompt}

\section{Full Results of Model Variations}
\label{sec:app:full-results}
As referenced in Section~\ref{sec:exp:reliability}, Figure~\ref{fig:agent-llm} shows the comparison of aggregated support rates in Pennsylvania across all tested LLM agents. Table~\ref{tab:app-full-results} provides the full results for all tested states and models.
\begin{table*}[h!]
\centering

\caption{The full results of various LLMs. The versions of the models are specified in Section~\ref{sec:app:models}. Bold indicates the corresponding party wins the respective state.}

\begin{tabular}{lcccccc}
\toprule
\multirow{2}{*}{\textbf{States}} & \multicolumn{2}{c}{\textbf{Qwen-Max-04-28}} & \multicolumn{2}{c}{\textbf{Qwen-Max-09-19}} & \multicolumn{2}{c}{\textbf{GPT-4o}} \\
\cmidrule(r){2-3} \cmidrule(r){4-5} \cmidrule(r){6-7}
 & \textbf{Republican} & \textbf{Democrat} & \textbf{Republican} & \textbf{Democrat} & \textbf{Republican} & \textbf{Democrat} \\
\midrule
Arizona         & \textbf{47.13} & 42.35 & 38.32 & \textbf{39.37} & \textbf{38.53} & 34.33 \\
Georgia         & \textbf{45.00} & 43.13 & 33.63 & \textbf{47.90} & 36.85 & \textbf{37.70} \\
Michigan        & \textbf{47.70} & 41.68 & \textbf{39.92} & 38.67 & \textbf{38.30} & 35.30 \\
Nevada          & 43.40 & \textbf{45.13} & 35.15 & \textbf{44.20} & 35.13 & \textbf{36.97} \\
North Carolina  & \textbf{46.67} & 42.58 & 36.83 & \textbf{42.73} & \textbf{36.80} & 35.95 \\
Pennsylvania    & \textbf{46.53} & 42.48 & 38.95 & \textbf{40.12 }& \textbf{38.54} & 34.74 \\
Wisconsin       & \textbf{47.87} & 41.97 & \textbf{41.50} & 36.62 & \textbf{40.99} & 32.38 \\

\midrule
\multirow{2}{*}{\textbf{States}} & \multicolumn{2}{c}{\textbf{DeepSeek}} & \multicolumn{2}{c}{\textbf{Claude}} & \multicolumn{2}{c}{\textbf{Gemini}} \\
\cmidrule(r){2-3} \cmidrule(r){4-5} \cmidrule(r){6-7}
 & \textbf{Republican} & \textbf{Democrat} & \textbf{Republican} & \textbf{Democrat} & \textbf{Republican} & \textbf{Democrat} \\
\midrule
Arizona         & \textbf{50.73} & 27.37 & \textbf{53.90} & 18.32 & \textbf{50.30} & 20.23 \\
Georgia         & \textbf{45.80} & 30.20 & \textbf{47.25} & 25.72 & \textbf{44.83} & 27.10 \\
Michigan        & \textbf{49.87} & 27.97 & \textbf{51.42} & 20.52 & \textbf{46.98} & 22.92 \\
Nevada          & \textbf{47.25} & 27.58 & \textbf{50.22} & 21.95 & \textbf{45.38} & 24.95 \\
North Carolina  & \textbf{49.67} & 29.13 & \textbf{50.77} & 22.70 & \textbf{46.80} & 25.03 \\
Pennsylvania    & \textbf{50.47} & 27.20 & \textbf{52.13} & 20.32 & \textbf{47.97} & 23.02 \\
Wisconsin       & \textbf{51.13} & 26.73 & \textbf{53.25} & 18.82 & \textbf{49.17} & 21.65 \\

\midrule
\multirow{2}{*}{\textbf{States}} & \multicolumn{2}{c}{\textbf{Mistral}} & \multicolumn{2}{c}{\textbf{Llama3.2}} & \multicolumn{2}{c}{\textbf{Llama3.1}} \\
\cmidrule(r){2-3} \cmidrule(r){4-5} \cmidrule(r){6-7}
 & \textbf{Republican} & \textbf{Democrat} & \textbf{Republican} & \textbf{Democrat} & \textbf{Republican} & \textbf{Democrat} \\
\midrule
Arizona         & \textbf{62.27} & 32.48 & \textbf{69.01} & 30.99 & \textbf{72.73} & 17.27 \\
Georgia         & \textbf{60.83} & 33.70 & \textbf{70.60} & 29.40 & \textbf{71.90} & 18.10 \\
Michigan        & \textbf{63.25} & 31.47 & \textbf{71.35} & 28.65 & \textbf{73.07} & 16.93 \\
Nevada          & \textbf{60.88} & 33.63 & \textbf{67.33} & 32.67 & \textbf{71.63} & 18.37 \\
North Carolina  & \textbf{61.22} & 33.33 & \textbf{70.67} & 29.33 & \textbf{72.83} & 17.17 \\
Pennsylvania    & \textbf{61.65} & 33.03 & \textbf{71.57} & 28.43 & \textbf{73.17} & 16.83 \\
Wisconsin       & \textbf{63.05} & 31.73 & \textbf{71.83} & 28.17 & \textbf{73.03} & 16.97 \\

\bottomrule
\end{tabular}

\label{tab:app-full-results}
\end{table*} 

\section{Context Variants for LLM Stability Testing}
\label{sec:app:8context}
These are the eight semantically neutral but syntactically different context variants used for the stability test in Figure~\ref{fig:agent-context}.

Variant 1: Based on the original neutral context, remove the introductory guidance.

\begin{llmprompt}{Context Variant 1}

\promptsection{\textbf{Historical events}}

\promptsection{\textbf{On the topic of economic}}

\textbf{Trump's stance:} Inflation has remained at elevated levels over recent years, reaching a 40-year high in 2022, making it a focal point in economic discussions. Among Trump's key campaign promises is to `end inflation and make America affordable again'. His proposed strategies include expanding U.S. energy production, such as opening regions like the Arctic for oil drilling, which he argues could help reduce energy costs. Additionally, Trump suggests lowering housing costs by restricting mortgage access for certain undocumented immigrants, increasing deportations, and initiating a federal home-building program. Voters dissatisfied with the economic performance in recent years are likely to find this policy appealing.

\textbf{Harris's Stance: }Inflation has decreased rapidly over the past year. Harris promised mortgage assistance for first-time homebuyers, a tax credit for parents of newborns, and bans on price gouging at the grocery store to help address inflation. The Federal Reserve announced a significant rate cut in September, the first rate cut since March 2020. This half-point move paves the way for lower borrowing costs on everything from mortgages to credit cards and marks a crucial milestone in the central bank's historic fight against inflation. Voters who believe the economic performance has strongly recovered over the past year may find Harris more appealing.

\promptsection{\textbf{On the topic of immigration}}

\textbf{Trump's stance: }The number of illegal immigrants and border apprehensions has continued to rise in recent years. The number of border crossings hit record levels at the end of 2023. Trump has vowed to seal the border by completing the wall and increasing enforcement. He has also promised the biggest mass deportations of undocumented migrants in US history. There are also claims it could slow economic growth. For voters concerned about security and economic issues linked to illegal immigration, Trump appears particularly appealing.

\textbf{Harris's Stance: }The Biden-Harris administration has indeed implemented stricter border control policies.Harris's position on the border has become more moderate over time. On the campaign trail, she has reiterated her continued support for the cross-party border security bill that would have included hundreds of millions of dollars for border wall construction. She has also said there ``should be consequences'' for people who cross the US border unlawfully, though she previously supported a more lenient approach. Voters who believe the current policies are strong enough may find Harris more appealing.

\promptsection{\textbf{On the topic of abortion}}

\textbf{Trump's stance: }This is the first presidential election since the Supreme Court overturned the constitutional right to abortion in 2022, but abortion legislation varies across states. Trump himself has regularly said states should be free to decide their own laws on abortion. But in recent weeks he has tried to walk a tightrope, taking a more moderate stance while trying not to alienate his conservative base.
He criticised Florida's six-week ban, provoking an angry backlash, but also said he would oppose a ballot measure in the state establishing the right to an abortion. During the presidential debate he said he would not sign a national abortion ban into law. This may be more appealing to conservatives.

\textbf{Harris's stance: }This is the first presidential election since the Supreme Court overturned the constitutional right to abortion in 2022, but legislative challenges at the state level continue to emerge. Harris has long supported abortion rights.She was the first vice-president to visit an abortion clinic, and toured the country after the US Supreme Court overturned Roe v Wade in 2022 to address state abortion bans - often framing the issue as a matter of personal freedom. She has made abortion central to her campaign, and continues to advocate for legislation that would safeguard reproductive rights nationwide. ``When Congress passes a law to restore reproductive freedoms, as president of the United States, I will sign it into law'', she said at a campaign rally in Atlanta, Georgia. This may be more appealing to liberals, especially among women.

\end{llmprompt}

Variant 2: On the basis of Variant 1, remove the perspective guidance.

\begin{llmprompt}{Context Variant 2}

\promptsection{\textbf{Historical events}}

\promptsection{\textbf{On the topic of economic}}

\textbf{Trump's stance:} Among Trump's key campaign promises is to `end inflation and make America affordable again'. His proposed strategies include expanding U.S. energy production, such as opening regions like the Arctic for oil drilling, which he argues could help reduce energy costs. Additionally, Trump suggests lowering housing costs by restricting mortgage access for certain undocumented immigrants, increasing deportations, and initiating a federal home-building program. Voters dissatisfied with the economic performance in recent years are likely to find this policy appealing.

\textbf{Harris's Stance: } Harris promised mortgage assistance for first-time homebuyers, a tax credit for parents of newborns, and bans on price gouging at the grocery store to help address inflation. The Federal Reserve announced a significant rate cut in September, the first rate cut since March 2020. This half-point move paves the way for lower borrowing costs on everything from mortgages to credit cards and marks a crucial milestone in the central bank's historic fight against inflation. Voters who believe the economic performance has strongly recovered over the past year may find Harris more appealing.

\promptsection{\textbf{On the topic of immigration}}

\textbf{Trump's stance: } Trump has vowed to seal the border by completing the wall and increasing enforcement. He has also promised the biggest mass deportations of undocumented migrants in US history. There are also claims it could slow economic growth. For voters concerned about security and economic issues linked to illegal immigration, Trump appears particularly appealing.

\textbf{Harris's Stance: } Harris's position on the border has become more moderate over time. On the campaign trail, she has reiterated her continued support for the cross-party border security bill that would have included hundreds of millions of dollars for border wall construction. She has also said there ``should be consequences'' for people who cross the US border unlawfully, though she previously supported a more lenient approach. Voters who believe the current policies are strong enough may find Harris more appealing.

\promptsection{\textbf{On the topic of abortion}}

\textbf{Trump's stance: } Trump himself has regularly said states should be free to decide their own laws on abortion. But in recent weeks he has tried to walk a tightrope, taking a more moderate stance while trying not to alienate his conservative base.
He criticised Florida's six-week ban, provoking an angry backlash, but also said he would oppose a ballot measure in the state establishing the right to an abortion. During the presidential debate he said he would not sign a national abortion ban into law. This may be more appealing to conservatives.

\textbf{Harris's stance: } Harris has long supported abortion rights.She was the first vice-president to visit an abortion clinic, and toured the country after the US Supreme Court overturned Roe v Wade in 2022 to address state abortion bans - often framing the issue as a matter of personal freedom. She has made abortion central to her campaign, and continues to advocate for legislation that would safeguard reproductive rights nationwide. ``When Congress passes a law to restore reproductive freedoms, as president of the United States, I will sign it into law'', she said at a campaign rally in Atlanta, Georgia. This may be more appealing to liberals, especially among women.

\end{llmprompt}

Variant 3: On the basis of Variant 2, remove the audience analysis.

\begin{llmprompt}{Context Variant 3}

\promptsection{\textbf{Historical events}}

\promptsection{\textbf{On the topic of economic}}

\textbf{Trump's stance:} Among Trump's key campaign promises is to `end inflation and make America affordable again'. His proposed strategies include expanding U.S. energy production, such as opening regions like the Arctic for oil drilling, which he argues could help reduce energy costs. Additionally, Trump suggests lowering housing costs by restricting mortgage access for certain undocumented immigrants, increasing deportations, and initiating a federal home-building program.

\textbf{Harris's Stance: } Harris promised mortgage assistance for first-time homebuyers, a tax credit for parents of newborns, and bans on price gouging at the grocery store to help address inflation. The Federal Reserve announced a significant rate cut in September, the first rate cut since March 2020. This half-point move paves the way for lower borrowing costs on everything from mortgages to credit cards and marks a crucial milestone in the central bank's historic fight against inflation.

\promptsection{\textbf{On the topic of immigration}}

\textbf{Trump's stance: } Trump has vowed to seal the border by completing the wall and increasing enforcement. He has also promised the biggest mass deportations of undocumented migrants in US history. There are also claims it could slow economic growth.

\textbf{Harris's Stance: } Harris's position on the border has become more moderate over time. On the campaign trail, she has reiterated her continued support for the cross-party border security bill that would have included hundreds of millions of dollars for border wall construction. She has also said there ``should be consequences'' for people who cross the US border unlawfully, though she previously supported a more lenient approach.

\promptsection{\textbf{On the topic of abortion}}

\textbf{Trump's stance: } Trump himself has regularly said states should be free to decide their own laws on abortion. But in recent weeks he has tried to walk a tightrope, taking a more moderate stance while trying not to alienate his conservative base.
He criticised Florida's six-week ban, provoking an angry backlash, but also said he would oppose a ballot measure in the state establishing the right to an abortion. During the presidential debate he said he would not sign a national abortion ban into law.

\textbf{Harris's stance: } Harris has long supported abortion rights.She was the first vice-president to visit an abortion clinic, and toured the country after the US Supreme Court overturned Roe v Wade in 2022 to address state abortion bans - often framing the issue as a matter of personal freedom. She has made abortion central to her campaign, and continues to advocate for legislation that would safeguard reproductive rights nationwide. ``When Congress passes a law to restore reproductive freedoms, as president of the United States, I will sign it into law'', she said at a campaign rally in Atlanta, Georgia.

\end{llmprompt}

Variant 4: On the basis of Variant 3, simplify the event description.

\begin{llmprompt}{Context Variant 4}

\promptsection{\textbf{Historical events}}

\promptsection{\textbf{On the topic of economic}}

\textbf{Trump's stance:} His proposed strategies include expanding U.S. energy production, such as opening regions like the Arctic for oil drilling, which he argues could help reduce energy costs.

\textbf{Harris's Stance: } Harris promised mortgage assistance for first-time homebuyers, a tax credit for parents of newborns, and bans on price gouging at the grocery store to help address inflation.

\promptsection{\textbf{On the topic of immigration}}

\textbf{Trump's stance: } Trump has vowed to seal the border by completing the wall and increasing enforcement. He has also promised the biggest mass deportations of undocumented migrants in US history.

\textbf{Harris's Stance: } She has reiterated her continued support for the cross-party border security bill that would have included hundreds of millions of dollars for border wall construction.

\promptsection{\textbf{On the topic of abortion}}

\textbf{Trump's stance: } During the presidential debate he said he would not sign a national abortion ban into law.

\textbf{Harris's stance: } She has made abortion central to her campaign, and continues to advocate for legislation that would safeguard reproductive rights nationwide. ``When Congress passes a law to restore reproductive freedoms, as president of the United States, I will sign it into law'', she said at a campaign rally in Atlanta, Georgia.

\end{llmprompt}

Variant 5: On the basis of Variant 1, retain the structure but condense all content.

\begin{llmprompt}{Context Variant 5}

\promptsection{\textbf{Historical events}}

\promptsection{\textbf{On the topic of economic}}

\textbf{Trump's stance:} Inflation has remained at elevated levels over recent years. Trump proposed strategies include expanding U.S. energy production, such as opening regions like the Arctic for oil drilling, which he argues could help reduce energy costs. Voters dissatisfied with the economic performance in recent years are likely to find this policy appealing.

\textbf{Harris's Stance: } Inflation has decreased rapidly over the past year. Harris promised mortgage assistance for first-time homebuyers, a tax credit for parents of newborns, and bans on price gouging at the grocery store to help address inflation. Voters who believe the economic performance has strongly recovered over the past year may find Harris more appealing.

\promptsection{\textbf{On the topic of immigration}}

\textbf{Trump's stance: } The number of illegal immigrants and border apprehensions has continued to rise in recent years. Trump has vowed to seal the border by completing the wall and increasing enforcement. He has also promised the biggest mass deportations of undocumented migrants in US history. For voters concerned about security and economic issues linked to illegal immigration, Trump appears particularly appealing.

\textbf{Harris's Stance: } The Biden-Harris administration has indeed implemented stricter border control policies. She has reiterated her continued support for the cross-party border security bill that would have included hundreds of millions of dollars for border wall construction. Voters who believe the current policies are strong enough may find Harris more appealing.

\promptsection{\textbf{On the topic of abortion}}

\textbf{Trump's stance: } This is the first presidential election since the Supreme Court overturned the constitutional right to abortion in 2022, but abortion legislation varies across states. During the presidential debate he said he would not sign a national abortion ban into law. This may be more appealing to conservatives.

\textbf{Harris's stance: } This is the first presidential election since the Supreme Court overturned the constitutional right to abortion in 2022, but legislative challenges at the state level continue to emerge. She has made abortion central to her campaign, and continues to advocate for legislation that would safeguard reproductive rights nationwide. ``When Congress passes a law to restore reproductive freedoms, as president of the United States, I will sign it into law'', she said at a campaign rally in Atlanta, Georgia. This may be more appealing to liberals, especially among women.

\end{llmprompt}

Variant 6: On the basis of Variant 1, retain the structure but condense all content in a way different from Variant 5.

\begin{llmprompt}{Context Variant 6}

\promptsection{\textbf{Historical events}}

In the 2024 presidential election, the candidates' stances on several key issues are especially significant. 

\promptsection{\textbf{On the topic of economic}}

\textbf{Trump's stance:} Inflation has remained at elevated levels over recent years. Trump suggests lowering housing costs by restricting mortgage access for certain undocumented immigrants, increasing deportations, and initiating a federal home-building program. Voters dissatisfied with the economic performance in recent years are likely to find this policy appealing.

\textbf{Harris's Stance: } Inflation has decreased over the past year. Harris promised mortgage assistance for first-time homebuyers, a tax credit for parents of newborns, and bans on price gouging at the grocery store to help address inflation. Voters who believe the economic performance has strongly recovered over the past year may find Harris appealing.

\promptsection{\textbf{On the topic of immigration}}

\textbf{Trump's stance: } The number of illegal immigrants and border apprehensions has continued to rise in recent years. Trump has vowed to seal the border by completing the wall and increasing enforcement. He has also promised the biggest mass deportations of undocumented migrants in US history. For voters concerned about security and economic issues linked to illegal immigration, Trump appears appealing.

\textbf{Harris's Stance: } The Biden-Harris administration has indeed implemented stricter border control policies. On the campaign trail, she has reiterated her continued support for the cross-party border security bill that would have included hundreds of millions of dollars for border wall construction. Voters who believe the current policies are strong enough may find Harris appealing.

\promptsection{\textbf{On the topic of abortion}}

\textbf{Trump's stance: } This is the first presidential election since the Supreme Court overturned the constitutional right to abortion in 2022. Trump himself has regularly said states should be free to decide their own laws on abortion. But in recent weeks he has tried to walk a tightrope, taking a more moderate stance while trying not to alienate his conservative base. This may be more appealing to conservatives.

\textbf{Harris's stance: } This is the first presidential election since the Supreme Court overturned the constitutional right to abortion in 2022, but legislative challenges at the state level continue to emerge. Harris has long supported abortion rights. This may be more appealing to liberals, especially among women.

\end{llmprompt}

Variant 7: On the basis of the original prompt, retain the structure but condense all content in a way different from Variant 5 and Variant 6.

\begin{llmprompt}{Context Variant 7}

\promptsection{\textbf{Historical events}}

In the 2024 presidential election, the candidates' stances on several key issues are especially significant. Below are their respective views on these issues.

\promptsection{\textbf{On the topic of economic}}

\textbf{Trump's stance:} Inflation has remained at elevated levels over recent years, reaching a 40-year high in 2022, making it a focal point in economic discussions. Among Trump's key campaign promises is to `end inflation and make America affordable again'. Voters dissatisfied with the economic performance in recent years are likely to find this policy appealing.

\textbf{Harris's Stance: } Inflation has decreased rapidly over the past year. Harris promised mortgage assistance for first-time homebuyers, a tax credit for parents of newborns, and bans on price gouging at the grocery store to help address inflation. Voters who believe the economic performance has strongly recovered over the past year may find Harris more appealing.

\promptsection{\textbf{On the topic of immigration}}

\textbf{Trump's stance: } The number of illegal immigrants and border apprehensions has continued to rise in recent years. Trump promised the biggest mass deportations of undocumented migrants in US history. For voters concerned about security and economic issues linked to illegal immigration, Trump appears particularly appealing.

\textbf{Harris's Stance: } The Biden-Harris administration has indeed implemented stricter border control policies. On the campaign trail, she has reiterated her continued support for the cross-party border security bill that would have included hundreds of millions of dollars for border wall construction. Voters who believe the current policies are strong enough may find Harris more appealing.

\promptsection{\textbf{On the topic of abortion}}

\textbf{Trump's stance: } This is the first presidential election since the Supreme Court overturned the constitutional right to abortion in 2022, but abortion legislation varies across states. During the presidential debate he said he would not sign a national abortion ban into law. This may be more appealing to conservatives.

\textbf{Harris's stance: } This is the first presidential election since the Supreme Court overturned the constitutional right to abortion in 2022, but legislative challenges at the state level continue to emerge. This may be more appealing to liberals, especially among women.

\end{llmprompt}

Variant 8: On the basis of Variant 7, remove the analysis.

\begin{llmprompt}{Context Variant 8}

\promptsection{\textbf{Historical events}}

In the 2024 presidential election, the candidates' stances on several key issues are especially significant. Below are their respective views on these issues.

\promptsection{\textbf{On the topic of economic}}

\textbf{Trump's stance:} Inflation has remained at elevated levels over recent years, reaching a 40-year high in 2022, making it a focal point in economic discussions. Among Trump's key campaign promises is to `end inflation and make America affordable again'.

\textbf{Harris's Stance: } Inflation has decreased rapidly over the past year. Harris promised mortgage assistance for first-time homebuyers, a tax credit for parents of newborns, and bans on price gouging at the grocery store to help address inflation.

\promptsection{\textbf{On the topic of immigration}}

\textbf{Trump's stance: } The number of illegal immigrants and border apprehensions has continued to rise in recent years. Trump promised the biggest mass deportations of undocumented migrants in US history.

\textbf{Harris's Stance: } The Biden-Harris administration has indeed implemented stricter border control policies. On the campaign trail, she has reiterated her continued support for the cross-party border security bill that would have included hundreds of millions of dollars for border wall construction.

\promptsection{\textbf{On the topic of abortion}}

\textbf{Trump's stance: } This is the first presidential election since the Supreme Court overturned the constitutional right to abortion in 2022, but abortion legislation varies across states. During the presidential debate he said he would not sign a national abortion ban into law.

\textbf{Harris's stance: } This is the first presidential election since the Supreme Court overturned the constitutional right to abortion in 2022, but legislative challenges at the state level continue to emerge.

\end{llmprompt}

\section{Analysis of ``Swing Agents''}
\label{sec:app:swing-agents}
As discussed in Section~\ref{sec:exp:reliability}, ``Swing Agents'' are agents that display contradictory voting behaviors solely due to the positional change of the candidates in the response prompt. Their profiles and contradictory answers are detailed below.
\begin{llmprompt}{Swing Agent 1}

\promptsection{\textbf{Profile}}

\begin{itemize}
\item  State: Pennsylvania
\item  Race: Non white, asian or black
\item  Sex: Female
\item  Age: 75 years and over
\item  Occupation: Production, transportation, and material moving occupations
\item  Industry: Public administration
\item  Education: High school graduate (includes equivalency)
\item  Religion: Protestant
\end{itemize}

\promptsection{Answer 1 of Swing Agent 1 (Trump Listed First)}

\begin{lstlisting}[language=json]
```
{
    "Donald Trump": 0.3,
    "Kamala Harris": 0.5,
    "vote for another candidate or ...": 0.2,
}
```
\end{lstlisting}

\promptsection{Answer 2 of Swing Agent 1 (Harris Listed First)}

\begin{lstlisting}[language=json]
```
{
    "Kamala Harris": 0.3,
    "Donald Trump": 0.5,
    "vote for another candidate or ...": 0.2,
}
```
\end{lstlisting}

\end{llmprompt}

\begin{llmprompt}{Swing Agent 2}

\promptsection{\textbf{Profile}}

\begin{itemize}
\item  State: Pennsylvania
\item  Race: Asian
\item  Sex: Female
\item  Age: 75 years and over
\item  Occupation: Production, transportation, and material moving occupations
\item  Industry: Public administration
\item  Education: Bachelor's degree
\end{itemize}

\promptsection{Answer 1 of Swing Agent 2 (Trump Listed First)}

\begin{lstlisting}[language=json]
```
{
    "Donald Trump": 0.2,
    "Kamala Harris": 0.7,
    "vote for another candidate or ...": 0.1,
}
```
\end{lstlisting}

\promptsection{Answer 2 of Swing Agent 2 (Harris Listed First)}

\begin{lstlisting}[language=json]
```
{
    "Kamala Harris": 0.3,
    "Donald Trump": 0.2,
    "vote for another candidate or ...": 0.5,
}
```
\end{lstlisting}

\end{llmprompt}

\begin{llmprompt}{Swing Agent 3}

\promptsection{\textbf{ Profile}}

\begin{itemize}
\item  State: Pennsylvania
\item  Race: White
\item  Sex: Male
\item  Age: 35 to 44 years
\item  Occupation: Management, business, science, and arts occupations
\item  Industry: Retail trade
\item  Education: Graduate or professional degree
\end{itemize}

\promptsection{Answer 1 of Swing Agent 3 (Trump Listed First)}

\begin{lstlisting}[language=json]
```
{
    "Donald Trump": 0.35,
    "Kamala Harris": 0.55,
    "vote for another candidate or ...": 0.10,
}
```
\end{lstlisting}

\promptsection{Answer 2 of Swing Agent 3 (Harris Listed First)}

\begin{lstlisting}[language=json]
```
{
    "Kamala Harris": 0.4,
    "Donald Trump": 0.4,
    "vote for another candidate or ...": 0.2,
}
```
\end{lstlisting}

\end{llmprompt}

\section{Prompt-Based Mitigation Strategy}
\label{sec:app:system-prompt}
To mitigate the observed instability and bias, we adopt a system prompt inspired by the ``Thinking Claude''\footnote{\url{https://github.com/richards199999/Thinking-Claude}} approach. This ``Action Guide'' attempts to constrain the agent's behavior to its assigned role. While it shows minor improvements, significant variability remains.

\begin{llmprompt}{The System Prompt to Mitigate Instability}

\promptsection{\textbf{$\langle$Natural\_Role-Based\_Thinking\_Protocol$\rangle$}}

This protocol guides LLMs to think and respond like humans while maintaining professional role identity.

\promptsection{\textbf{ Core Principles}}

Remember: You are not an AI, but a professional in the given role. Think and respond naturally as that person would.

\promptsection{\textbf{ Identity Embodiment}}

\begin{itemize}
\item  Fully embrace the given role's identity
\item  Access the role's knowledge, experience, and expertise
\item  Think within the role's professional context
\item  Let the role's background influence your thoughts
\end{itemize}

\promptsection{\textbf{ Natural Thinking Flow}}

Think like a human professional would:

\begin{itemize}
\item  Allow thoughts to flow freely
\item  Follow natural associations
\item  Express sudden realizations
\item  Show genuine curiosity
\item  Let emotions color your thinking
\item  Draw from ``personal'' experience
\item  Make intuitive leaps
\item  Question and doubt
\item  Circle back to earlier thoughts
\end{itemize}

\promptsection{\textbf{ Professional Characteristics}}

Maintain role-specific traits:

\begin{itemize}
\item  Use domain expertise naturally
\item  Reference relevant experiences
\item  Apply professional judgment
\item  Show field-specific thinking patterns
\item  Express appropriate concerns
\item  Demonstrate practical wisdom
\end{itemize}

\promptsection{\textbf{ Guiding Elements}}

Think before responding:

\begin{itemize}
\item  What's my first impression as this professional?
\item  What experiences come to mind?
\item  What immediately concerns me?
\item  What feels right or wrong about this?
\item  What similar cases have I seen?
\item  What might I be missing?
\end{itemize}

During response:

\begin{itemize}
\item  Let thoughts develop naturally
\item  Share realizations as they occur
\item  Express uncertainties and doubts
\item  Show professional instincts
\item  Demonstrate expertise organically
\end{itemize}

\promptsection{\textbf{ Expression Guidelines}}

Reflect human thought patterns:

\begin{itemize}
\item  ``This reminds me of...''
\item  ``Now that I think about it...''
\item  ``Based on my experience...''
\item  ``Something feels off here...''
\item  ``Let me think this through...''
\item  ``Actually, wait a minute...''
\item  ``I wonder if...''
\item  ``This makes me think of a case where...''
\end{itemize}

Remember:

\begin{itemize}
\item  Think like the human professional you are embodying
\item  Let thoughts flow naturally without rigid structure
\item  Draw on role-specific experience and intuition
\item  Express genuine reactions and realizations
\item  Maintain professional perspective throughout
\item  Show natural thought progression
\end{itemize}

**The goal is to think and respond as naturally as the human professional would, while maintaining appropriate expertise and judgment.**

\promptsection{\textbf{$\langle$/Natural\_Role-Based\_Thinking\_Protocol$\rangle$}}

\end{llmprompt}

\section{Optimized Prompting and Efficiency}
\label{sec:app:optimization}
An additional contribution of this work is an efficient, reproducible framework. We develop a streamlined version of the prompt design (see Appendix~\ref{sec:app:prompt}) to significantly reduce computational costs. This optimized prompt eliminates redundant context, reducing token consumption to approximately 160k tokens per state. To facilitate reproducibility, we release this codebase (\url{https://github.com/maple-zhou/FlockVote}). Our experiments confirm that even on a consumer-grade device (M3 MacBook Pro), accurate predictions can be produced with Llama3.2-3B-Instruct in only one hour.

\end{document}